\documentclass[12pt,aps,pre,amsmath,amssymb]{revtex4}
\usepackage{xcolor}
\usepackage{setspace}
\usepackage{graphicx}
\newcommand{\citen}[1]{%
  \begingroup
    \romannumeral-`\x 
    \setcitestyle{numbers}%
    \cite{#1}%
  \endgroup
}
\usepackage{mathrsfs}

\DeclareSymbolFont{matha}{OML}{txmi}{m}{it}
\DeclareMathSymbol{\varw}{\mathord}{matha}{119}   
\newcommand{\kB}{k_{\rm B}} 
\newcommand{\lb}{l_{\rm B}} 

\newcommand{\bra}[1]{\langle #1 |} 
\newcommand{\ket}[1]{| #1 \rangle} 

\renewcommand{\i}{{\rm i}}  
\newcommand{\np}{{n_{\rm p}}}  
\newcommand{\nc}{{n_{\rm c}}}   
\newcommand{\ns}{{n_{\rm s}}}   

\newcommand{\zs}{z_{\rm s}} 
\newcommand{\zc}{z_{\rm c}} 

\newcommand{\nw}{n_{\rm w}}   

\newcommand{\R}{{\bf R}} 
\newcommand{\rr}{{\bf r}}   
\newcommand{\kk}{{\bf k}} 
\newcommand{\kzero}{{\bf 0}} 

\newcommand{\T}{\mathscr{T}}    
\newcommand{\U}{\mathscr{U}}   
\newcommand{\Hp}{{\cal H}_{\rm p}}      

\newcommand{\Z}{{\cal Z}} 
\newcommand{\Zp}{{\cal Z}_{\rm p}} 
\newcommand{\Qp}{{\cal Q}_{\rm p}} 
\newcommand{\Qs}{{\cal Q}_{\rm s}} 
\newcommand{\Qc}{{\cal Q}_{\rm c}} 

\newcommand{\fion}{f_{\rm ion}} 
\newcommand{\fzero}{f_{\rm 0}} 

\newcommand{\w}{\varw} 
 
\newcommand{\avg}[1]{\left\langle #1 \right\rangle} 
\newcommand{\DD}{\mathscr{D}} 

\newcommand{\gmm}[1]{{ g^{\rm mm}_{#1} }}
\newcommand{\gmc}[1]{ { g^{\rm mc}_{#1}}}
\newcommand{\gcc}[1]{{ g^{\rm cc}_{#1}}}

\newcommand{\LGdd}{\bar{g}}   
\newcommand{\LGss}{\bar{\xi}}   
\newcommand{\LGds}{\bar{\zeta}}   

\begin{document}

$\null$
\hfill {\bf December 20, 2024}
\vskip 0.3in

\begin{center}

{\Large\bf 
Electrostatics of Salt-Dependent Reentrant}\\

\vskip 0.3cm

{\Large\bf 
Phase Behaviors Highlights Diverse Roles of}\\

\vskip 0.3cm

{\Large\bf 
ATP in Biomolecular Condensates}\\

\vskip .5in
{\bf Yi-Hsuan L{\footnotesize{\bf{IN}}}},$^{1,2,\dagger,\S}$
{\bf Tae Hun K{\footnotesize{\bf{IM}}}},$^{1,2,3,4,\ddagger,\S}$
{\bf Suman D{\footnotesize{\bf{AS}}}},$^{1,5,\S}$
{\bf Tanmoy P{\footnotesize{\bf{AL}}}},$^{1}$\\
{\bf Jonas W{\footnotesize{\bf{ESS\'EN}}}},$^{1}$
{\bf Atul Kaushik R{\footnotesize{\bf{ANGADURAI}}}},$^{1,2,3,4}$
{\bf Lewis E. K{\footnotesize{\bf{AY}}}},$^{1,2,3,4}$
\\
{\bf Julie D. F{\footnotesize{\bf{ORMAN}}}-K{\footnotesize{\bf{AY}}}}$^{2,1}$
and
{\bf Hue Sun C{\footnotesize{\bf{HAN}}}}$^{1,*}$

$\null$

$^1$Department of Biochemistry,
University of Toronto, Toronto, Ontario M5S 1A8, Canada\\
$^2$Molecular Medicine, Hospital for Sick Children, Toronto, 
Ontario M5G 0A4, Canada\\
$^3$Department of Molecular Genetics,
University of Toronto,\\ Toronto, Ontario M5S 1A8, Canada\\
$^4$Department of Chemistry,
University of Toronto, Toronto, Ontario M5S 3H6, Canada\\
$^5$Department of Chemistry, Gandhi Institute of Technology and
Management, Visakhapatnam, Andhra Pradesh 530045, India\\

%
\end{center}
$\null$\\

\noindent
$^\dagger$Present address: HTuO Biosciences, 1001 West Broadway, Suite 300,
Vancouver, British\\ 
{\phantom{$^\dagger$Present address:\ }} 
Columbia V6H 4B1, Canada.

\vskip 0.1cm

\noindent
$^\ddagger$Present address: Department of Biochemistry, School of Medicine,\\
{\phantom{$^\dagger$Present address:\ }} 
Case Western Reserve University, Cleveland, Ohio 44106, U.S.A. 


\vskip 0.2cm

\noindent
$^\S$Contributed equally.

\vskip 1.3cm

\noindent
$^*$Correspondence information:\\
{\phantom{$^\dagger$}}
Hue Sun C{\footnotesize{HAN}}.$\quad$
E-mail: {\tt huesun.chan@utoronto.ca}\\
{\phantom{$^\dagger$}}
Tel: (416)978-2697; Fax: (416)978-8548\\
{\phantom{$^\dagger$}}
Department of Biochemistry, University of Toronto,
Medical Sciences Building -- 5th Fl.,\\
{\phantom{$^\dagger$}}
1 King's College Circle, Toronto, Ontario M5S 1A8, Canada.\\

\vfill\eject

\noindent
{\large\bf Abstract}\\

\noindent
Liquid-liquid phase separation (LLPS) involving intrinsically disordered
protein regions (IDRs) is a major physical mechanism for biological
membraneless compartmentalization. The multifaceted electrostatic effects in
these biomolecular condensates are exemplified here by experimental
and theoretical investigations of the different salt-
and ATP-dependent LLPSs of an IDR of messenger RNA-regulating
protein Caprin1 and its phosphorylated variant pY-Caprin1, exhibiting, e.g.,
reentrant behaviors in some instances but not others. 
Experimental data are rationalized by
physical modeling using analytical theory, molecular dynamics, and polymer
field-theoretic simulations, indicating that interchain ion bridges
enhance LLPS of polyelectrolytes such as Caprin1 and the high
valency of ATP-magnesium is a significant factor for its colocalization
with the condensed phases, as similar trends are observed for 
other IDRs. 
The electrostatic nature of these features complements
ATP's involvement in $\pi$-related interactions and as an amphiphilic 
hydrotrope, 
underscoring a general role of biomolecular condensates 
in modulating ion concentrations and its functional ramifications.
\\

\vfill\eject

\noindent
{\large\bf Impact Statement}\\

A combination of polymer physics theories consistently rationalizes
comprehensive experimental effects of salt, ATP, phosphorylation, and sequence
charge and arginine/lysine patterns on the
reentrant phase behaviors of intrinsically disordered proteins.

\vfill\eject

\noindent
{\Large\bf Introduction}

Broad-based recent efforts have uncovered
many intriguing features of biomolecular condensates, revealing
and suggesting myriad known and potential biological
functions \cite{cliff2017,BrianJulie2020,rosen2021}.
These assemblies are underpinned substantially, though not exclusively,
by liquid-liquid phase
separation (LLPS) of intrinsically disordered regions (IDRs) as well as
folded domains of proteins and
nucleic acids \cite{Roland2019,TanjaRohitNatChem2022}, while more complex equilibrium and
non-equilibrium mechanisms also
contribute \cite{RosenPappu2017,biochemrev,McKnight2018,CFLee2018,Tjian2019,YHLin2022,Rohit2022,MZ2023,Pappu2023}.

Electrostatics plays major roles in biophysical and biochemical 
processes~\cite{Barry1995,HXZhouReview2018}.
Because of the relatively high compositions of charge residues
in IDRs, 
electrostatics is particularly important for IDR 
LLPS \cite{Nott15,linPRL}, which is often also facilitated by
$\pi$-related interactions \cite{moleculargrammar,robert},
hydrophobicity, hydrogen bonding \cite{Fawzi2019,Cai2022},
and is modulated by temperature \cite{biochemrev,jeetainACS},
hydrostatic pressure \cite{Roland2019,roland20}, osmolytes \cite{Roland2019},
RNA \cite{AlbertiSci2018,BrianTsang2019,Feig2021,Potoyan2022},
salt, pH \cite{IP5-2017}, 
molecular crowding \cite{Spruijt2020,Mukherjee2022,Rakshit2024}
and post-translational modifications
(PTMs) \cite{cliff2017,julie2019,Shewmaker2019,Gladfelter2019}.
Multivalency underlies many aspects of IDR
properties \cite{borg2007,Julie2012,kaw2013,NatPhys,cosb15}.
Here, we focus primarily on how PTM- and salt-modulated multivalent
charge-charge interactions might alter IDR condensate behaviors and their
possible functional ramifications. In general, electrostatic effects on IDR
LLPS \cite{Nott15,IP5-2017,Feig2021,LaiLuhua2021}
are dependent upon their sequence charge
patterns \cite{rohit2013,kings2015,linJML,lin2017,singperry2017,Alan2020,BenSabari2023,Paletal_JPCL2024}.
Intriguingly, some IDRs undergo reentrant
phase separation \cite{Roland2019} or dissolution \cite{Deniz2017}
when temperature,
pressure \cite{Roland2019}, salt \cite{KnowlesNatComm2021,Joan2022},
RNA \cite{Deniz2017,Banerjee2019}, or concentrations of small molecules
such as heparin \cite{Surewicz2020} is varied.
Reentrance, especially when induced by salt and RNA, suggest a
subtle interplay between multivalent sequence-specific charge-charge
interactions and hydrophobic, non-ionic \cite{KnowlesNatComm2021,Joan2022},
cation-$\pi$ \cite{Deniz2017,Banerjee2019}, or
$\pi$-$\pi$ interactions.

An important modulator of biomolecular LLPS is adenosine triphosphate (ATP).
As energy currency, ATP hydrolysis is utilized to synthesize
or break chemical bonds and drive transport to regulate ``active liquid''
properties such as concentration gradients and droplet
sizes \cite{CFLee2018,CFLee2022}. Examples include ATP-driven
assembly of stress granules \cite{parker2016}, splitting of bacterial
biomolecular
condensates \cite{Nollmann2020}, and destabilization of nucleolar
aggregates \cite{Weeks2018}.
ATP can also influence biomolecular LLPS without hydrolysis,
akin to other LLPS promotors or suppressors \cite{HXZhou2018,HXZhou2019}
that are effectively ligands of the
condensate scaffold \cite{Rohit2021}, or through ATP's effect on lowering
free [Mg$^{2+}$] \cite{Beato2019}.
Notably, as an amphiphilic hydrotrope \cite{Seishi2022}
with intracellular concentrations much higher than
that required for an energy source, ATP is also seen to
afford an important function independent of hydrolysis
by solubilizing proteins, preventing LLPS
and destabilizing aggregates, as exemplified by measurements
on several proteins including fused in sarcoma (FUS) \cite{HymanSci2017}.

Subsequent investigations indicate, however, that hydrolysis-independent
[ATP] effects on biomolecular LLPS are neither invariably monotonic for a
given system nor universal across different systems.
For instance, ATP promotes, not
suppresses, LLPS of an IgG1 antibody \cite{Qian2021}, 
basic IDPs \cite{HXZhou2024},
and enhances LLPS of full-length and the C-terminal domain (CTD) of FUS at
low [ATP] but prevents LLPS at high [ATP] \cite{SongBBRC2018}.
The latter reentrant behavior has been surmised to arise from
ATP binding bivalently \cite{SongBBRC2018,SongPLoS2019}
or trivalently \cite{MaSciAdv2022}
to charged residues arginine (R) or lysine (K)
by a combination of cation-$\pi$ and electrostatic interactions, an effect
also seen in the ATP-mediated LLPS of basic IDPs \cite{HXZhou2024}.
A similar scenario
was invoked for the reentrant phase behavior of transactive response
DNA-binding protein of 43 kDa (TDP-43) \cite{SongCommBiol2021}.
Most recently, ATP-mediated assembly-disassembly reentrant behavior
similar to that of FUS CTD was also observed for the RG/RRG-rich IDR motif 
with a positive net charge from the heterogeneous nuclear 
ribonucleoprotein G~\cite{Zhuetal_JMCB2024}.

While $\pi$-related interactions are important for biomolecular
LLPS in general \cite{robert,moleculargrammar} and their
interplay with electrostatics likely underlies
reentrant biomolecular phase behaviors modulated by
RNA \cite{Deniz2017,Banerjee2019} or simple salts \cite{KnowlesNatComm2021},
the degree to which electrostatics alone can, in large measure, rationalize
hydrolysis-independent ATP-modulated biomolecular phase reentrance has
not been sufficiently appreciated. This question deserves attention.
For instance, the suppression of cold-inducible RNA-binding protein
condensation by ATP has been suggested to be
electrostatically driven \cite{Madl2021}. The aforementioned
ATP-modulated reentrant phase behavior of FUS \cite{SongBBRC2018,SongPLoS2019}
is reminiscent of the 236-residue N-terminal IDR
of DEAD-box RNA helicase Ddx4's lack of LLPS at low
[NaCl] ($<15$--$20$ mM), LLPS at higher [NaCl] \cite{kings2020}
and decreasing LLPS propensity when [NaCl] is further
increased \cite{Nott15,linPRL}. Indeed, the finding that FUS CTD
(net charge per residue (NCPR) $=15/156=0.096$) exhibits ATP-dependent reentrant
phase behaviors while the N-terminal domain (NCPR $=3/267=0.011$)
does not \cite{SongPLoS2019} is consistent
with electrostatics-based theory for the difference in
salt-dependent LLPS of polyelectrolytes and
polyampholytes \cite{kings2020} 
and a recent atomic simulation
study of direct and indirect salt effects on LLPS~\cite{MacAinsh2024}.

With this in mind, we seek to delineate the degree to which theories
focusing primarily on electrostatics
can rationalize experimental ATP-related LLPS data on
the 103-residue C-terminal IDR of human cytoplasmic
activation/proliferation-associated protein-1 (Caprin1).
Full-length Caprin1 (709 amino acid residues) is a ubiquitously expressed
phosphoprotein that regulates
stress \cite{Schrader2007,Dawson2011,interactome2021,Gong2022}
and neuronal \cite{Khandjian2012} granules, is necessary for normal
cellular proliferation \cite{Luzio1995,Schrader2005},
and may be essential for long-term memory \cite{Noda2017}.
Caprin1 dysfunction leads to multiple diseases such as
nasopharyngeal carcinoma \cite{Mai2022} as well as language impairment and
autism spectrum disorder \cite{Brain2022}, via, e.g., Caprin1's
modulation of the function of the fragile X mental retardation protein (FMRP)
\cite{Khandjian2012,BrianTsang2019,julie2019}.
The C-terminal 607--709 Caprin1 IDR,
referred to simply as Caprin1 below, is biophysically and
functionally significant: It is sufficient for LLPS in
vitro \cite{julie2019}, important for assembling
stress granules in the cell \cite{Schrader2007,Dawson2011}, and has
a substantial body of
experiments \cite{julie2019,LewisJACS2020,LewisPNAS2021,LewisPNAS2022}
for comparison with theory. Since tyrosine phosphorylations of Caprin1 in
vivo \cite{Skrzypek2015} may regulate translation in neurons \cite{julie2019},
the Caprin1 system is also useful for gaining insights into phosphorregulation 
of biomolecular condensates \cite{Fawzi2017,dignon18,Morgan2020}.

Recent advances in theory and computation enable
modeling of sequence-specific IDR LLPS \cite{linPRL,MacAinsh2024,dignon18,RauscherPomes2017,suman1,joanJPCL2019,joanJCP,lassi,SumanPNAS,koby2020,Mpipi}.
Among the approaches, 
polymer chain models of IDRs are inherently more 
realistic in capturing sequence properties than models without a chain 
description such as patchy particle theory~\cite{HXZhou2018}. For chain models,
all-atom simulation offer a high degree of geometric and energetic 
realism~\cite{RauscherPomes2017} but its high computational cost often 
makes it difficult to achieve sufficient sampling and equilibration 
for the large system sizes that are needed for modeling 
biomolecular LLPS processes~\cite{MacAinsh2024}.
However, even coarse-grained explicit-chain simulation affords 
more realistic geometric and energetic representations than analytical theory,
but analytical theory offers significant advantages in numerical 
tractability \cite{MiMB2023}. For our present purposes,
the analytical rG-RPA formulation \cite{kings2020},
which synthesizes Kuhn-lengh renormalization (renormalized Gaussian, rG)
and random phase approximation (RPA) \cite{linPRL}
to treat both high-net-charge polyelectrolytes
and essentially net-neutral polyampholytes \cite{kings2020},
is particularly well suited for Caprin1
and its phosphorylated variant pY-Caprin1. 
To gain deeper insights into
the pertinent physical principles and to assess possible limitations 
of this analytical approximation, 
we further leverage a
methodological combination of rG-RPA \cite{kings2020}, field-theoretic
simulation (FTS) \cite{joanJPCL2019,Pal2021}, and coarse-grained explicit-chain
molecular dynamics (MD) \cite{dignon18,SumanPNAS}
to better elucidate the effects of salt, phosphorylation, and
ATP on LLPS of Caprin1 and pY-Caprin1.
\\


\noindent
{\Large\bf Results}

\noindent
{\bf Overview of key observations from complementary approaches}\\
The complementary nature of our multiple methodologies allows us to 
focus sharply on the electrostatic aspects of hydrolysis-independent role 
of ATP in biomolecular condensation by comparing ATP's effects with 
those of simple salt. Here, Caprin1 and pY-Caprin1 are modeled 
minimally as heteropolymers of charged and neutral beads in rG-RPA and FTS.
ATP and ATP-Mg are modeled as 
simple salts (single-bead ions) in rG-RPA whereas they are modeled 
with more structural complexity as short charged polymers (multiple-bead 
chains) in FTS, though the latter models are still highly coarse-grained. 
Despite this modeling difference, rG-RPA and FTS both rationalize
experimentally observed ATP- and NaCl-modulated reentrant LLPS of 
Caprin1 and a lack of a similar reentrance for pY-Caprin1 as well as
a prominent colocalization of ATP with the Caprin1 condensate. Consistently,
the same contrasting trends in the effect of NaCl on Caprin1 
and pY-Caprin1 are also seen in our coarse-grained MD simulations,
though polymer field theories tend to overestimate LLPS 
propensity \cite{Shen2017}.
The robustness of the theoretical trends across different modeling 
platforms underscores electrostatics as a significant component in 
the diverse roles of ATP in the context of its well-documented
ability to modulate biomolecular
LLPS via hydrophobic and $\pi$-related 
effects \cite{HymanSci2017,SongPLoS2019,HXZhou2024}.
Analyses of these other nonelectrostatic effects are mostly beyond
the scope of the present work but their impact are nevertheless 
illustrated by the Flory-Huggins interactions augmented to rG-RPA 
to quantitatively account for experimental data
and our MD simulation of the arginine-to-lysine Caprin1 mutants.
These findings are detailed below.
\\

\noindent
{\bf Physical theories of Caprin1 and phosphorylated Caprin1 LLPSs
as those of polyelectrolytes and polyampholytes}\\
The 103-residue Caprin1 is a highly charged IDR with
19 charged residues [Fig.~1a and {\it Supplementary Information -- Appendix} 
({\it SI Appendix}), SI Appendix-figure 1]: 15 R, 1 K,
and 3 aspartic acids (D); fraction of charged residues $=19/103=0.184$ and
NCPR $=13/103=0.126$. With a substantial positive net
charge, Caprin1's phase behaviors are markedly different from those of
polyampholytic IDRs with nearly zero net charge such as Ddx4 to which early
sequence-specific LLPS theories were targeted \cite{Nott15,linPRL}.
Instead, Caprin1 behaves like chemically synthesized
polyelectrolytes \cite{Dobrynin2005}. In contrast, when most or
all of the 7 tyrosines (Y) in the Caprin1 IDR are phosphorylated (pY),
negative charges are added to produce
a near-net-neutral polyampholyte. Mass spectrometry indicates
that the experimental sample of highly phosphorylated Caprin1 consists 
mainly of a mixture of IDRs with 6 or 7 phosphorylations
(SI Appendix-figure 2).
We refer to this experimental sample as pY-Caprin1 below.
For simplicity, we use only the Caprin1 IDR with 7 pYs
to model the behavior of this
experimental sample in our theoretical/computational 
formulations, partly to avoid the combinatoric complexity of
sequences with 5 or 6 pYs.
Accordingly, since the charge of a pY
is $\approx -2$ at the experimental pH $=7.4$,
$-14$ charges are added to Caprin1 for
our model pY-Caprin1, resulting in a polyampholyte
with a very small NCPR $=-1/103=-0.00971$ (Fig.~1b).
Both the experimental pY-Caprin1
(NCPR $\approx \pm 1/103=\pm 0.00971$) and model pY-Caprin1
are expected to exhibit phase properties similar to other polyampholytic IDRs.

While sequence-specific RPA has been applied successfully to model
electrostatic effects on the LLPSs of various polyampholytic IDRs
\cite{linPRL,lin2017,biochemrev,SumanPNAS,Wessen2021},
RPA is less appropriate for polyelectrolytes with large
NCPR \cite{Mahdi2000,Ermoshkin2003,Orkoulas2003} because
of its treatment of polymers as ideal Gaussian chains \cite{MuthuMacro2017}.
Traditionally, theories for polyelectrolytes tackle their peculiar
conformations by various renormalized blob
constructs \cite{Mahdi2000,Dobrynin2005},
two-loop polymer field theory \cite{Muthu1996},
modified thermodynamic perturbation theory \cite{Budkov2015}, and
renormalized Gaussian fluctuation (RGF) theory \cite{Shen2017,Shen2018},
among others. As such, these formulations are
mostly designed for homopolymers,
making it difficult to apply directly
to heteropolymeric biopolymers.
In order to analyze Caprin1 and pY-Caprin1 LLPSs, we utilize 
rG-RPA \cite{kings2020}, which combines Gaussian chains of effective 
(renormalized) Kuhn length with the key idea of RGF \cite{kings2015}.
\\


\noindent
{\bf Phase properties predicted by rG-RPA theory for Caprin1 and pY-Caprin1 
with monovalent counterions and salt are in agreement
with experiment}\\
Fig.~1c and d show that the salt- and temperature ($T$)-dependent phase
diagrams predicted by rG-RPA with an augmented
Flory-Huggins (FH) mean-field 
$\chi(T) = \epsilon_h / T^* + \epsilon_s$
parameter for nonelectrostatic interactions, where $\epsilon_h$
and $\epsilon_s$ are the enthalpic and entropic contributions, respectively,
and $T^*$ is reduced temperature \cite{linPRL,kings2020}
(Eq.~10 of ref.~\cite{linPRL} and
``rG-RPA+FH'' theory 
in {\it SI Appendix}),
are in reasonable agreement with experiment
using bulk [Caprin1] 
(initial overall concentration)
$\approx 200$ $\mu$M.
(Concentrations are provided in molarity and also as mass density in
Fig.~1 and subsequent figures).
The rG-RPA+FH results in Fig.~1c indicate that
(i) Caprin1 undergoes LLPS below 20$^\circ$C with 100 mM
NaCl, and that (ii) LLPS propensity, quantified by the upper critical
solution temperature (UCST), increases with [NaCl].
These predictions are consistent with experimental data,
including the observation that Caprin1 does not phase
separate at room temperature without salt, ATP, RNA or other proteins,
though Caprin1 LLPS can be triggered by adding
wildtype (WT) and phosphorylated FMRP
and/or RNA (overall [Caprin1] $\gtrsim 10$ $\mu$M) \cite{julie2019},
NaCl \cite{LewisJACS2020}, or ATP (overall [Caprin1] $=400$ $\mu$M)
\cite{LewisPNAS2021}.
The trend here is also in line with
other theories of polyelectrolytes \cite{Shen2018}.
In contrast, rG-RPA+FH results in Fig.~1d for pY-Caprin1 shows
decreasing LLPS propensity with increasing [NaCl],
consistent with experimental data
and the expected salt dependence of LLPS of nearly net-neutral
polyampholytic IDRs such as Ddx4 \cite{linPRL}.

Interestingly, the decrease in some of the condensed-phase [pY-Caprin1]s
with decreasing $T$ (orange and green symbols for
$\lesssim 20^\circ$C in Fig.~1d trending toward slightly lower [pY-Caprin1])
may suggest a 
hydrophobicity-driven
lower critical solution 
temperature (LCST)-like reduction of LLPS propensity as temperature 
approaches $\sim 0^\circ$C 
as in cold denaturation of globular
proteins \cite{biochemrev,jeetainACS} though the hypothetical LCST is
below $0^\circ$C and therefore not experimentally accessible. 
If that is the case, the LLPS region would resemble
those with both an UCST and a LCST \cite{Roland2019}.
As far as simple modeling is concerned, such a feature may be captured
by a FH model wherein interchain contacts are favored by entropy
at intermediate to low temperatures and by enthalpy at high temperatures,
thus entailing a heat capacity contribution in $\chi(T)$, with
$\epsilon_h\rightarrow\epsilon_h(T)$, 
$\epsilon_s\rightarrow\epsilon_s(T)$ \cite{biochemrev,Dilletal1989,Kaya2003}, 
beyond the temperature-independent $\epsilon_h$ and 
$\epsilon_s$ used in Fig.~1c,d and Fig.~2.
Alternatively, a reduction in overall condensed-phase 
concentration can also be caused by formation of heterogeneous 
locally organized structures with large voids at low temperatures
even when interchain interactions are purely
enthalpic (Fig.~4 of ref.~\cite{Panag2020}).
\\

\noindent
{\bf Salt-IDR two-dimensional phase diagrams are instrumental for
exploring broader phase properties}\\
Fig.~1c and d, though informative, are computed
by a restricted rG-RPA+FH that assumes a spatially uniform [Na$^+$].
For a more comprehensive physical picture, we now examine
possible differences in salt concentration between the IDR-dilute
and condensed phases by applying unrestricted rG-RPA+FH to
compute two-dimensional salt-Caprin1/pY-Caprin1 phase
diagrams (Fig.~2).

As stated in {\it Materials and Methods} and {\it SI Appendix},
here we define ``counterions'' and ``salt ions'', respectively, as the
small ions with charges opposite and identical in sign to that
of the net charge, $Q$, of a given polymer.
For the Caprin1/NaCl system, since Caprin1's net charge is
positive, Na$^+$ is salt ion and Cl$^{-}$ is counterion.
Overall electric neutrality of the system implies that
the concentrations ($\rho$'s) of polymer ($\rho_{\rm p}$),
counterions ($\rho_{\rm c}$), and salt ions ($\rho_{\rm s}$) are related by
\begin{equation}
\label{eq:Qzrho}
|Q| \rho_{\rm p} + z_{\rm s} \rho_{\rm s} = z_{\rm c} \rho_{\rm c} \; ,
\end{equation}
where $z_{\rm s}$ and $z_{\rm c}$ are,
respectively, the valencies of salt ions and counterions.
For Caprin1 and pY-Caprin1, $Q=+13$ and $-1$, respectively, and
$(z_{\rm s}, z_{\rm c}) = (1,1)$, $(1,2)$, and $(2,4)$ are models
for different small-ion species in the system.
Specifically, in Fig.~2, we identify the $z_{\rm s}=1$ salt ion
as Na$^+$ (Fig.~2a--f) and the $z_c=1$ counterion as Cl$^-$ (Fig.~2a--d),
the $z_{\rm c}=2$ counterion as (ATP-Mg)$^{2-}$ (Fig.~2g,h),
the $z_{\rm s}=2$ salt ion
as Mg$^{2+}$ and the $z_c=4$ counterion as ATP$^{4-}$ (Fig.~2i--l).
As mentioned above, in the present rG-RPA formulation, (ATP-Mg)$^{2-}$
and ATP$^{4-}$ are modeled minimally as a single-bead ion. They are
represented by charged polymer models with more structural complexity
in the FTS models below. 
\\

\noindent
{\bf Behavioral trends of rG-RPA-predicted Na$^+$-Caprin1
two-dimensional phase\\ diagrams are consistent with experiment}\\ 
Notably,
Fig.~2a,b ($z_{\rm s}=z_{\rm c}=1$)
predicts that Caprin1 does not phase separate
without Na$^+$, consistent with
experiment, indicating that monovalent
counterions alone (Cl$^-$ in this case) are insufficient for Caprin1 LLPS.
When [Na$^+$] is increased, the system starts to phase separate
at a small [Na$^+$] $\lesssim 0.1$ M, with LLPS propensity
increasing to a maximum at [Na$^+$] $\sim$1 M before decreasing
at higher [Na$^+$],
in agreement with experiment (Fig.~3a, blue data points) and
consistent with Caprin1 LLPS
propensity increasing with [NaCl] from 0.1 to 0.5 M (Fig.~1c).
The predicted reentrant dissolution of Caprin1 condensate at
high [Na$^+$] in Fig.~2a is consistent with measurement
up to [Na$^+$] $\approx 4.6$ M indicating a significant decrease in
LLPS propensity when [Na$^+$] $\gtrsim 2.5$ M (Fig.~3a), though the
gradual decreasing trend suggests that complete
dissolution of condensed droplets is not likely even when NaCl
reaches its saturation concentration of $\sim 6$ M.

The negative tieline slopes in Fig.~2a,b predict that
Na$^+$ is partially excluded
from the Caprin1 condensate. This ``salt partitioning''
is most likely caused by Caprin1's net positive charge and is
consistent with published research on polyelectrolytes with
monovalent salt \cite{Shen2018,Eisenberg1959,Zhang2016}.
Here, the rG-RPA predicted trend is consistent with our experiment showing
significantly reduced [Na$^+$] in the Caprin1-condensed phase
compared to the Caprin1-dilute phase (Table~1),
although the larger experimental reduction of [Na$^+$] in the Caprin1 
condensed droplet relative to our theoretical prediction
remains to be elucidated. 
In this regard, a similar experimental trend of Na$^+$ tielines
was observed recently for the IDP A1-LCD (WT) with a positive (+8) net
charge~\cite{TanjaRohitJACS2024}.
In contrast, for the near-neutral, very slightly negative
model pY-Caprin1 (Fig.~2c,d), rG-RPA predicts LLPS
at [Na$^{+}$] $\approx 0$, and the positive tieline slopes
indicate that [Na$^{+}$] is higher in the condensed
than in the dilute phase. Consistent with Fig.~1d,
Fig.~2c shows that pY-Caprin1 LLPS propensity always decreases with
increasing [Na$^{+}$].
\\

\noindent
{\bf rG-RPA-predicted salt-IDR two-dimensional phase diagrams
underscore effects of counterion valency on LLPS}\\
Interestingly, a different salt dependence of Caprin1 LLPS is predicted
when the salt ion remains monovalent but
the monovalent counterion
Cl$^-$ is replaced by a divalent $z_c=2$ anion modeling (ATP-Mg)$^{2-}$
(as a one-bead ion)
under the simplifying assumption that ATP$^{4-}$ and Mg$^{2+}$ do not
dissociate in solution. The corresponding rG-RPA results (Fig.~2e--h)
indicate that, in the present of divalent counterions (needed for overall
electric neutrality of the Caprin1 solution),
Caprin1 can undergo LLPS without the monvalent salt (Na$^{+}$) ions
(LLPS regions extend to [Na$^{+}$] $=0$ in Fig.~2e,f; 
i.e., 
$\rho_{\rm s}=0$, $\rho_{\rm c}>0$ in Eq.~\eqref{eq:Qzrho}), 
because the configurational
entropic cost of concentrating counterions in the Caprin1 condensed
phase is lesser for divalent ($z_{\rm c}=2$) than for monovalent 
($z_{\rm c}=1$) counterions as only half of the former 
are needed for approximate electric neutrality in the condensed phase.

Other predicted differences between monovalent (Fig.~2a,b) and divalent
(Fig.~2e,f) counterions' impact on Caprin1 LLPS include: (i) The maximum
condensed-phase [Caprin1] at low [Na$^{+}$] is lower with monovalent
than with divalent counterions ([Caprin1] 
$\sim 40$ mM vs. $\sim 70$ mM).
(ii) The [Na$^{+}$] at the commencement of reentrance (i.e., at the
maximum condensed-phase [Caprin1]) is much higher with monovalent than
with divalent counterions ([Na$^+$] $\sim 1$ M vs. $\sim 0.1$ M).
(iii) [Na$^{+}$] is depleted in the Caprin1 condensate with
both monovalent and divalent counterions when overall [Na$^+$] is high
(negative tieline slopes for [Na$^+$]$\gtrsim$ 2 M in Fig.~2a,e).
However, for lower overall [Na$^+$], [Na$^+$] is slightly higher in
the Caprin1 condensate with divalent but not with
monovalent counterions (slightly positive tieline slopes for
[Na$^+$]$\lesssim$ 2 M in Fig.2e,f).
This prediction suggests that under physiological
[Na$^+$]=150$\sim$170 mM, monovalent positive salt ions such
as Na$^+$ can be attracted, somewhat counterintuitively, into biomolecular
condensates scaffolded by positively-charged polyelectrolytic IDRs in the
presence of divalent counterions. 
This phenomenon most likely arises from the attraction of the
positively charge monovalent salt ions to the negatively charged
divalent counterions in the protein-condensed phase because
although the three negatively charged D residues in Caprin1 
can attract Na$^+$, it is notable that Na$^+$ is depleted in 
condensed Caprin1 when the counterion is monovalent (Fig.~2a).
\\

\noindent
{\bf rG-RPA is consistent with experimental
[ATP-Mg]-dependent Caprin1\\ reentrant phase behaviors}\\
For the $z_{\rm s}=2$, $z_{\rm c}=4$ case in Fig.~2i--l
modeling (ATP-Mg)$^{2-}$ complex dissociating completely in solution into
Mg$^{2+}$ salt ions and ATP$^{4-}$ counterions (modeled as single-bead
ions), rG-RPA predicts Caprin1
LLPS with ATP$^{4-}$ (Fig.~2k,l) in the absence of Mg$^{2+}$
(the LLPS region includes the horizontal axes in Fig.~2i,j),
likely because the configurational entropy loss of tetravalent counterions
in the Caprin1 condensate is less than that of divalent and monovalent
counterions.
Tetravalent counterions also increase the 
theoretical maximum condensed-phase [Caprin1] to 
$\gtrsim$ 120 mM. At the commencement of reentrance
(maximum condensed-phase [Caprin1] in Fig.~2i,j),
[Mg$^{2+}$] $\sim$0.4 M, which is intermediate between the
corresponding [Na$^+$] $\sim 1.0$ and $0.1$ M, respectively,
for monovalent and divalent counterions with $(z_{\rm s}, z_{\rm c})=(1,2)$
and $(1,1)$.
All tieline slopes for Mg$^{2+}$ and ATP$^{4-}$ in Fig.~2i--l are significantly
positive, except in an extremely high-salt region with
[Mg$^{2+}$]$> 8$M, indicating that [(ATP-Mg)$^{2-}$] is
almost always substantially enhanced in the Caprin1 condensate.
These observations from analytical theory will be corroborated 
by FTS below 
with the introduction of structurally more realistic models of 
(ATP-Mg)$^{2-}$, ATP$^{4-}$ together with the possibility of
simultaneous inclusion of Na$^+$, Cl$^-$, and Mg$^{2+}$ in the FTS models of
Caprin1/pY-Caprin1 LLPS systems.
Despite the tendency for polymer field theories to overestimate LLPS propensity
and condensed-phase concentrations quantitatively because they do not
account for ion condensation \cite{Shen2017}---which can be severe
for small ions with more than $\pm 1$ charge valencies 
as in the case of condensed [Caprin1] $\gtrsim$ 120 mM in Fig.~2i--l,
our present rG-RPA-predicted semi-quantitative trends are
consistent with experiments indicating
[ATP-Mg]-dependent reentrant phase behavior of Caprin1
(Fig.~3a, red data points, and Fig.~3b)
and that [Mg$^{2+}$] as well as [ATP$^{4-}$] are significantly enhanced
in the Caprin1 condensate by a factor of $\sim 5$--$60$  for overall
[ATP-Mg] = 3--30 mM (Table~2).
\\

\noindent
{\bf Coarse-grained MD with explicit small ions is useful for investigating
subtle salt dependence in biomolecular LLPS}\\
To gain deeper insights, 
we extend the widely-utilized coarse-grained explicit-chain MD model
for biomolecular condensates \cite{SumanPNAS,dignon18,panag2017}
to include explicit small cations and anions
({\it Materials and Methods}). 
ATP-mediated LLPS of short basic peptides was studied recently
using all-atom simulations indicating ATP engaging in
electrostatic and cation-$\pi$ bridging interactions \cite{HXZhou2024}. 
Here, we limit the small ions in our coarse-grained MD simulations
of Caprin1 and pY-Caprin1 LLPS to Na$^+$ and Cl$^-$, focusing on
the physical origins of reentrance or lack thereof as well as the
effects of ariginine-to-lysine (RtoK) mutations on Caprin1.
Coarse-grained models allow for the study of larger systems (IDPs of
longer chain lengths and more IDPs in the system), though they cannot
provide insights into more subtle structural and energetic effects
as in all-atom simulations \cite{KnowlesNatComm2021,HXZhou2024,MacAinsh2024}.
For computational efficiency, here we neglect 
solvation effects that can arise from the directional
hydrogen bonds among water molecules (see, e.g., ref.~\citen{Seishi2001}) by
treating other aspects of the aqueous solvent implicitly as in most, though
not all \cite{Wessen2021,MiMB2023} applications of the
methodology \cite{dignon18}.
Several coarse-grained interaction schemes were used in recent
MD simulations of biomolecular LLPS 
\cite{dignon18,SumanPNAS,Mpipi,Urry-Mittal,FB,Kresten,WessenJPCB2022}.
Since we are primarily interested in general principles
rather than quantitative details of the phase behaviors of Caprin1 and its
RtoK mutants, here we adopt
the Kim-Hummer (KH) energies for pairwise amino acid
interactions derived from contact statistics of
folded protein structures \cite{dignon18}, which can largely capture the
experimental effects of R vs K on LLPS \cite{SumanPNAS}.
\\

\noindent
{\bf Explicit-ion MD rationalizes experimentally observed [NaCl]-dependent 
{\hbox{Caprin1~reentrant~phase~behaviors~and~depletion~of~Na$^+$~in~Caprin1~condensate}}}\\
Consistent with experiment (Fig.~3)
and rG-RPA (Fig.~2a--d),
explicit-ion coarse-grained MD results in Fig.~4 show [NaCl]-dependent
reentrant phase behavior for Caprin1 but not for pY-Caprin1
(non-monotonic and monotonic trends indicated,
respectively, by the grey arrows in Fig.~4a,b).
In other
words, the critical temperature $T_{\rm cr}$, which is defined as the
maximum temperature (UCST)
of a given phase diagram (binodal, or coexistence curve),
increases then decreases with addition of NaCl for Caprin1 but $T_{\rm cr}$
always decreases with increasing [NaCl] for pY-Caprin1.
Moreover, consistent with the rG-RPA-predicted tielines in
Fig.~2a--d (negative slopes for Caprin1 and positive slopes for pY-Caprin1),
Fig.~4e,g show that Na$^+$ is slightly
depleted in the Caprin1 condensed droplet, exhibiting the
same trend as that in experiment (Fig.~3a, blue data points; and Table~1)
but is enhanced in the pY-Caprin1 droplet (Fig.~4f,h).
Because model temperatures in Fig.~4a,b and subsequent MD results are
given in units of the MD-simulated $T_{\rm cr}$ of WT Caprin1
at [NaCl] $=0$ (denoted as $T^0_{\rm cr}$ here), the $T_{\rm cr}$s of
systems with higher or lower LLPS propensities than WT Caprin1 at
zero [NaCl] is characterized, respectively,
by $T_{\rm cr}/T^0_{\rm cr}>1$ or $<1$.

Fig.~4e,g show that [Cl$^-$] is enhanced while [Na$^+$] is depleted
in the Caprin1 droplet. 
By comparison, Fig.~4f,h show that both [Cl$^-$] 
and [Na$^+$] are enhanced in the pY-Caprin1 droplet with an excess
of [Na$^+$] to balance the negatively charged pY-Caprin1 (Fig.~4h).
The enhancement of [Cl$^-$] in the Caprin1 condensed phase depicted 
in Fig.~4f,h
is further illustrated in Fig.~5a-d by comparing the entire simulation box
with a condensed droplet in the middle
(Fig.~5a) with individual distributions of the Caprin1 IDR (Fig.~5b),
Na$^{+}$ (Fig.~5c), and Cl$^{-}$ (Fig.~5d).
A similar trend, also attributed to charge effects, was observed in
explicit-water, explicit-ion MD simulations in the presence of a preformed
condensate of the N-terminal RGG domain of LAF-1 with a positive net
charge \cite{Zhengetal-ions2020}.
For Caprin1, Fig.~5e,f suggests that, as counterion, Cl$^-$ can
coordinate two positively charged R residues and thereby stabilize
indirect counterion-bridged interchain contacts among polycationic
Caprin1 molecules to promote LLPS, consistent with an early lattice-model
analysis of generic polyelectrolytes \cite{Orkoulas2003}
and a recent atomic simulation study of A1-LCD \cite{MacAinsh2024}.
\\

\noindent
{\bf Explicit-ion MD offers insights into counterion-mediated
interchain bridging interactions among condensed Caprin1 molecules}\\
To assess the extent to which Cl$^-$-mediated bridging interactions 
(as illustrated in Fig.~5f) contribute to condensation of
polyelectrolytic IDRs, we examine the relative positions of positively 
charged arginine residues (Arg$^+$) and negatively charged 
counterions (Cl$^-$) of a Caprin1 solution under phase-separation
conditions in which essentially all Caprin1 molecules are in the
condensed phase, using 4,000 frames (MD snapshots) of an equilibrated
salt-free ([NaCl] $=0$) ensemble of 100 WT Caprin1 chains (net charge 
per chain $=+13$) with 1,300 Cl$^-$ counterions at $T < T^0_{\rm cr}$ as
an example (Fig.~6). For simplicity, we focus on Arg$^+$--Cl$^-$ interactions
because the overwhelming majority (15/16) of the positively charged residues
in Caprin1 are arginines. The computed radial distribution function,
$\rho(r)$, of Cl$^-$ around a given Arg$^+$ 
exhibits a sharp peak at small $r$ that drops to a minium at 
$r\approx 11$~\AA~(Fig.~6a), indicating a strong spatial association
between the oppositely charged Arg$^+$ and Cl$^-$ as expected. Indeed,
within the ensemble we analyze, $5,121,148/(4,000\times 1,300)=98.5\%$ 
of the Cl$^-$ ions are within 11~\AA~of an Arg$^+$. We next enumerate 
putative bridging interactions involving two Arg$^+$s on different
Caprin1 chains and one Cl$^-$
(Fig.~6b) by identifying three-bead configurations in which the 
distance of Cl$^-$ to each of the two Arg$^+$ is $\le 11$~\AA~(within 
the dominant small-$r$ peak of $\rho(r)$ in Fig.~6a), which implies 
that the distance between the two Arg$^+$s is $\le 22$~\AA.  In our 
ensemble, $4,519,387/(4,000\times 1,300)=86.9\%$ of the Cl$^-$ counterions
are identified to be in one or more of a total of 25,112,331 such
putative bridging interaction configurations. This means that, on average,
each Cl$^-$ is involved in $25,112,331/4,519,387=5.56$ configurations, and
thus are coordinating $\approx 4$ Arg$^+$s because there are 6 ($\approx 5.56$)
ways of pairing 4 Arg$^+$s. Fig.~6c shows the distribution of putative 
bridging configurations with respect to Arg$^+$--Arg$^+$ distance $R$.
Spatial distributions of Cl$^-$ in these configurations are provided
in Fig.~6d,e, which are quite similiar to those of isolated 
Arg$^+$--Cl$^-$--Arg$^+$ systems for $R\lesssim 14$~\AA~(Fig.~6f,g).
Among the putative bridging configurations, we make an energetic distinction
between true bridging and neutralizing (screening) configurations.
Physically, a true bridging configuration may be defined by an overall
favorable ($<0$) sum of (i) unfavorable Coulomb potential between two 
Arg$^+$ and (ii) the favorable Coulomb potential between the Cl$^-$ and 
one of the Arg$^+$s that is farther away from the Cl$^-$. By the same token,
a neutralizing (screening) configuration may be defined by a 
corresponding overall
unfavorable or neutral ($\ge 0$) sum of these two Coulomb potentials (i.e.,
the farther Arg$^+$--Cl$^-$ distance is larger than the Arg$^+$--Arg$^+$
distance). 
In this regard, and in more general terms, 
Cl$^-$ ions in bridging and neutralizing
interactions may be considered, respectively, as a ``strong-attraction 
promoter'' and a ``weak-attraction suppressor'' 
of LLPS~\cite{HXZhou2018,HXZhou2019}.

In the present analysis, we group putative bridging configurations
by $R$ in bins of 2~\AA~(Fig.~6c). Accordingly,
we may classify Cl$^-$ positions satisfying the above
condition of favorable ($<0$) sum of Coulomb potentials
for all $R$ values within the 2~\AA~range of the bin
as in true bridging configurations (79.6\%),
those Cl$^-$ positions satisfying the above condition of unfavorable ($\ge 0$) 
sum of Coulomb potentials for all $R$ values in the 2~\AA~range
as in neutralizing configurations (7.4\%), and those that 
satisfy neither as ``intermediate'' configurations (13.0\%).
Even with this more stringent criterion, $\approx 80\%$ of putative
bridging configurations are true bridging configurations. Because on average
a Cl$^-$ counterion known to be involved in at least one putative bridging
configuration is on average participating in $\sim 5$--$6$ such
configurations, the probability that it is involved in at least one
true bridging configuration is very high, at $\approx 1.0 - (0.2)^5=99.97\%$.
Thus, even without taking into consideration bridging interactions involving
lysines, we may reasonably conclude that 
an overwhelming majority of $\approx 87\%$ 
of Cl$^-$ counterions in the coarse-grained MD system considered
are engaged in condensation-driving true bridging interactions
coordinating pairs of Arg$^+$ on different Caprin1 chains. 
Similar extensive Cl$^-$ and Na$^+$ bridging interactions are observed in a 
recent all-atom molecular study of LLPS of short peptides under a 
variety of overall salt concentrations~\cite{MacAinsh2024}.
\\

\noindent
{\bf Explicit-ion MD rationalizes [NaCl]-dependent phase 
properties of arginine-lysine mutants of Caprin1}\\ 
We apply our MD methodology also to four RtoK Caprin1 variants,
termed 15Rto15K, 4Rto4K$_{\rm N}$, 4Rto4K$_{\rm M}$, and 4Rto4K$_{\rm C}$
(SI Appendix-figure 1),
which involve 15 or 4 RtoK substitutions \cite{LewisJACS2020}.
The simulated phase diagrams in Fig.~7
exhibit reentrant phase behaviors for all three 4Rto4K variants.
While these results are consistent with experiments
showing LLPS of these 4Rto4K variants commencing at different
nonzero [NaCl]s \cite{LewisJACS2020},
the simulated reentrant dissolution is not observed experimentally,
probably because the actual [NaCl] needed
is beyond the experimentally investigated or physically possible range
of salt concentration.
Simulated reentrant phase behaviors are also seen
for 15Rto15K; but as will be explained below, its much lower
simulated UCST is consistent with no experimental
LLPS for this variant \cite{LewisJACS2020}.
Since our main focus here is on general physical principles,
we do not attempt to fine-tune the MD parameters
for a quantitative match between simulation and experiment.
Experimentally,
only WT exhibits a clear trend toward reentrant dissolution
of condensed droplets (with a LLPS propensity plateau at [NaCl]
$\approx 1.55$--$2.5$ M, Fig.~3a, blue data points), whereas
the LLPS of 4Rto4K$_{\rm M}$ and 4Rto4K$_{\rm C}$ commences at
[NaCl] $\approx 1.3$ M, LLPS propensity then increases with [NaCl]
(a trend consistent with the MD-predicted increasing LLPS propensity at
low [NaCl]s in Fig.~7b,c), but no sign
of reentrant dissolution is seen up to the maximum [NaCl] = 2 M
investigated experimentally for the RtoK variants
(Fig.~9B of ref.~\citen{LewisJACS2020}).
In contrast, the MD phase diagrams
in Fig.~7 show a maximum LLPS propensity
(highest $T_{\rm cr}$) at
[NaCl] $\approx 0.5$ M. This qualitative agreement with quantitative mismatch
suggests that real Caprin1 LLPS is somewhat less sensitive to small monovalent
ions than that stipulated by the present MD model. This question should be
tackled in future studies by considering, for example,
alternate pairwise amino acid interaction
energies \cite{dignon18,SumanPNAS,Mpipi,Urry-Mittal,FB,Kresten,WessenJPCB2022}
and their temperature dependence \cite{jeetainACS,Roland2019}.

Limitations notwithstanding, the MD-simulated trend
agree largely with experiment.  Predicted LLPS propensities quantified
by the $T_{\rm cr}$s in Fig.~7 follow the rank order of
WT $>$ 4Rto4K$_{\rm M}$ $>$ 4RtoK$_{\rm N}$ $\approx$ 4Rto4K$_{\rm C}$
$>$ 15Rto15K, which is essentially identical to that measured
experimentally, viz., WT $>$ 4Rto4K$_{\rm M}$ $>$ 4RtoK$_{\rm C}$ $>$
4Rto4K$_{\rm N}$ $>$ 15Rto15K (Fig.~9B of ref.~\citen{LewisJACS2020}).
In comparing
theoretical and experimental LLPS,
a low theoretical $T_{\rm cr}$ can practically mean no experimental
LLPS when the theoretical $T_{\rm cr}$ is below the freezing
temperature of
the real system \cite{linPRL,jacob2017}.
Fig.~7a shows that
even the highest $T_{\rm cr}$ for 15Rto15K (at model [NaCl] = 480 mM) is
essentially at the same level as $T^0_{\rm cr}$ for WT at [NaCl] = 0
($T_{\rm cr}/T^0_{\rm cr}\approx 1$). This MD prediction
is consistent with the combined experimental observations of no LLPS
for 15Rto15K up to at least [NaCl] = 2 M and no LLPS for WT Caprin1 at
[NaCl] = 0 (Fig.~9B,C of ref.~\citen{LewisJACS2020}).
\\

\noindent
{\bf Field-theoretic simulation (FTS) is an efficient tool for studying
multiple-component phase properties}\\
We next turn to modeling of Caprin1 or pY-Caprin1 LLPS modulated by both
ATP-Mg and NaCl.
Because tackling such
many-component LLPS systems using rG-RPA or explicit-ion MD is
numerically challenging, here we adopt the
complementary FTS approach \cite{Fredrickson2006}
outlined in {\it Materials and Methods}
for this aspect of our investigation.
FTS is based on complex Langevin dynamics \cite{Parisi1983,Klauder1983},
which is related to an earlier formulation for stochastic
quantization \cite{ParisiWu1981,MBH_Ghostfree} and has been applied
extensively to polymer solutions \cite{Fredrickson2002,Fredrickson2006}.
Recently, FTS
has provided insights into charge-sequence-dependent
LLPS of IDRs \cite{joanJPCL2019,joanElife,Pal2021,WessenJPCB2022,MiMB2023}.
The starting point of FTS
is identical to that of rG-RPA. FTS invokes no RPA and is thus
advantageous over rG-RPA in this regard, though it is still
limited by the lattice size used for simulation and its
restricted treatment of excluded volume \cite{Pal2021}.
Here we apply the protocol detailed in refs.~\citen{Pal2021,MiMB2023}.
\\

\noindent
{\bf A simple model of ATP-Mg for FTS}\\
Going beyond the single-bead model for (ATP-Mg)$^{2-}$ in our
analytical rG-RPA theory (Fig.~2),
we now adopt a 6-bead polymeric representation of (ATP-Mg)$^{2-}$ (Fig.~8a) in
which four negative and two positive charges serve to model ATP$^{4-}$
and Mg$^{2+}$ respectively. Modeling (ATP-Mg)$^{2-}$ as a
short charged polymer enables
application of existing FTS formulations for multiple charge sequences
to systems with IDRs and (ATP-Mg)$^{2-}$. While the model in Fig.~8a does not
capture structural details, its charge distribution does
correspond roughly to that of the chemical structure of (ATP-Mg)$^{2-}$.
In developing FTS models involving IDR, (ATP-Mg)$^{2-}$, and NaCl, we first
assume for simplicity that (ATP-Mg)$^{2-}$ does not dissociate and consider
systems
consisting of any given
overall concentrations of IDR and (ATP-Mg)$^{2-}$
wherein all positive and negative charges
on the IDR and (ATP-Mg)$^{2-}$ are balanced, respectively, by Cl$^-$ and Na$^+$
to maintain overall electric neutrality (Fig.~8a).
\\

\noindent
{\bf Phase behaviors can be probed by FTS  density
correlation functions}\\
LLPS of FTS systems can be
monitored by correlation functions \cite{Pal2021}.
Here, we compute intra-species
IDR self-correlation functions $G_{\rm pp}(r)$ (Fig.~8b,c)
and inter-species cross-correlation functions $G_{\rm pq}(r)$
between the IDR and (ATP-Mg)$^{2-}$ or NaCl (Fig.~8d,e) at three different
overall [(ATP-Mg)$^{2-}$] $=$
$10^{-4}b^{-3}$, $0.03b^{-3}$, and $0.5 b^{-3}$,
where
$b$ may be taken as
the peptide virtual bond length $\approx 3.8$\AA~({\it Materials and Methods}).
The correlation functions in Fig.~8b--e are normalized by overall densities
$\rho^0_{\rm p}$ of the IDR and $\rho^0_{\rm q}$ for
(ATP-Mg)$^{2-}$, Na$^+$ or Cl$^-$, wherein density is
the bead density for the given molecular species in units of $b^{-3}$.
LLPS of the IDR is signaled by $G_{\rm pp}(r)/(\rho^0_{\rm p})^2$
in Fig.~8b,c dropping below the unity
baseline
(dashed) at large distance $r$ because it implies
a spatial region with depleted IDR
below the overall concentration, which is possible only if the IDR
is above the overall concentration in at least another spatial
region.
In other words, $G_{\rm pp}(r)/(\rho^0_{\rm p})^2<1$ for large $r$
indicates that IDR concentration is heterogeneous
and thus the system is phase separated.
For small $r$, $G_{\rm pp}(r)/(\rho^0_{\rm p})^2$
is generally expected to increase because IDR chain connectivity
facilitates correlation among residues local along
the chain. On top of this, LLPS propensity may be quantified by
$G_{\rm pp}(r)/(\rho^0_{\rm p})^2$ for small $r$
because a higher value indicates a higher tendency for different
chains to associate and thus a higher LLPS propensity \cite{Pal2021}.
\\

\noindent
{\bf FTS rationalizes [ATP-Mg]-modulated
Caprin1 reentrant phase behaviors and their colocalization 
in the condensed phase}\\
$\null$[(ATP-Mg)$^{2-}$]-modulated reentrance
is predicted by FTS
for Caprin1 but not for pY-Caprin1: When
[(ATP-Mg)$^{2-}$]/$b^{-3}$ varies from $10^{-4}$ to $0.03$ to $0.5$, small-$r$
values of the Caprin1 $G_{\rm pp}(r)$ in Fig.~8b initially increase
then decrease, whereas the corresponding small-$r$ values of the
pY-Caprin1 $G_{\rm pp}(r)$ in Fig.~8c decrease monotonically,
consistent with rG-RPA (Fig.~2g,h,k,l)
and experiment (Fig.~3).
The inter-species cross-correlations in Fig.~8d,e show
further that when an IDR condensed phase is present
at [(ATP-Mg)$^{2-}$] = 0.03$b^{-3}$ (as indicated by
large-$r$
behaviors of $G_{\rm pp}(r)/(\rho^0_{\rm p})^2$
in Fig.~8b,c), (ATP-Mg)$^{2-}$
is colocalized with Caprin1 or pY-Caprin1 (high value
of $G_{\rm pq}/\rho^0_{\rm p}\rho^0_{\rm q}$ for small $r$) in the
IDR-condensed droplet.  By comparison, the variation of [Na$^+$] and [Cl$^-$]
is much weaker.
For Caprin1, Cl$^-$ is enhanced over Na$^+$ in the Caprin1 condensed phase
(small-$r$ $G_{\rm pq}/\rho^0_{\rm p}\rho^0_{\rm q}$ of the former
larger than the latter in Fig.~8d), but the reverse is seen
for pY-Caprin1 (Fig.~8e). This FTS-predicted difference, most likely arising
the positive net charge on Caprin1 and the smaller negative net charge
on pY-Caprin1, is consistent with the MD results in Fig.~4e--h and
SI Appendix-figure 3.
\\

\noindent
{\bf FTS rationalizes experimentally
observed residue-specific binding of Caprin1 with ATP-Mg}\\
The propensities for (ATP-Mg)$^{2-}$, Na$^+$, and Cl$^-$ to associate with
each residue $i$ along the Caprin1 IDR ($i=1,2,\dots,103$) in
FTS are quantified by the residue-specific integrated correlation
${\cal G}^{(i)}_{\rm pq}/\rho^0_{{\rm p},i}\rho^0_{\rm q}$ in Fig.~8f, which
is the integral of the
corresponding $G^{(i)}_{\rm pq}(r)$ from $r=0$
to a relative short cutoff distance $r=r_{\rm contact}$
to provide a relative contact frequency for residue $i$ and ionic
species q to be in spatial proximity ({\it Materials and Methods} and {\it SI
Appendix}). Notably, the residue-position-dependent integrated
correlation for (ATP-Mg)$^{2-}$ varies significantly, exhibiting much larger
values near the N-terminal and a little before the C-terminal but weaker
correlation elsewhere (Fig.~8f, red symbols). The two regions of high
integrated correlation (i.e., favorable association) coincide with
regions with high sequence concentration
of positively charged residues. This FTS prediction is remarkably similar
to the experimental NMR finding that binding between (ATP-Mg)$^{2-}$ and Caprin1
occurs strongly at the arginine-rich N- and C-terminal regions, as
indicated by the volume ratio $V/V_0$ data in Fig.~1C of
ref.~\citen{LewisPNAS2021} that quantifies the ratio of peaks in NMR
spectra in the presence and absence of trace amounts of ATP-Mn.
For comparison with the FTS results, this
set of experimental data is replotted as $1-V/V_0$ in Fig.~8f (grey symbols,
right vertical axis) to illustrate the similarity in experimental and
theoretical trends because $1-V/V_0$ is expected to trend with
contact frequency.  Corresponding FTS results for Na$^+$
and Cl$^-$ in Fig.~8f exhibit much less residue-position-dependent
variation, with Cl$^{-}$ displaying only slightly enhanced association in
the same arginine-rich regions, and Na$^+$ showing even less variation,
presumably because the positive charges on Caprin1 are already essentially
neuralized by the locally associated (ATP-Mg)$^{2-}$ or Cl$^-$ ions.
The theory-experiment agreement in Fig.~8f regarding ATP-Caprin1 interactions
indicates once again that electrostatics is an important driving force
underlying many aspects of experimentally observed Caprin1--(ATP-Mg)$^{2-}$
association.
\\

\noindent
{\bf FTS snapshots of [ATP-Mg]-modulated
reentrant phase behaviors and Caprin1-ATP-Mg colocalization}\\
The above FTS-predicted trends
are further illustrated in Fig.~9 by field snapshots.
Such FTS snapshots are generally useful for visualization and
heuristic understanding
\cite{joanJPCL2019,Pal2021,WessenJPCB2022},
including insights into subtler aspects of spatial arrangements
exemplified by recent studies of 
subcompartmentalization entailing either co-mixing or
demixing in multiple-component LLPS that are verifable by
explicit-chain MD \cite{Pal2021,WessenJPCB2022}.
Now,
trends deduced from the correlation functions in Fig.~8 are
buttressed by the representative snapshots in Fig.~9: As the bead density
of (ATP-Mg)$^{2-}$ is increased from $10^{-4}b^{-3}$ to $0.03 b^{-3}$
to $0.5 b^{-3}$,
the spatial distribution of Caprin1 evolves from an initially
dispersed state to a concentrated droplet to a (reentrant) dispersed state
again (Fig.~9a), whereas the initial dense pY-Caprin1 droplet becomes
increasingly dispersed monotonically (Fig.~9b).
Colocalization of (ATP-Mg)$^{2-}$ with both the Caprin1 (Fig.~9c)
and pY-Caprin1 (Fig.~9d) droplets is clearly visible
at [(ATP-Mg)$^{2-}$] $=0.03b^{-3}$, though the degree of colocalization is
appreciably higher for Caprin1 than for pY-Caprin1. This
is likely because the positive net charge of Caprin1
is more attractive to (ATP-Mg)$^{2-}$.
By comparison, variations in Na$^+$ and Cl$^-$ distribution
between Caprin1/pY-Caprin1 dilute and condensed phases are not so discernible
in Fig.~9e--h, consistent with the small
differences in the corresponding FTS correlation functions (Fig.~8d,e).
\\

\noindent
{\bf Robustness of general trends predicted by FTS}\\
We have also assessed the generality of the results in Figs.~8 and 9
by considering three variations in the molecular species treated by FTS:
(i) Caprin1 or pY-Caprin1 with only Na$^+$ and Cl$^-$ but
no (ATP-Mg)$^{2-}$ (SI Appendix-figure 4),
(ii) Caprin1 with (ATP-Mg)$^{2-}$ and either Na$^+$ or Cl$^-$
(but not both) to maintain overall charge neutrality or
pY-Caprin1 with (ATP-Mg)$^{2-}$ and Na$^+$ as counterion but no Cl$^-$
(SI Appendix-figure~5), and
(iii) Caprin1 or pY-Caprin1 with ATP$^{4-}$, Mg$^{2+}$,
Na$^+$ and Cl$^-$ (SI Appendix-figure~6).
Despite these variations in FTS models, SI Appendix-figures~4--6
consistently show reentrant behavior for Caprin1 but not pY-Caprin1
and Appendix-figures~5 and 6 both exhibit
colocalization of ATP with condensed Caprin1,
suggesting that these features are robust consequences of the
basic electrostatics at play in Caprin1/pY-Caprin1 + ATP-Mg + NaCl systems.
\\



\noindent
{\Large\bf Discussion}

It is reassuring that, in agreement with experiment,
all of our electrostatics-based theoretical approaches
consistently predict salt-dependent reentrant phase behaviors for Caprin1,
whereas pY-Caprin1 LLPS propensity
decreases monotonically with increasing salt
(Figs.~2, 4, 8, and 9). This effect applies to small monovalent salts
exemplified by Na$^+$ and Cl$^-$ as well as to our electrostatics-based
single- and multiple-bead models of (ATP-Mg)$^{2-}$ or ATP$^{4-}$, 
with ATP exhibiting a significant colocalization with the Caprin1
condensed phase (Figs.~2g,h,k,l and 9c) attributable to the higher
valency of (ATP-Mg)$^{2-}$ and ATP$^{4-}$ than that of monovalent ions.
As mentioned above,
the difference in salt-dependent LLPS of Caprin1 and pY-Caprin1 originates
largely from the polyelectrolytic nature of Caprin1 and the
polyampholytic nature of pY-Caprin1 \cite{kings2020} corresponding,
respectively, to the ``high net charge'' and ``screening''
classes of IDPs in a more recent analysis \cite{MacAinsh2024}.
\\


\noindent
{\bf Related studies of electrostatic effects on biomolecular condensates}\\
Our theoretical predictions are also largely
in agreement with recent computational studies on
salt concentrations in the dilute versus condensed
phases \cite{Zhengetal-ions2020} and salt-dependent reentrant
behaviors \cite{KnowlesNatComm2021} of other biomolecular condensates,
including explicit-water, explicit-ion atomic simulations with
preformed condensates of the N-terminal RGG domain of
LAF-1 \cite{Zhengetal-ions2020}
and of the highly positive proline-arginine 25-repeat dipeptide
PR$_{25}$ \cite{Espinosa2021}.

A recent study examines
salt-dependent reentrant LLPSs of
full-length FUS (WT and G156E mutant), TDP-43, bromodomain-containing
protein 4 (Brd4), sex-determining region Y-box 2 (Sox2), and
annexin A11 \cite{KnowlesNatComm2021}.
Unlike the requirement of a nonzero monovalent salt
concentration for Caprin1 LLPS, LLPS is observed for all these six proteins
with KCl, NaCl or other salts at concentrations
as low as 50 mM.
Also unlike Caprin1,
their protein condensates dissolve at intermediate salt then re-appear
at higher salt, a phenomenon the authors
rationalize by a tradeoff between decreasing favorability of cation-anion
interactions and increasing favorability of cation-cation, cation-$\pi$,
hydrophobic, and other interactions with increasing monovalent salt
\cite{KnowlesNatComm2021}.

Two reasons may account for this difference. First,
Caprin1 does not phase separate at low salt because it is
a relatively strong polyelectrolyte
(NCPR $=+13/103=+0.126$). By comparison, five of the six proteins in
ref.~\citen{KnowlesNatComm2021} are much weaker polyelectrolytes or
not at all, with NCPR
$=+14/526=+0.0266$, $+13/526=+0.0247$, $-7/80=-0.0875$, $0$, and
$+3/326=+0.00920$, respectively, for FUS (WT, mutant), TDP-43, Brd4,
and A11. Apparently, their weak electrostatic repulsions can be
overcome by favorable nonelectrostatic interactions alone to enable LLPS.

Second, compared to Caprin1, the proteins in ref.~\citen{KnowlesNatComm2021}
are either significantly larger (WT and mutant FUS) or significantly more
hydrophobic and aromatic (the other four proteins), both properties
are conducive to LLPS. For instance,
although Sox2's NCPR $=+14/88=+0.159$ is higher than
that of Caprin1, among Sox2's amino acid residues,
$21/88=23.9\%$ are large hydrophobic or aromatic residues leucine (L),
isoleucine (I), valine (V), methionine (M), phenylalanine (F), or
tryptophan (W), and $17/88=19.3\%$ are large aliphatic residues L, I, V, or M.
This amino acid composition suggests that hydrophobic or $\pi$-related
interactions in Sox2 can be sufficient to overcome electrostatic
repulsion to effectuate LLPS at zero salt.
In contrast, the Caprin1 IDR contains
merely one L; only $10/103=9.7\%$ of the residues of Caprin1 are in
the L, I, V, M, F, W hydrophobic/aromatic category and only $6/103=5.8\%$ are
in the L, I, V, M aliphatic category.
The corresponding aliphatic fractions of TDP-43, Brd4 and A11, at
$21/80=26.3\%$, $33/132=25\%$, and $90/326=27.6\%$,
respectively, are also significantly higher than that of Caprin1.
\\


\noindent
{\bf Effects of salt on biomolecular LLPS}\\
Effects of salts on LLPS, including partition of salt
into polymer-rich phases, are of long-standing interest in
polymer physics \cite{riggleman2023}.
In the biomolecular condensate context,
the versatile functional roles of salts are highlighted by
the interplay between electrostatic and cation-$\pi$
interactions \cite{kobyJPCL2023,Jho2017},
salts' modulating effects on heat-induced LLPSs of
RNAs \cite{banerjee2023},
their regulation of condensate liquidity \cite{Ichijo2023},
and even their potential impact in extremely high-salt exobiological
environments \cite{roland2021}.
While some of these recent studies focus primarily on salts' electrostatic
screening effects without changing the signs of the effective polymer 
charge-charge interaction \cite{kobyJPCL2023},
effective attractions between like charges bridged by salt or other
oppositely-charged ions \cite{Orkoulas2003} as illustrated by
Caprin1 (Fig.~5f, Fig.~6) 
and a recent study of A1-LCD \cite{MacAinsh2024}
are likely needed to account for phenomena such as salt-induced 
dimerization of highly charged, medically relevant arginine-rich 
cell-penetrating short peptides \cite{LundPNAS2017,LeeTW2022}.
In this regard, it should be noted that positively and negatively charged 
salt ions can also coordinate with backbone carbonyls and amides,
respectively, in addition to coordinating with charged amino 
acid sidechains~\cite{MacAinsh2024}.
The impact of such effects, which are not considered in the present
coarse-grained models, should be ascertained by further
investigations using atomic 
simulations~\cite{MacAinsh2024,RauscherPomes2017,Zhengetal-ions2020}.
\\

\noindent
{\bf Tielines in protein-salt phase diagrams}\\
In view of Caprin1's polyelectrolytic nature,
the mildly negative tieline slopes in
Fig.~2a,b are consistent with rG-RPA predictions for
a fully charged polyelectrolyte (Fig.~10a of ref.~\citen{kings2020}).
This depletion of monovalent salt in the condensed phase is similar to that
observed in the complex coacervation of oppositely charged polyelectrolytes
\cite{CSing2017,dePablo2018,Azzaroni2023}.
By comparison, the positive rG-RPA tieline slopes
for polyampholytic pY-Caprin1 (Fig.~2c,d), confirmed by
MD in Fig.~4f,h, are appreciably steeper than that predicted
for fully charged ($\pm 1$) diblock polyampholytes
by rG-RPA and the essentially flat tielines predicted by FTS
(Fig.~10b of ref.~\citen{kings2020} and Fig.~7 of ref.~\citen{joanJCP}).
Whether this difference originates from the presence of divalently
charged ($-2$) phosphorylated sites in pY-Caprin1 remains to be elucidated.
In any event, tieline analysis is generally instrumental for revealing
details, such as stoichiometry, of the interactions driving
multiple-component biomolecular 
LLPSs~\cite{YHLin2022,TanjaRohitJACS2024,KnowlesPRX2022},
rG-RPA should be broadly useful as a
computationally efficient tool for this purpose \cite{kings2020}.
\\

\noindent
{\bf Counterion valency}\\
Our rG-RPA prediction that the maximum condensed-phase [Caprin1]
at low [Na$^{+}$] is substantially higher with divalent
than with monovalent counterions is in line with early findings
that higher-valency counterions
are more effective in bridging polyelectrolyte interactions to
favor LLPS \cite{OdlCruz1995} and recent observations
that salt ions with higher valencies enhance biomolecular
LLPS \cite{Skepo2021,Nott2023}.
The possibility that this counterion/salt effect on LLPS may be exploited
more generally for biological functions and/or biomedical applications
remains to be further explored.
In this regard, while recognizing that ATP can engage in
$\pi$-related interactions \cite{HXZhou2024,SongBBRC2018,SongPLoS2019},
our electrostatics-based perspective of ATP-dependent reentrant phase behaviors
is consistent with recent observations on polylysine LLPS modulated by
enzymatically catalyzed ATP turnovers \cite{Azzaroni2023,Spruijt2018}.
More broadly, differential effects of salt ions on biomolecular LLPS can 
also arise from the sizes and charge densities of the ions---properties 
related to the Hofmeister 
phenomena~\cite{HribarDillJACS2002,NinhamChemRev2012}---even for
ions with the same valency~\cite{TanjaRohitJACS2024}. These features 
should be addressed in future theoretical models as well.
\\


\noindent
{\bf Prospective extensions of the present theoretical methodology}\\
Beyond the above comparisons,
further experimental testing of other aspects of our theoretical
predictions should be pursued, especially those pertaining to pY-Caprin1.
Future theoretical efforts should
address a broader range of scenarios by independent variations
of [ATP$^{4-}$], [Mg$^{2+}$], [Na$^+$], [Cl$^-$]
and to account for nonelectrostatic aspects of ATP-Mg
dissociation \cite{WessenJCP2022}
with predictions such as tieline slopes analyzed in detail
to delineate effects of
sizes, charge densities~\cite{TanjaRohitJACS2024}, and 
configurational entropy of salt ions \cite{Muthu2018}
as well as solvent quality \cite{dePablo2021}.
In addition to our basic modeling constructs,
the impact of excluded volume and solvent/cosolute-mediated
temperature-dependent effective interactions
should be incorporated.
Excluded volume is known to affect
LLPS \cite{joanJCP}, demixing of
IDP species in condensates \cite{Pal2021},
and partition of salt ions in polymer LLPS \cite{dePablo2018}.
Moreover, LCST can be driven not only
by hydrophobicity \cite{biochemrev,jeetainACS,Roland2019}
but also by electrostatics, as suggested by
experiment on complex coacervates
of oppositely charged polyelectrolytes \cite{Prabhu2019}.
Bringing together these features into a comprehensive
formulation will afford a more accurate physical picture.
\\


\noindent
{\bf Summary}\\
To recapitulate, we have employed three complementary theoretical
and computational approaches
to account for the interplay between sequence pattern, phosphorylation,
counterion, and salt in the phase behaviors of IDPs.
Application to the Caprin1 IDR and its
phosphorylated variant pY-Caprin1 provides physical rationalization for
a variety of trends observed in experiments,
including reentrance behaviors and very substantial ATP colocalization.
These findings support a
significant---albeit not exclusive---role of electrostatics in these
biophysical phenomena, providing physical insights into effects of
sequence-specific charge-charge interactions on ATP-modulated physiological
functions of biomolecular condensates such as regulation of ion
concentrations.
The approach developed here should be of general
utility as a computationally efficient tool for hypothesis generation,
design of new experiments, exploration and testing of biophysical
scenarios, as well as a starting point for more sophisticated
theoretical/computational modeling.
\\

\noindent
{\Large\bf Materials and Methods}\\

\noindent
Further details of the experimental and theoretical/computational methodologies
outlined below are provided in {\it SI Appendix}.
\\

\noindent
{\bf Experimental sample preparation}\\
The low complexity 607--709 domain of Caprin1 was expressed and purified
as before \cite{julie2019,LewisPNAS2021}.
WT Caprin1 was used in all experiments except those on [NaCl] dependence
reported in Table~1 and Fig.~3a, for which a double mutant was used
because residue pairs N623-G624 and N630-G631 in WT Caprin1 form
isoaspartate (IsoAsp) glycine linkages over time which alters the
charge distribution of the IDR \cite{LewisJACS2020}.
\\

\noindent
{\bf Phosphorylation of the Caprin1 IDR}\\
Phosphorylation of the WT Caprin1 IDR was performed as described in our
prior study \cite{julie2019} by using the kinase
domain of mouse Eph4A (587-896) \cite{sicheri2006} with an N-terminal
His-SUMO tag.
\\

\noindent
{\bf Determination of phase diagrams}\\
We established phase diagrams for Caprin1 and pY-Caprin1 by measuring the
protein concentrations in dilute and condensed phases across a range of
[NaCl]s (Fig.~1c,d). Initially homogenizing the two
phases of the demixed samples into a milky dispersion through vortexing,
$\sim 200$ $\mu$L aliquots were then incubated
in a PCR thermocycler with a heated lid at 90$^\circ$C, in triplicate,
for a minimum of one hour. During incubation, the condensed
phase settled and formed a clear phase at the bottom.
For concentration measurements, the samples
were diluted in 6 M GdmCl and 20 mM NaPi (pH 6.5). The dilute phase
(top layer) was analyzed through a tenfold dilution of 10 $\mu$L samples,
and the condensed phase (bottom layer) was analyzed through 250- to 500-fold
dilution of 2 or 10 $\mu$L samples.
\\

\noindent
{\bf Concentrations of salt and ATP-Mg in dilute and condensed phases}\\
Inductively coupled plasma optical emission spectroscopy (ICP-OES)
measurements of [Na$^+$] were performed using a Thermo
Scientific iCAP Pro ICP-OES instrument in axial mode.
ICP-OES was also used to
determine [ATP] and [Mg$^{2+}$] (Table~2). The detection of
phosphorus and magnesium served as proxies for quantifying ATP and Mg$^{2+}$
levels, respectively.  Standard curves were prepared using solutions
with known [ATP] and [Mg$^{2+}$], ranging from 0 to 90 ppm
for ATP and 0 to 25 ppm for Mg$^{2+}$.
\\

\noindent
{\bf Caprin1 phase separation propensity at high salt concentrations}\\
A 6 mM solution of double-mutant Caprin1 IDR (see above) in buffer
(25 mM sodium phosphate, pH 7.4) was
prepared by exchanging (3 times) the purified protein after size exclusion
chromatography using centrifugal concentrators (3 kDa, EMD Millipore).
Caprin samples for turbidity measurements were prepared by taking 0.5 $\mu$L
of the above solution and diluting it into buffer (25 mM sodium phosphate,
pH 7.4) containing varying [NaCl]s ranging from 0 to 4.63 M,
in a sample volume of 9 $\mu$L, so as to achieve [Caprin1]
of 300 $\mu$M.
After rigorous mixing, 5 $\mu$L samples were loaded into a $\mu$Cuvette G1.0
(Eppendorf). OD600 measurements (Fig.~3a) were recorded three times using a
BioPhotometer D30 (Eppendorf).
\\

\noindent
{\bf [ATP-Mg]-dependent Caprin1 phase behaviors}\\
Turbidity assays were conducted using the method we
described previously \cite{LewisJACS2020}.
\\

\noindent
{\bf Sequence-specific theory of heteropolymer phase separation}\\
As detailed in refs.~\citen{MiMB2023,WessenJPCB2022},
an example of the sequence-specific polymer
theories \cite{linPRL,kings2020} is that
for a solution with a single species of charged
heteropolymers in $\np$ copies, $\nc$ counterions (same type),
and $\ns$ salt ions (same type, but different from the counterions).
Each polymer chain has $N$ monomers (residues) with charge sequence
$\ket{\sigma} = [\sigma_1, \sigma_2, ...  \sigma_N ]^{\rm T}$ in
vector notation, where $\sigma_i \in \{ 0, \pm 1, -2\}$ is the charge
of the $i$th residue.  The counterions and salt ions are monomers
carrying $\zc$ and $\zs$ charges, respectively.
The particle-based partition function is given by
\begin{equation}
\Z = \frac{1}{\np!\nc!\ns!\nw!}\int
                \prod_{\alpha=1}^{\np}\prod_{i=1}^N d\R_{\alpha,i}
                \prod_{a=1}^{\ns+\nc} d\rr_a
                e^{-\T[\R ]-\U[\R,\rr ]}
\; ,
        \label{eq:ZiniSI}
\end{equation}
where $\nw$ denotes the number of water (solvent) molecules,
$\R_{\alpha,i}$ is the position vector of the $i$th residue of the
$\alpha$th polymer, $\rr_a$ is the position vector of the $a$th small ion.
$\T$ accounts for polymer chain connectivity modeled
by a Gaussian elasticity potential with Kuhn length $l$.
$\U$ describes the interactions among all molecular components
of the system, here consisting only of Coulomb electrostatics (el)
and excluded-volume (ex) for simplicity, viz.,
$\U = \U_{\rm el} + \U_{\rm ex}$.  Their interaction strengths are
governed by the Bjerrum length $\lb$
and the two-body excluded volume parameter $v_2$.
By introducing conjugate fields $\psi(\rr)$,
$\w(\rr)$
and applying the Hubbard-Stratonovich
transformation, the system defined by the particle-based partition
function in Eq.~\eqref{eq:ZiniSI} is recast as a field theory of $\psi,\w$
in which their interactions with polymer, salt, and
counterion are described, respectively,
by single-molecule partition functions
$\Qp$, $\Qs$, and $\Qc$.
For instance,
\begin{equation}
\Qp[\psi, \w] =  \int \prod_{i=1}^N d\R_i \exp\Big( -\Hp\left[ \R; \psi,
\w \right] \Big)
\label{eq:Qp}
\end{equation}
where $\Hp$ is the single-polymer Hamiltonian
and the chain label $\alpha$ is dropped.
\\

\noindent
{\bf Renormalized-Gaussian random-phase-approximation (rG-RPA)}\\
Following refs.~\citen{kings2015,kings2020},
$\Hp$ can be separated
into a Gaussian-chain Hamiltonian with an effective (renormalized)
Kuhn length $l_1 = x l $ and a remaining term,
$\Hp = \Hp^0 + \Hp^1$, where
\begin{subequations}
\begin{align}
\Hp^0 = &  \frac{3}{2l^2 x}\sum_{i=1}^{N-1}\left( \R_{i+1} - \R_i\right)^2
\; , \\
\Hp^1 = & \frac{3}{2l^2}\left(1-\frac{1}{x}\right)\sum_{i=1}^{N-1}
\left( \R_{i+1} - \R_i\right)^2
                - \i\sum_{i=1}^N\left[  \sigma_i\psi(\R_i) + \w(\R_i) \right]
\; ,
\end{align}
\end{subequations}
with $\i^2=-1$.
By requiring the observable polymer square end-to-end distance
be properly quantified by
$\Hp^0$, $x$ can be approximated by
variational theory \cite{kings2015}. RPA can then be
applied to the renormalized Gaussian (rG) chain system with
$l\rightarrow xl$ and a corresponding scaling of the contour length
to arrive at an improved theory, rG-RPA, for sequence-specific LLPS.
\\

\noindent
{\bf Explicit-ion coarse-grained molecular dynamics (MD)}\\
The MD model in this work augments a class of implicit-water
coarse-grained models \cite{dignon18,SumanPNAS}
that utilize a ``slab'' approach for efficient
equilibration \cite{panag2017} by
incorporating explicit small ions.
As before \cite{SumanPNAS}, the total MD potential energy
$U_{\rm T}$ is the sum of long-spatial-range
electrostatic (el) and short-spatial-range (sr) interactions of the
Lennard-Jones (LJ) type as well as bond interactions, i.e.,
$U_{\rm T} = U_{\rm el} + U_{\rm sr} + U_{\rm bond}$.
With small ions, the electrostatic component is given by
a sum of polymer-polymer (pp), polymer-ion (pi), and ion-ion (ii)
contributions: $U_{\rm el} = U_{\rm el,pp} + U_{\rm el,pi} + U_{\rm el,ii}$.
Details of these terms are provided in {\it SI Appendix}.
\\

\noindent
{\bf Field-theoretic simulation (FTS)}\\
FTS is useful for sequence-specific
multiple-component LLPSs encountered in biomolecular settings.
The new applications developed here are based on recent
advances (see, e.g., refs.~\citen{joanJPCL2019,joanJCP,MiMB2023,Pal2021,Fredrickson2006,Fredrickson2002}).
Consider the field theoretic Hamiltonian
\begin{equation} \label{eq:H_FTS}
H[\w, \psi] = \int d\rr \left( \frac{\left[\nabla \psi(\rr) \right]^2}{8\pi\lb}
+ \frac{\w(\rr)^2}{2v_2}\right)
        - \sum_m n_m \ln {\cal Q}_m[\breve{\w}, \breve{\psi}],
\end{equation}
where ${\cal Q}_m$ is single-molecule partition function
[here $m$ labels the components in the system, cf. Eq.~\eqref{eq:Qp}] and
the breves denote convolution with $\Gamma$, i.e., for a generic field $\phi$,
$\breve{\phi}(\rr) = \Gamma \star \phi(\rr) \equiv \int d \rr'\,
\Gamma(\rr - \rr') \phi(\rr')$; here $\phi = \w, \psi$, and $\Gamma$
is a Gaussian smearing function \cite{MiMB2023}.
FTS utilizes the Complex-Langevin (CL) method \cite{Parisi1983,Klauder1983}
by introducing an artifical CL time variable ($t$), viz.,
$\w(\rr)\rightarrow \w(\rr,t)$, $\psi(\rr)\rightarrow \psi(\rr,t)$
and letting the system evolve in CL time in accordance with
a collection of Langevin equations
\begin{equation}
\label{eq:CL_evolution}
\frac{\partial \phi(\rr,t)}{\partial t} = - \frac{\delta H}{\delta \phi(\rr,t) } + \eta_{\phi}(\rr,t), \quad \phi=\w,\psi \; ,
\end{equation}
where the Gaussian noise $\eta_{\phi}(\rr,t)$
satisfies $\langle \eta_{\phi}(\rr,t) \eta_{\phi'}(\rr',t') \rangle =
2 \delta_{\phi,\phi'} \delta(\rr - \rr') \delta(t-t')$.
Thermal averages of thermodynamic observables are then computed as
asymptotic CL time averages of the corresponding field operators.
Spatial information about condensation and proximity of various components
is readily gleaned from density-density
correlation functions \cite{Pal2021,MiMB2023},
\begin{equation}
\label{eq:Gij}
G_{m,n}(| \rr - \rr' |) = \langle \hat{\rho}_m(\rr )
\hat{\rho}_n(\rr ') \rangle \, ,
\end{equation}
where $m,n$ are labels for the components in the model system.
For instance,
$m$ may represent all polymer beads (denoted
``p'') irrespective of the sequence positions of the beads
[$\hat{\rho}_{\rm p}(\rr) = \sum_{i=1}^{n_{\rm p}}
\sum_{\alpha=1}^N \Gamma (\rr - \R_{\alpha,i})$],
and
$n$ may represent
all six beads in our
ATP-Mg model (Fig.~8a).
One may also define
\begin{equation}
\label{eq:bead_spec_corr}
G^{(i)}_{\mathrm{pq}}(| \rr - \rr' |) \equiv
\left\langle \hat{\rho}_{\mathrm{p},i}(\rr)
\hat{\rho}_{\rm q}(\rr') \right\rangle \; ,
\end{equation}
where $(i)$ represents the $i$th residue along a protein chain
[$\hat{\rho}_{\mathrm{p},i}(\rr) \equiv
\sum_{\alpha=1}^N \Gamma (\rr - \R_{\alpha,i})$ is the density of
the $i$th residue among all the protein chains],
and ${\rm q}=$ (ATP-Mg)$^{2-}$, Na$^+$, or Cl$^-$.
With this definition,
residue-specific relative contact frequencies are
estimated by integrating
Eq.~\eqref{eq:bead_spec_corr} over a spherical volume within
a small inter-component distance $r_{\rm contact}$:
\begin{equation}
\label{eq:calG}
\mathcal{G}^{(i)}_{\mathrm{pq}} \equiv
4\pi\int_{0}^{r_{\rm contact}} dr\, r^2  G^{(i)}_{\mathrm{pq}}(r)
\; .
\end{equation}
For the normalized
$\mathcal{G}^{(i)}_{\mathrm{pq}}/\rho^0_{{\mathrm{p}},i}\rho^0_{\mathrm{q}}$
plotted in Fig.~8f,
$\rho^0_{\mathrm{p},i}$ and $\rho^0_{\mathrm{q}}$ are bulk (overall)
densities, respectively, of
the $i$th protein residue and of (ATP-Mg)$^{2-}$ or small ions,
and $r_{\rm contact}\approx 1.5b$ is used to characterize contacts.
Further details are provided in {\it SI Appendix}.
\\



\noindent
{\large\bf Acknowledgements}\\
This work was supported by Canadian Institutes of Health Research (CIHR) grant
NJT-155930 and Natural Sciences and Engineering Research Council of 
Canada (NSERC) grant RGPIN-2018-04351 to H.S.C., CIHR grant 
FDN-148375, NSERC grant RGPIN-2016-06718, and Canada Research Chairs
Program to J.D.F.-K. as well as CIHR grant FDN-503573 to L.E.K.
A.K.R. was supported by a CIHR postdoctoral fellowship.
We are grateful for the computational resources provided generously 
by Compute/Calcul Canada and the Digital Research Alliance of Canada.

\vfill\eject

\noindent
{\bf Table~1.} Sodium ions are depleted in the Caprin1-condensed phase
relative to the Caprin1-dilute phase. Consistent with theory, [Na$^+$]
is consistently lower in the Caprin1-condensed phase for two temperatures
at which the measurements were performed.
\begin{center}
\begin{tabular}{cccc}
\hline
Bulk & $T$ &
Caprin1-Dilute & Caprin1-Condensed\\
{} [Na$^+$] (mM) & ($^\circ$C) & [Na$^+$] (mM) & [Na$^+$] (mM)\\
\hline
$300$ & $25$ & $341.3\pm 45.5$ & $140.7\pm 6.0$\\
$300$ & $35$ & $289.5\pm 21.9$ & $149.0\pm 2.5$\\ 
\hline
\end{tabular}
\end{center}
\noindent
uncertainty ($\pm$) is standard deviation of
triplicate measurements.

$\null$\\
$\null$\\

\noindent
{\bf Table~2.} Colocalization of ATP-Mg in the Caprin1-condensed phase.
For three overall ATP-Mg concentrations at room temperature,
the concentrations of ATP$^{4-}$ and Mg$^{2+}$ are all
significantly higher in the Caprin1-condensed than in the Caprin1-dilute
phase.
\begin{center}
\begin{tabular}{c|ccc|ccc}
\hline
& \multicolumn{3}{c|} {Caprin1-Dilute} &
\multicolumn{3}{c} {Caprin1-Condensed}\\
\cline{2-7} 
{[ATP-Mg]} &
[Caprin1] & [Mg$^{2+}$] & [ATP$^{4-}$] &
[Caprin1] & [Mg$^{2+}$] & [ATP$^{4-}$] \\
(mM) &
($\mu$M) & (mM) & (mM) & (mM) & (mM) & (mM) \\
\hline
{3} &
$\,$ $67.7\pm 5.0$ $\,$ & $\,$ $2.85\pm 0.05$ $\,$ & $\,$ $2.76\pm 0.07$ $\,$ &
$\,$ $29.9\pm 3.8$ $\,$ & $\,$ $70.7\pm 6.0$ $\,$ & $\,$ $143\pm 30$ $\,$ \\
{10} &
$26.4\pm 1.2$ & $8.57\pm 0.14$ & $8.53\pm 0.97$ &
$35.3\pm 3.5$ & $137\pm 12$ & $197\pm 11$ \\
{30} &
$117\pm 3$ & $28.2\pm 0.3$ & $27.6\pm 0.8$ &
$28.0\pm 2.0$ & $134\pm 7$ & $174\pm 22$ \\
\hline
\end{tabular}
\end{center}

$\null$\\ $\null$\\ $\null$\\

\vfill\eject

\setcounter{figure}{0}
\renewcommand{\figurename}{{\bf Figure}}
\renewcommand{\thefigure}{{\bf \arabic{figure}}}
\centerline{\Large\bf Figures (Main Text)}

$\null$\\

\begin{figure*}[ht]
{\includegraphics[width=\columnwidth,angle=0]{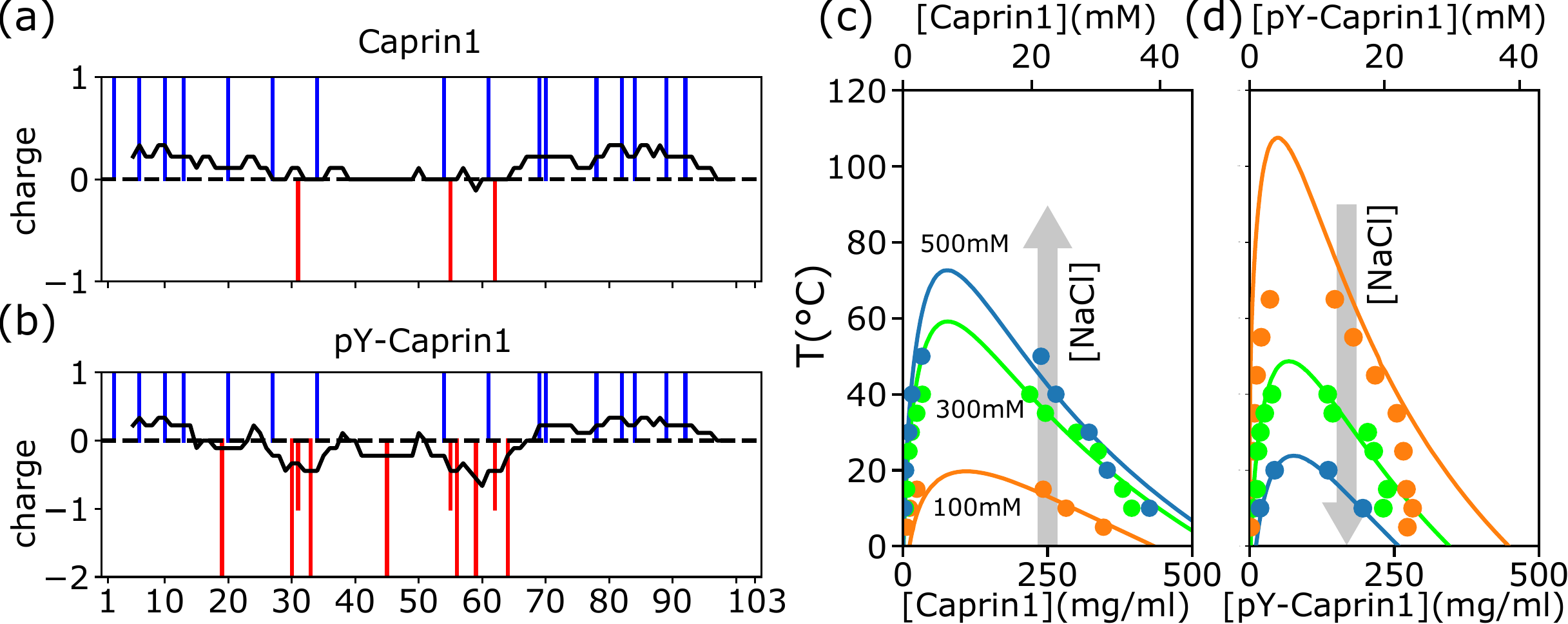}}
\caption{rG-RPA+FH theory predictions rationalize different salt dependence
of Caprin1 and pY-Caprin1 LLPS. (a,b) Vertical lines indicate the sequence
positions (horizontal variable) of positively charged residues (blue)
and negatively charged residues or phosphorylated tyrosines (red) for
(a) Caprin1 and (b) pY-Caprin1. (c,d) rG-RPA+FH coexistence 
curves (phase diagrams, continuous curves color-coded for the NaCl 
concentrations indicated) agree reasonably well with experiment
(dots, same color code). The grey arrows in (c,d) highlight
that when [NaCl] increases, LLPS propensity increases for (c) Caprin1
but decreases for (d) pY-Caprin1.
As described in our prior RPA+FH and rG-RPA+FH 
formulations \cite{linPRL,kings2020}, the theoretical coexistence curves
shown in (c,d) are determined by fitting an effective relative 
permittivity $\epsilon_{\rm r}$ as well as the enthalpic and 
entropic parts of a FH parameter $\chi(T) = \epsilon_h/T^* + \epsilon_s$ 
to experimental data.  For the present Caprin1 and pY-Caprin1 systems, 
the fitted $\epsilon_{\rm r}=80.5$, which is remarkably close to that 
of bulk water ($\epsilon_{\rm r}\approx 78.5)$.
The fitted $(\epsilon_h,\epsilon_s)$ is $(1.0,0.0)$ for Caprin1 
and $(1.0,-1.5)$ for pY-Caprin1. These fitted energetic parameters 
are equivalent \cite{linPRL}
to $\Delta H\approx -1.1$ kcal mol$^{-1}$ and $\Delta S =0.0$
for forming a residue-residue contact in the Caprin1 system (c) (i.e., it is
enthalpically favorable), and
$\Delta H\approx -1.1$ kcal mol$^{-1}$ and 
$\Delta S \approx -3.0$ cal mol$^{-1}$K$^{-1}$
for forming a residue-residue contact in the pY-Caprin1 system (d) (i.e., it
is enthalpically favorable and entropically unfavorable).
}
\label{fig1}
\end{figure*}

\vfill\eject

\begin{figure}[ht]
{\includegraphics[width=\columnwidth,angle=0]{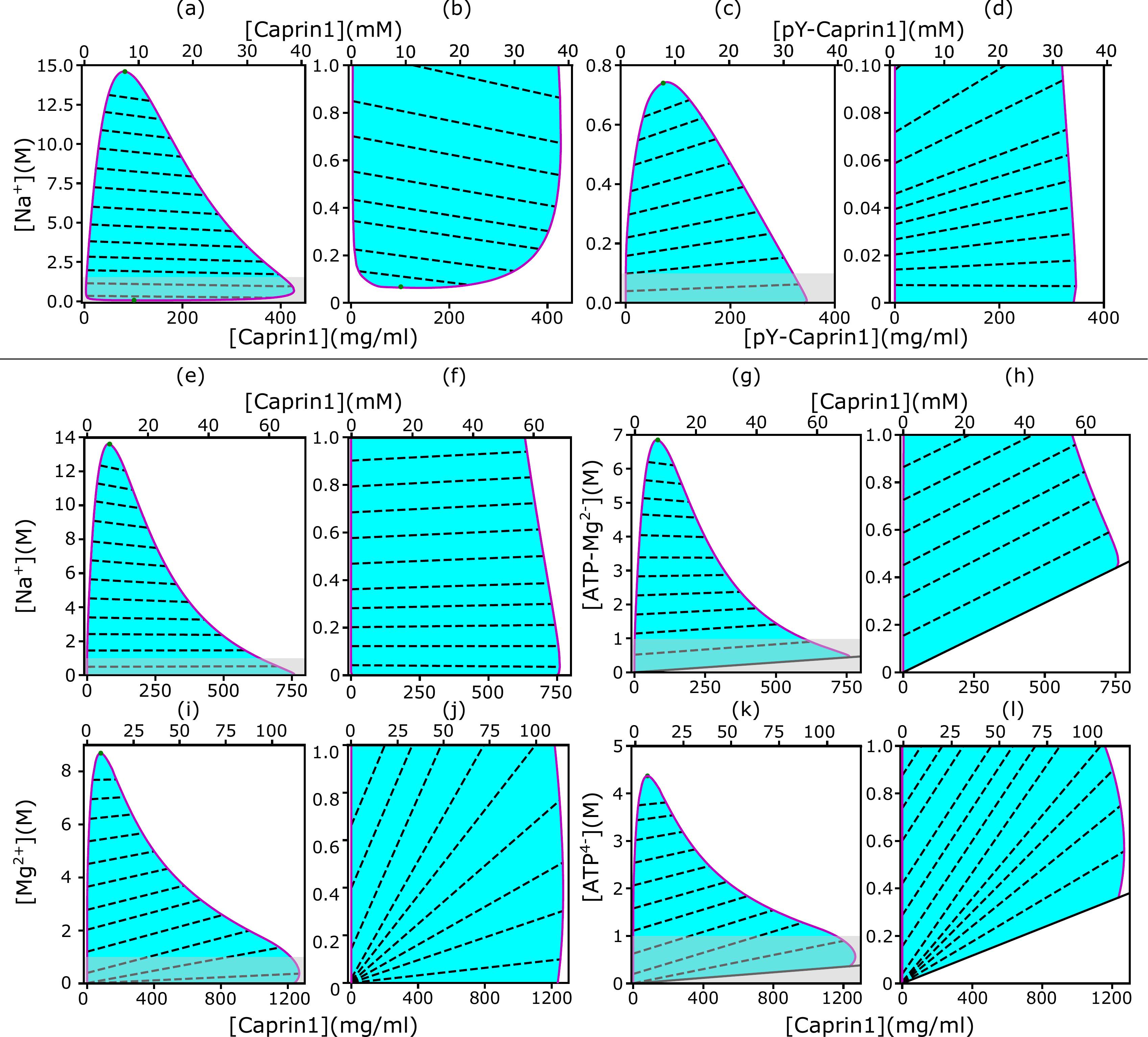}}
\caption{rG-RPA+FH theory rationalizes [NaCl]-modulated
reentrant phase behavior of Caprin1.  
In each salt-protein phase
diagram ($T=300$ K), tielines (dashed) connect coexisting phases on the
boundary (magenta curve)
of the cyan-shaded coexistence region.
For clarity, zoomed-in views of the grey-shaded part in
(a, c, e, g, i, k) are provided by the plots to
the right, i.e., (b, d, f, h, j, l), respectively.
The solid inclined lines in (g, h, k, l) mark the minimum counterion
concentrations required for overall electric neutrality.
Results are shown for monovalent cation and anion with
Caprin1 (a, b) or pY-Caprin1 (c, d); or monovalent cation
and divalent anion with Caprin1 (e--h); or divalent cation and tetravalent
anion with Caprin1 (i--l).
Cation-modulated reentrant phase behaviors is seen for a wide
concentration range for Caprin1 in (a, b) but only a very narrow range of high
Caprin1 concentrations in (e, f, i, j).
The $(\epsilon_h,\epsilon_s)$ values for computing the phase diagrams here
for Caprin1 and pY-Caprin1, respectively,
are the same as those used for Fig.~1c and Fig.~1d.
}
\label{fig2}
\end{figure}

\vfill\eject

\begin{figure}[ht]
{\includegraphics[width=\columnwidth,angle=0]{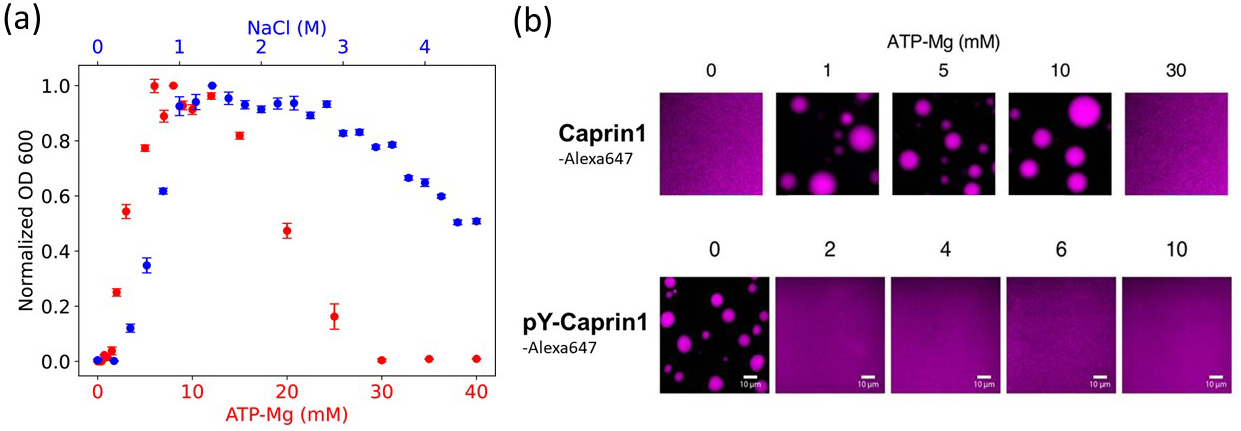}}
\caption{Experimental demonstration of [ATP-Mg]- and [NaCl]-modulated
reentrant phase behavior for Caprin1.
(a) Turbidity quantified by optical density at 600 nm (OD600, normalized
by peak value) to assess Caprin1 LLPS propensity
at [Caprin1] =
200 $\mu$M [for ATP-Mg dependence (red), bottom
scale] or [Caprin1] = 300 $\mu$M [for NaCl dependence (blue), top
scale], measured at room temperature ($\sim 23^\circ$C).
Error bars are one standard deviations of triplicate measurements,
which in most cases was smaller than the plotting symbols.
The ATP-Mg dependence seen here for 200 $\mu$M Caprin1
is similar to the results for 400 $\mu$M Caprin1 (Fig.~6C
of ref.~\cite{LewisPNAS2021}).
(b) Microscopic images of Caprin1 and pY-Caprin1 at varying [ATP-Mg]
at room temperature,
showing reentrant behavior for Caprin1 but not for pY-Caprin1.
Each sample contains 200 $\mu$M of either Caprin1 or
pY-Caprin1, with 1\% of either Caprin1-Cy5 or pY-Caprin1-Cy5 (labeled
with Cyanine 5 fluorescent dye) added for visualization, in a 25 mM
HEPES buffer at pH 7.4. Scale bars represent 10 $\mu$m.
}
\label{fig3}
\end{figure}

\vfill\eject

\begin{figure}[ht]
{\includegraphics[width=0.68\columnwidth,angle=0]{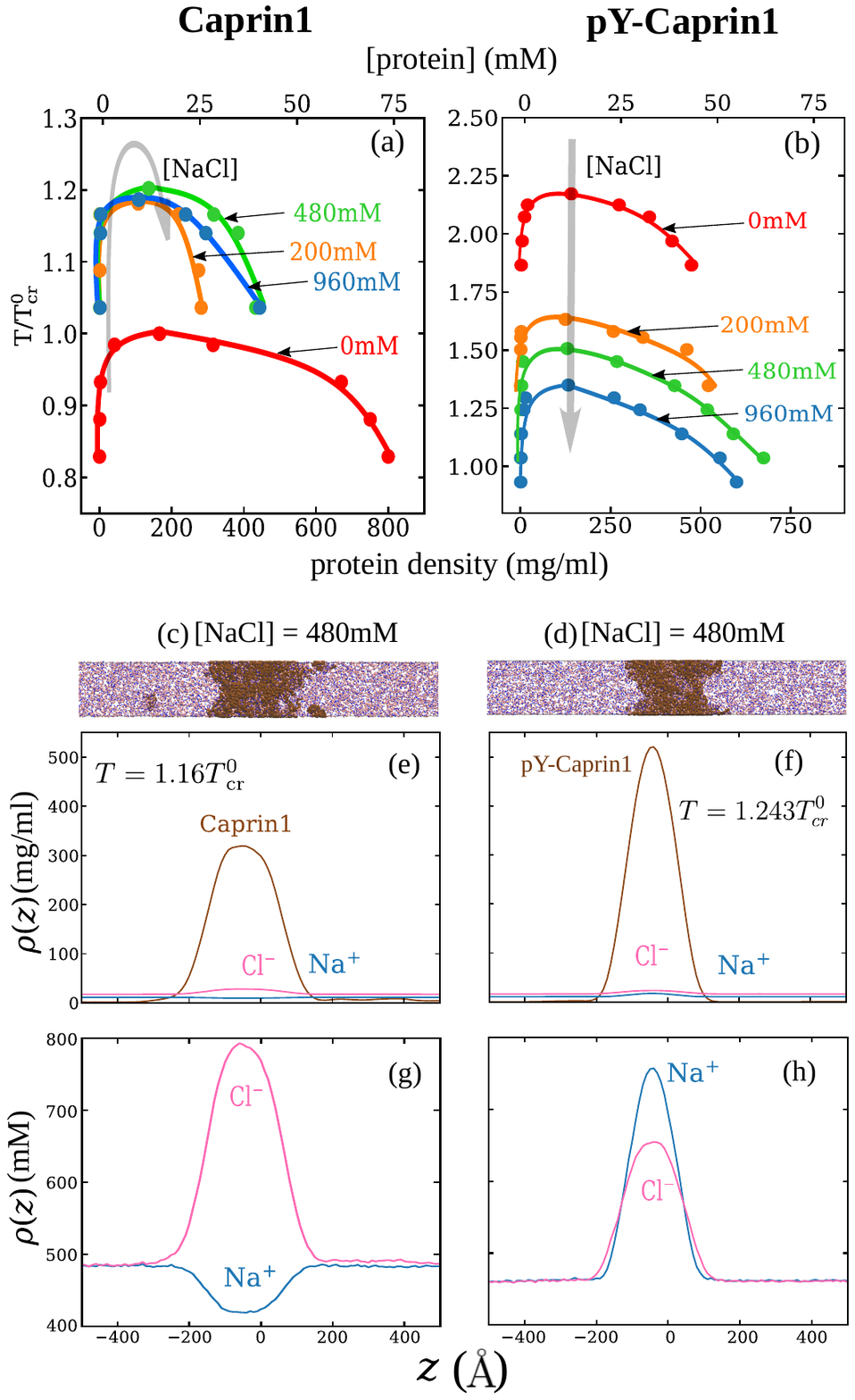}}
\caption{Explicit-ion coarse-grained MD
rationalizes [NaCl]-modulated reentrant behavior for Caprin1
and lack thereof for pY-Caprin1. (a) Simulated phase diagrams (binodal curves)
of Caprin1 at different temperatures plotted in units of 
$T^0_{\rm cr}$ (see text). Symbols are simulated data points.
Continuous curves are guides for the eye. Grey arrow indicates variation
in [NaCl]. (b) Same as (a) but for pY-Caprin1.
(c) A snapshot showing phase equilibrium between dilute and 
condensed phases of Caprin1 (brown chains) immersed in Na$^+$ (blue)
and Cl$^-$ (red) ions simulated at [NaCl] = 480 mM.
(d) A similar snapshot for pY-Caprin1.
(e, f) Mass density profiles, $\rho(z)$ (in units of mg/ml), 
of Na$^+$, Cl$^-$, and (e) Caprin1
or (f) pY-Caprin1 along the elongated dimension $z$ of the simulation box
showing variations of Na$^+$ and Cl$^-$ concentrations between the
protein-dilute phase (low $\rho$ for protein) and protein-condensed
phase (high $\rho$ for protein) at the simulation temperatures indicated.
(g, h) Corresponding zoomed-in concentration profiles $\rho(z)$ in units 
of mM for Na$^+$ and Cl$^-$. Additional mass density profiles for
[NaCl]$=200$ mM and 400 mM are provided in Appendix-figure~3.
}
\label{fig4}
\end{figure}

\vfill\eject

\begin{figure}[ht]
{\includegraphics[width=\columnwidth,angle=0]{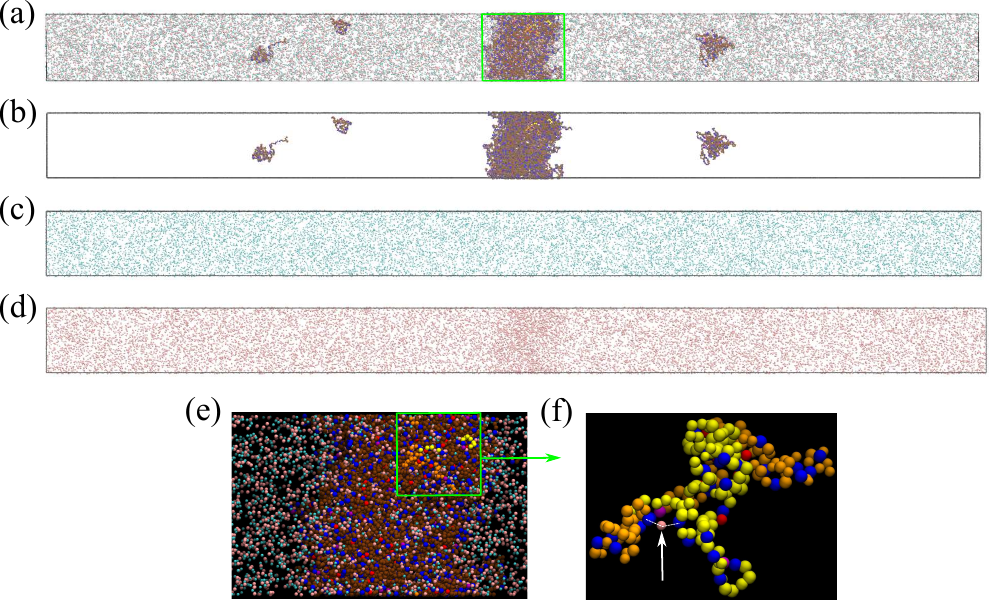}}
\caption{Counterions can stabilize Caprin1 condensed phase by favorable bridging
interactions. 
(a) Snapshot from explicit-ion coarse-grained MD
under LLPS conditions for Caprin1, showing the spatial
distributions of Caprin1, Na$^+$, and Cl$^-$ (as in Fig.~4c).
The three components of the same snapshot are also shown separately
in (b) Caprin1, (c) Na$^+$, and (d) Cl$^-$.
(e) A zoomed-in view of the condensed droplet [corresponding to the green
box in (a)], now with a black background and a different color scheme.
(f) A further zoomed-in view of the part enclosed by the green box in
(e) focusing on two interacting Caprin1 chains. A Cl$^-$ ion
(pink bead indicated by the arrow) is seen interacting favorably with 
two arginines residues (blue beads) on the two Caprin1 
chains (whose uncharged residues are colored differently by yellow or orange,
lysine and aspartic acids in both chains are depicted, respectively,
in magenta and red).
}
\label{fig5}
\end{figure}

\vfill\eject

\begin{figure}[ht]
\vskip -1.0cm
{\includegraphics[width=0.68\columnwidth,angle=0]{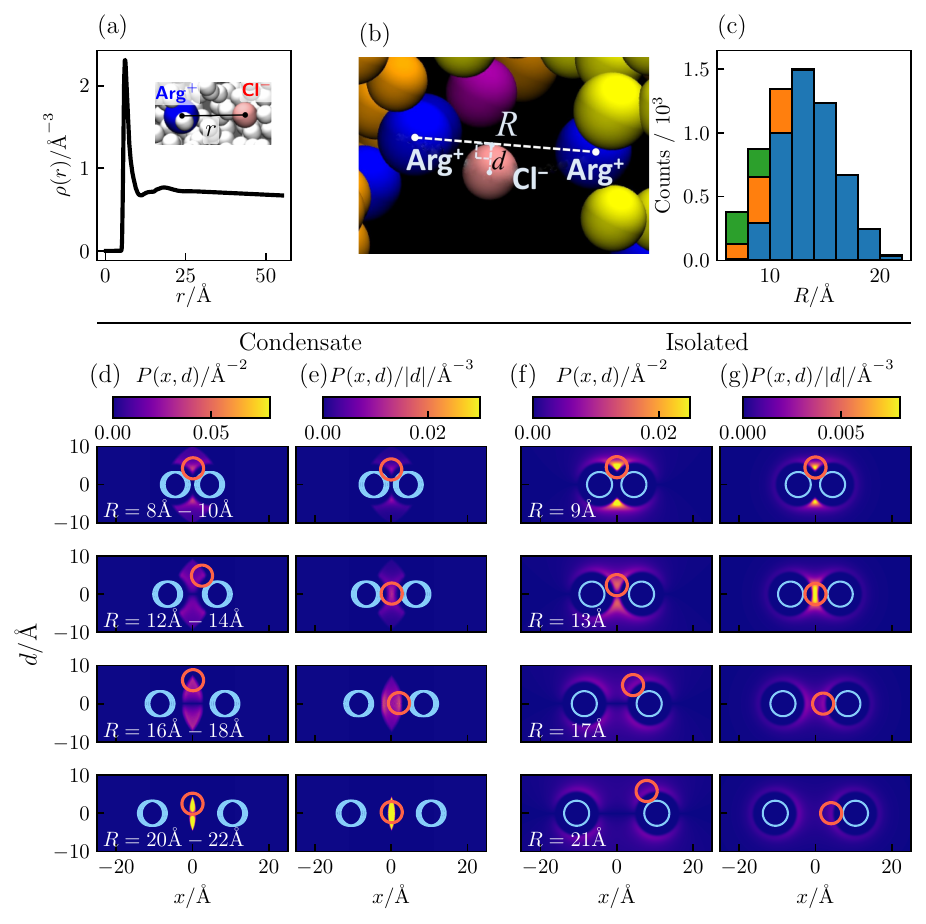}}
\vskip -0.3cm
\caption{Counterion interactions in polyelectrolytic Caprin1. Shown 
distributions are averaged from 4,000 equilibrated coarse-grained MD 
snapshots of 100 Caprin1 chains and 1,300 Cl$^-$ counterions under 
phase-separation conditions ($T/T^0_{\rm cr}=160/193=0.829$) in a 
$115\times 115\times 1610$~\AA$^3$ simulation box in which essentially
all Caprin1 chains are in a condensed droplet. 
(a) Radial distribution function of Cl$^-$ around a positively
charged arginine residue (Arg$^+$). (b) A zoomed-in view of Fig.~5f
showcasing a putative bridging configuration with a Cl$^-$ interacting
favorably with a pair of Arg$^+$s on two different Caprin1 chains. 
Configurational geometry is characterized by Arg$^+$--Arg$^+$ distance $R$
and the distance $d$ of the Cl$^-$ from the line connecting the two Arg$^+$s.
(c) Distribution of putative bridging interaction configurations
with respect to $R$. Numbers of true bridging, neutralizing, and intermediate
are, respectively, in blue, green and orange. 
(d, e) Heat maps of two-dimensional projections of
spatial distributions of Cl$^-$ around 
two Arg$^+$s satisfying the putative bridging interaction conditions among 
the MD snapshots. (f, g) Corresponding projected 
distributions of isolated Arg$^+$--Cl$^-$--Arg$^+$ Boltzmann-averaged
systems at model temperature $T$.
Here, $P(x,d)$ is the total density of Cl$^-$ on a circle of radius $|d|$ 
perpendicular to the heat map at horizontal position $x$ (d, f); thus
the average Cl$^-$ density at a given point $(x,d)$ 
is $P(x,d)/2\pi|d|$,
the patterns of which are exhibited by $P(x,d)/|d|$ heat maps in (e, g). 
$P(x,d)$ is symmetric with respect to $d\leftrightarrow -d$ by 
construction, i.e., $P(x,d)=P(x,-d)$.
In each heat map, the size and (ranges of) positions of model Arg$^+$s 
are indicated by blue circles;
the size and the position or one of two positions (at $\pm d$)
of maximum Cl$^-$ density is indicated 
by a magenta circle.
The MD-simulated distributions of the condensed system (d, e) are quite
similar to the theory-computed isolated system (f, g) for $R\lesssim 14$~\AA,
indicating that individual bridging interactions in the crowded Caprin1 
condensates may be understood approximately by the electrostatics of 
an isolated, three-bead Arg$^+$--Cl$^-$--Arg$^+$ system.
For larger $R$, the heat maps in (f, g) and (d, e) are not as similar 
because some of the configurations in the isolated 
system (f, g) are precluded by the requirement that Arg$^+$--Cl$^-$ 
distance $< 11$~\AA~for putative bridging interactions in (d, e).
}
\label{fig6}
\end{figure}

\vfill\eject

\begin{figure}[ht]
{\includegraphics[width=\columnwidth,angle=0]{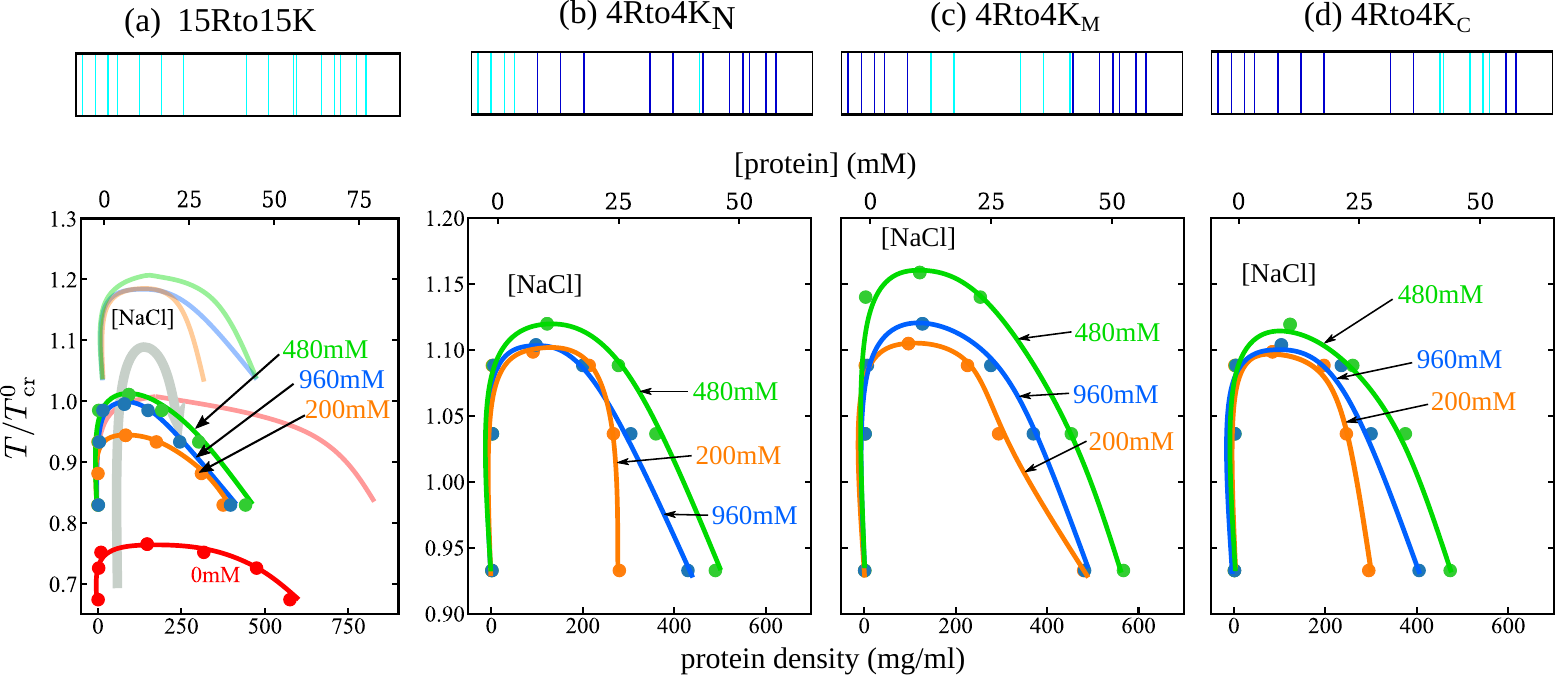}}
\caption{Explicit-ion coarse-grained MD
rationalizes [NaCl]-modulated phase behavior for RtoK
variants of Caprin1. Four variants studied
experimentally \cite{LewisJACS2020} are simulated: (a) 15Rto15K, in which
15 R's in the WT Caprin1 IDR are substituted by K,
(b) 4RtoK$_{\rm N}$, (c) 4RtoK$_{\rm M}$, and (d) 4RtoK$_{\rm C}$,
in which 4 R's are substituted by K
in the (b) N-terminal, (c) middle, and (d) C-terminal regions, respectively.
Top panels show positions of the R (dark blue) and
K (cyan) along the Caprin1 IDR sequence.
Lower panels are phase diagrams in the same style as Fig.~4.
The phase diagrams for WT Caprin1 from Fig.~4a are included
as continuous curves with no data points in (a) for comparison.
}
\label{fig7}
\end{figure}

\vfill\eject

\begin{figure}[ht]
{\includegraphics[width=0.85\columnwidth,angle=0]{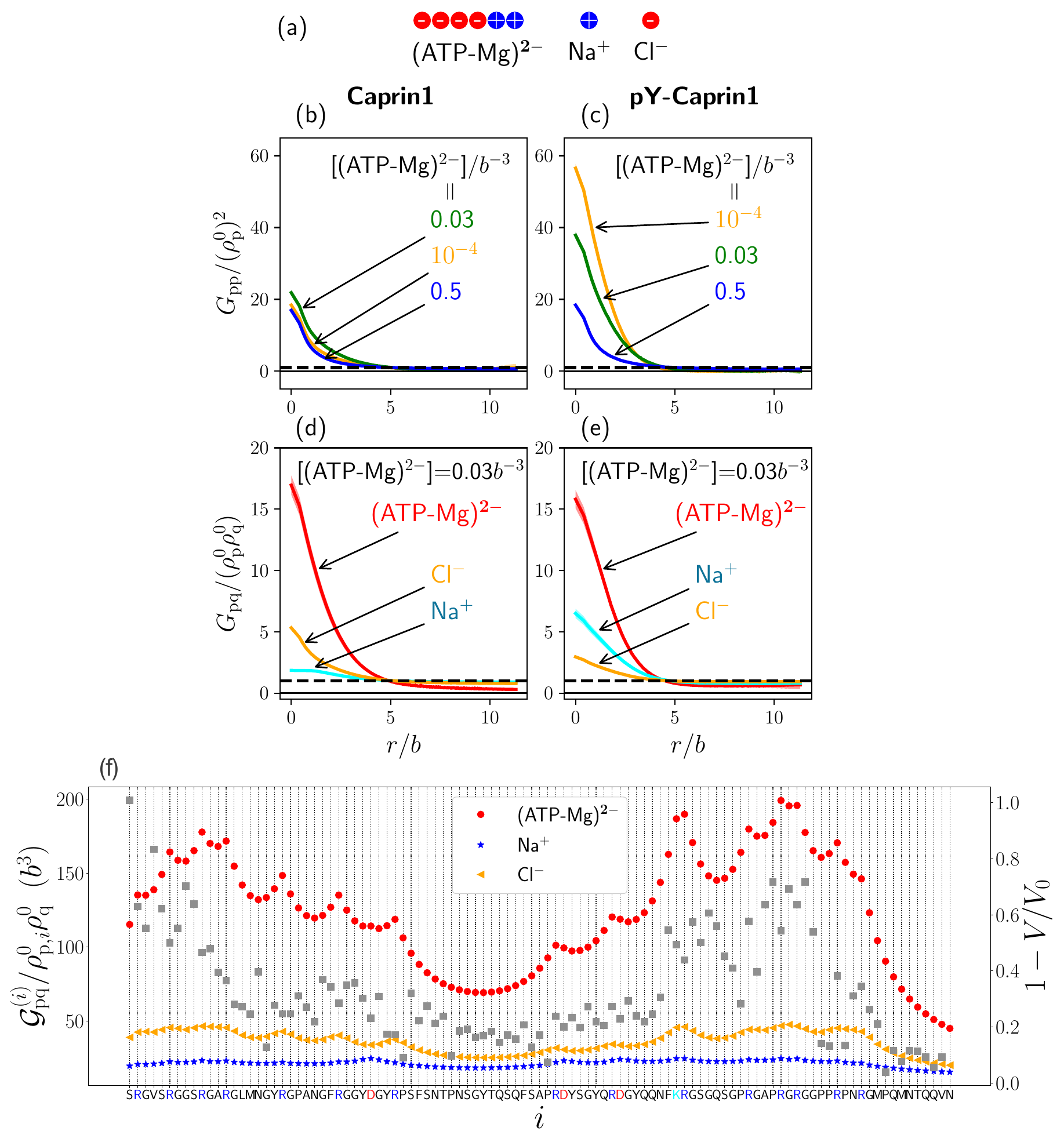}}
\caption{FTS rationalizes experimental 
observation of Caprin1-ATP interactions.
(a) The 6-bead model for (ATP-Mg)$^{2-}$ and the single-bead
models for monovalent salt ions used in the present FTS.
(b--e) Normalized protein-protein correlation functions at three
[(ATP-Mg)$^{2-}$] values (b, c)
and protein-ion correlation functions [Eq.~\eqref{eq:Gij}]
at [(ATP-Mg)$^{2-}]/b^{-3}$ = 0.03 (d, e) for Caprin1 (b, d)
and pY-Caprin1 (c, e),
computed for Bjerrum length $l_{\rm B}=7b$. Horizontal dashed lines
are unity baselines (see text).
(f) Values of position-specific integrated correlation
${\cal G}^{(i)}_{\rm pq}/\rho^0_{{\rm p},i}\rho^0_{\rm q}$ (left vertical
axis) correspond to the relative contact frequencies between individual
residues ($i$) along the Caprin1 IDR sequence with ${\rm q}=$
(ATP-Mg)$^{2-}$, Na$^+$, or Cl$^-$ under the same conditions
as (d) [Eq.~\eqref{eq:calG}] (color symbols).
Included for comparison are experimental NMR volume ratios $V/V_0$
data on site-specific Caprin1-ATP association \cite{LewisPNAS2021}.
$V/V_0$ decreases with increased contact probability, although a precise
relationship is yet to be determined.
Thus, the plotted $1-V/V_0$ (grey data points, right vertical scale)
is expected to correlate with contact frequency.
}
\label{fig8}
\end{figure}

\vfill\eject

\begin{figure}[ht]
{\includegraphics[width=\columnwidth,angle=0]{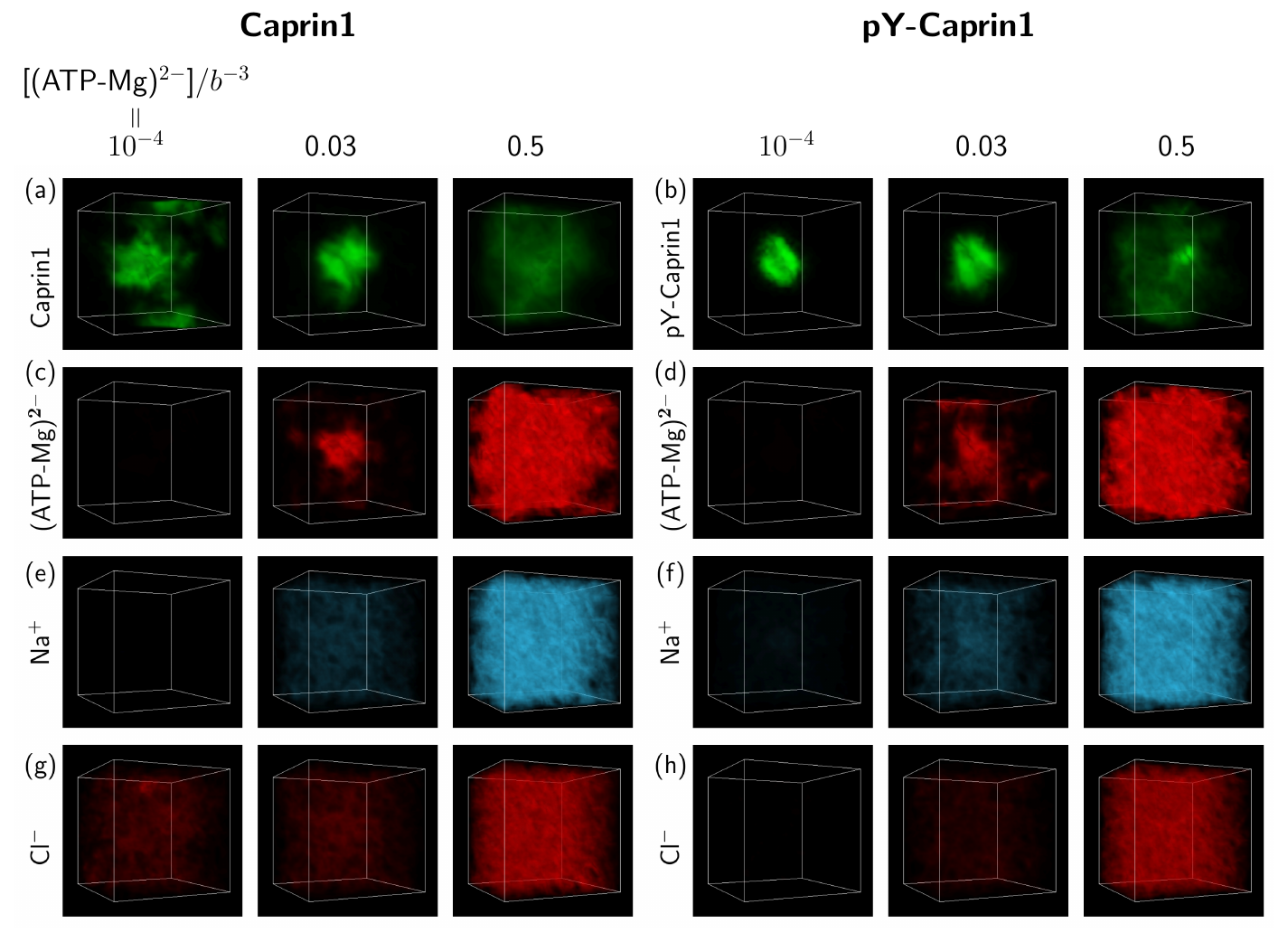}}
\caption{FTS rationalizes colocalization of ATP-Mg with the Caprin1
condensate.
FTS snapshots are from simulations at $l_{\rm B}=7b$ (same as that for Fig.~8).
Spatial distributions of real positive parts of the density fields for
the protein (a, b),
(ATP-Mg)$^{2-}$ (c, d), Na$^+$ (e, f), and Cl$^-$ (g, h) components 
are shown by three snapshots each for
Caprin1 (left panels) and pY-Caprin1 (right panels) at 
different [(ATP-Mg)$^{2-}$] values as indicated.
Colocalization of (ATP-Mg)$^{2-}$ with the Caprin1 condensed droplet
is clearly seen in the [(ATP-Mg)$^{2-}]/b^{-3}$ = 0.03 panel of (c).
}
\label{fig9}
\end{figure}

\vfill\eject

\clearpage

$\null$
\hfill {\bf December 20, 2024}
\vskip 0.3in

\begin{center}

{\Huge\bf Supplementary Information}\\

\vskip 0.3cm

{\Huge ({\it SI Appendix})}\\

\vskip 0.3cm

{\Large\it for}

\vskip 0.3cm

{\Large\bf 
Electrostatics of Salt-Dependent Reentrant}\\

\vskip 0.2cm

{\Large\bf 
Phase Behaviors Highlights Diverse Roles of}\\

\vskip 0.2cm

{\Large\bf 
ATP in Biomolecular Condensates}\\

\vskip .5in
{\bf Yi-Hsuan L{\footnotesize{\bf{IN}}}},$^{1,2,\dagger,\S}$
{\bf Tae Hun K{\footnotesize{\bf{IM}}}},$^{1,2,3,4,\ddagger,\S}$
{\bf Suman D{\footnotesize{\bf{AS}}}},$^{1,5,\S}$
{\bf Tanmoy P{\footnotesize{\bf{AL}}}},$^{1}$\\
{\bf Jonas W{\footnotesize{\bf{ESS\'EN}}}},$^{1}$
{\bf Atul Kaushik R{\footnotesize{\bf{ANGADURAI}}}},$^{1,2,3,4}$
{\bf Lewis E. K{\footnotesize{\bf{AY}}}},$^{1,2,3,4}$
\\
{\bf Julie D. F{\footnotesize{\bf{ORMAN}}}-K{\footnotesize{\bf{AY}}}}$^{2,1}$
and
{\bf Hue Sun C{\footnotesize{\bf{HAN}}}}$^{1,*}$

\vskip 0.3cm

$^1$Department of Biochemistry,
University of Toronto, Toronto, Ontario M5S 1A8, Canada\\
$^2$Molecular Medicine, Hospital for Sick Children, Toronto, 
Ontario M5G 0A4, Canada\\
$^3$Department of Molecular Genetics,
University of Toronto,\\ Toronto, Ontario M5S 1A8, Canada\\
$^4$Department of Chemistry,
University of Toronto, Toronto, Ontario M5S 3H6, Canada\\
$^5$Department of Chemistry, Gandhi Institute of Technology and
Management, Visakhapatnam, Andhra Pradesh 530045, India\\

%
\end{center}
$\null$\\

\noindent
$^\dagger$Present address: HTuO Biosciences, 1001 West Broadway, Suite 300,
Vancouver, British\\
{\phantom{$^\dagger$Present address:\ }} 
Columbia V6H 4B1, Canada.

\vskip 0.1cm

\noindent
$^\ddagger$Present address: Department of Biochemistry, School of Medicine,\\
{\phantom{$^\dagger$Present address:\ }} 
Case Western Reserve University, Cleveland, Ohio 44106, U.S.A. 


\vskip 0.2cm

\noindent
$^\S$Contributed equally.

\vskip 1.3cm

\noindent
$^*$Correspondence information:\\
{\phantom{$^\dagger$}}
Hue Sun C{\footnotesize{HAN}}.$\quad$
E-mail: {\tt huesun.chan@utoronto.ca}\\
{\phantom{$^\dagger$}}
Tel: (416)978-2697; Fax: (416)978-8548\\
{\phantom{$^\dagger$}}
Department of Biochemistry, University of Toronto,
Medical Sciences Building -- 5th Fl.,\\
{\phantom{$^\dagger$}}
1 King's College Circle, Toronto, Ontario M5S 1A8, Canada.\\

\vfill\eject
\renewcommand{\theequation}{{\rm S}\arabic{equation}}
\setcounter{equation}{0}

\noindent{\Large\bf Supplementary Materials and Methods}
\\

\noindent
{\large\bf Experimental 
information additional to that in the maintext}
\\

{\bf Sample preparation -- Wildtype (WT) Caprin1.}
As stated in the maintext, WT Caprin1 was used in all reported 
experiments except those on [NaCl] dependence presented in 
maintext Table~1 and figure~3a.
The amino acid sequence of WT Caprin1 is given in SI Appendix-figure~1
(all supporting figures with figure numbers prefixed by ``S'' 
are provided in this {\it SI Appendix}).
The preparation of sample WT Caprin1 is now briefly described as follows: 
Caprin1 with an
N-terminal His-SUMO tag was produced in BL21 (DE3)-RIPL Codon Plus {\it E.
coli} cells. These cells were cultured until an optical density at 600
nm (OD600) of 0.6 at 37$^\circ$C and then
induced with 0.5 mM IPTG for overnight expression at 23$^\circ$C.
The harvested cells were suspended in a lysis buffer containing 6 M
guanidine hydrochloride (GuHCl), 25 mM Tris, 500 mM NaCl, 20 mM imidazole,
2 mM $\beta$-mercaptoethanol
(BME), at pH 8.0, and lysed via sonication. The supernatant, post-sonication,
was applied to Ni-NTA (Cytiva) and washed with lysis, wash (25 mM Tris, 500 mM
NaCl, 20 mM imidazole, 2 mM BME, at pH 8.0), and elution (25 mM Tris, 500 mM
NaCl, 300 mM imidazole, 2 mM BME, at pH 8.0) buffers. Post-elution, the sample
was treated with ULP1 during dialysis against a dialysis buffer (25 mM Tris,
250 mM NaCl, and 2 mM BME at pH 8.0). This step was followed by His-SUMO tag
removal through Ni-NTA column chromatography. Final purification of Caprin1 was
performed using FPLC with a Superdex 75 16/60 column, equilibrated with a gel
filtration buffer (3 M GuHCl, 25 mM Tris, 500 mM NaCl, 2 mM BME, pH 8.0). The
protein fractions were then dialyzed twice to remove GuHCl before use in
experiments.
\\

{\bf Sample preparation -- Double-mutant (N623T, N630T) variant of Caprin1.}
Our sample preparation for the double-mutant variant used in [NaCl]
dependence studies reported in maintext Table~1 and figure~3a proceeded 
as follows.
To abolish IsoAsp formation for the salt concentration measurements (Table~1)
and  the [NaCl]-dependent turbidity measurements in figure~3a, we used a double
mutant of the Caprin1 IDR (N623T,N630T) in which the two asparagine residues
are mutated to threonine. This double mutant has been shown to exhibit a
similar propensity to phase separate as the WT Caprin1 IDR\cite{LewisPNAS2022}
purified as described previously.\cite{LewisJACS2020,LewisPNAS2022} 

Purified Caprin1 was first exchanged into buffer (25mM sodium phosphate, pH
7.4) via dialysis and was concentrated to $\sim 6$ mM using 3 kDa centrifugal
Amicon
concentrators (EMD Millipore). The pH of the concentrated protein was adjusted
to 7.4 using concentrated hydrochloric acid. 
Phase separated samples of Caprin1 were prepared by addition of a concentrated
stock solution of NaCl (25mM sodium phosphate, pH 7.4, 4M NaCl) to achieve a
bulk (overall) salt concentration of 300 mM NaCl. 
Condensed and dilute phases of Caprin1
were transferred all together into an Eppendorf tube using a syringe.
After rigorous vortexing, the phase separated samples were incubated at the
desired temperatures using a thermocycler with a heated lid (95$^\circ$C).
At least one hour was required to allow droplets to form a large condensed
phase droplet at the bottom of the tube.

2 $\mu$L of condensed and dilute phases were pipetted into 
48 $\mu$L of a 2.8 M urea
solution (U4883, Sigma) in MilliQ water in a 15 mL falcon tube, using a positive
and an air displacement pipette, respectively. The outside of the tips were
wiped with a KimWipe to remove excess protein, prior to transferring into the
urea solutions. Following transfer, the samples were digested for 
inductively coupled plasma optical emission spectroscopy (ICP-OES)
measurements by the addition of 630 $\mu$L of concentrated nitric acid (67\%,
NX0407, Sigma) and 630 $\mu$L hydrogen peroxide (95321, Sigma), and incubated 
in an oven at 60$^\circ$C for 54 hours. Post digestion, the sample tubes were
cooled at room temperature and centrifuged. 40 $\mu$L of hydrogen peroxide was
then added to the samples. No bubbles were observed, indicating the
completion of the digestion process. The samples were then bought up to
12 mL using MilliQ water, to achieve a final nitric acid concentration
of 3.5\%. Blank samples for the condensed and dilute phases were prepared
by pipetting 2 $\mu$L of MilliQ water using a positive and an air displacement
pipette, respectively, and subsequently following the digestion protocol
described above. Sodium standards (0.1, 0.2, 0.5, 1, 2, 4, 8 and 10 ppm)
in 3.5\% nitric acid for ICP-OES measurements were prepared by dilution
of sodium standard solution (00462, Sigma) with MilliQ water and
concentrated nitric acid (67\%, NX0407, Sigma). All samples were filtered
using 0.22 $\mu$m syringe filters prior to ICP-OES. Condensed
and dilute phases were drawn in triplicate at each temperature.
\\

{\bf Phosphorylation of the Caprin1 IDR.}
The purified protein was initially concentrated to 25-50 $\mu$M in a reaction
buffer comprising 25 mM Tris pH 7.4, 50 mM KCl, 10 mM MgCl$_2$, 3 mM ATP and 2
mM DTT. This mixture was then placed into a dialysis tubing with a 3 kDa
cut-off. To the protein sample, purified His-SUMO-Eph4A was added to 5--10
$\mu$M, and the reaction mixture was subsequently dialyzed against 4 liters of
the same reaction buffer, either at room temperature or at 4$^\circ$C
overnight.  Mass spectrometry indicates that the resulting sample consists
mainly of a mixture of Caprin1 IDRs with six or seven phosphorylations and a
very small fraction of IDRs with five phosphorylations 
(SI Appendix-figure~2). 
\\

{\bf Determination of phase diagrams.}
The initial homogenization described in the maintext 
ensures that the condensed phase in small droplets
can rapidly equilibrate with the dilute phase.
Absorbance at 280 nm was measured and converted to concentration using the
Beer-Lambert law, with an extinction coefficient ($\epsilon$) of 10,430
M$^{-1}$cm$^{-1}$, based on the molecular masses of 11,108 Da for Caprin1 and
11,668 Da for pY-Caprin1.  The reported concentration values and uncertainties,
calculated as means and standard deviations, were derived from triplicate
measurements.
\\

{\bf Concentrations of salt and ATP-Mg in dilute and condensed phases.}
ICP-OES measurements were performed in triplicate for each sample.  Mean value
and uncertainty for the salt concentration were obtained by taking the average
and standard deviation over the triplicate samples at the given temperature
(maintext Table~1). Specific details of sample preparation for the set of 
ATP-Mg--dependent experiments
are provided above in this {\it SI Appendix}.
As for salt-dependent experiments,
these measurements were performed in triplicate and standard deviations were
calculated to assess experimental uncertainties (maintext Table~2).
\\

{\bf Caprin1 phase separation propensity at high salt concentrations.}
Averages and standard deviations over the three OD600 measurements 
were reported by the blue symbols in maintext figure~3a.
\\

{\bf [ATP-Mg]-dependent Caprin1 phase behaviors.} A brief summary of the
turbility assays in ref.~\citen{LewisJACS2020} that we utilized is as follows:
The WT Caprin1 IDR was diluted to a 200 $\mu$M concentration using a buffer
composed of 25 mM HEPES and 2 mM DTT at pH 7.4, with varying levels of ATP-Mg.
Samples were prepared with ATP-Mg concentrations ranging from 0 to 40 mM.
Following thorough mixing, 5 $\mu$L of each sample was placed into a
$\mu$Cuvette G1.0 (Eppendorf), and OD600 was measured using a BioPhotometer D30
(Eppendorf).  This procedure was performed three times for analysis of
experimental uncertainties (red symbols in maintext figure~3a).
\\


\noindent
{\large\bf Sequence-specific theory of heteropolymer phase separation --\\
Summary of key steps in the field-theoretic formulation}
\\

The following is a more extensive summary to supplement the brief 
outline in {\it Materials and Methods} of the maintext provided under the
heading ``squence-specific theory of heteropolymer phase separation''.
In general, 
sequence-specific polymer field theories\cite{linPRL,kings2020}
are constructed to model systems of polymers with
various salt and counterions. 
Further details are available from our recent
publications (e.g., refs.~\citen{MiMB2023,WessenJPCB2022} and 
references therein).

Using the same notation for the partition function $\Z$ in Eq.~(2), 
with $\T+\U$ being the Hamiltonian in units of 
the product $\kB T$ of
Boltzmann's constant $\kB$ and absolute temperature $T$, 
the connectivity term $\T$ is given by 
\begin{equation}
\T[\R] = \frac{3}{2l^2}\sum_{\alpha=1}^{\np}\sum_{i=1}^{N-1}
\left(\R_{\alpha,i+1}-\R_{\alpha,i}\right)^2 \; 
        \label{eq:TiniSI}
\end{equation}
with $[\R]$ being shorthand for $[\{\R_{\alpha,i}\}]$. 
Considering the case when the total potential energy $\U$ 
is taking one of its simplest forms, in that it serves only to model
the two-body (pairwise) interactions among polymer residues, salt ions, and
counterions, we further confine the interaction types in our
formulation to Coulomb electrostatics (el) and excluded-volume (ex).
As stated in the maintext, their interaction strengths are
governed by the Bjerrum length $\lb$
($\lb = e^2/4\pi\epsilon_0\epsilon_{\rm r}\kB T$, where $e$ is protonic 
charge, $\epsilon_0$ is vacuum permittivity, and $\epsilon_{\rm r}$ 
is relative permittivity), and the two-body excluded volume parameter $v_2$.
We now provide the precise field-theoretic forms 
for the Coulomb electrostatic potential $\U_{\rm el}$ and two-body 
excluded volume interaction $\U_{\rm ex}$ terms in 
$\U = \U_{\rm el} + \U_{\rm ex}$ by introducing the number density 
operators
\begin{subequations} \label{eq:micro_density}
\begin{align}
\hat{\rho}_{\rm p}(\rr) = & \sum_{\alpha=1}^\np\sum_{i=1}^N
\delta(\rr-\R_{\alpha,i}) \; ,\\
\hat{\rho}_{\rm s}(\rr) = & \sum_{a=1}^\ns \delta(\rr - \rr_a) \; , \\
\hat{\rho}_{\rm c}(\rr) = & \sum_{a=\ns+1}^{\ns+\nc} \delta(\rr - \rr_a) \; ,
\end{align}
\end{subequations}
for the monomers (residues or beads) of the polymer (p), salt (s), and 
counterion (c), 
respectively, where $\delta$ represents the Dirac $\delta$ distribution.
Solvent (water) degrees of freedom [$n_{\rm w}$ in Eq.~(2) for $\Z$ in 
the maintext] are not included in
Eq.~\eqref{eq:micro_density} above because they are only used for the 
incompressibility constraint in our rG-RPA formulation as an approximate 
treatment of excluded volume, and solvents are not treated
explicitly at all in the present field-theoretic simulation (FTS), i.e.,
the $n_{\rm w}$ factor is dropped for FTS. 
The corresponding charge density operators for the number density
operators in Eq.~\eqref{eq:micro_density} are
\begin{subequations}  \label{eq:micro_charge_density}
\begin{align}
\hat{c}_{\rm p}(\rr) = & \sum_{\alpha=1}^\np\sum_{i=1}^N\sigma_i
\delta(\rr-\R_{\alpha,i}) \; , \\
\hat{c}_{\rm s}(\rr) = & \zs \sum_{a=1}^\ns \delta(\rr - \rr_a) = 
\zs\hat{\rho}_{\rm c} \; , \\
\hat{c}_{\rm c}(\rr) = & \zc \sum_{a=\ns+1}^{\ns+\nc} \delta(\rr - \rr_a) = 
\zc\hat{\rho}_{\rm s} \; .
\end{align}
\end{subequations}
For $\hat{c}_{\rm p}$, $\sigma_i=+1$ for arginine and lysine,
$\sigma_i=-1$ for aspartic and glutamic acids, $\sigma_i=-2$ for
phosphorylated tyrosine, and $\sigma_i=0$ for
all other amino acid residues in the Caprin1/pY-Caprin1 sequences 
studied (as stated in the maintext).
$\U_{\rm el}$ and $\U_{\rm ex}$ are now given by
\begin{subequations}
\begin{align}
\U_{\rm el}  = &  \frac{\lb}{2}\int d\rr d\rr'
        \Big[  \hat{c}_{\rm p}(\rr) + \hat{c}_{\rm s}(\rr) + 
\hat{c}_{\rm c}(\rr)  \Big]
        \frac{1}{|\rr-\rr'|}
        \Big[   \hat{c}_{\rm p}(\rr') + \hat{c}_{\rm s}(\rr') + 
\hat{c}_{\rm c}(\rr')  \Big], \\
\U_{\rm ex}  = &  \frac{v_2}{2}\int d\rr 
        \Big[  \hat{\rho}_{\rm p}(\rr) + \hat{\rho}_{\rm s}(\rr) + 
\hat{\rho}_{\rm c}(\rr)  \Big]^2 \; .
\end{align}
\end{subequations}

As mentioned in the maintext, two conjugates fields, $\psi(\rr)$ 
for Coulomb interaction and $\w(\rr)$ for excluded volume, 
are then introduced to linearize the density operators that 
are quadratic in $\U_{\rm el}$ and $\U_{\rm ex}$
by applying the Hubbard-Stratonovich transformation,\cite{Strat57,Hub59}
resulting in a reformulated partition function 
$\Z' \equiv (\np!\ns!\nc!\nw!)\Z$ [with $\Z$ given by Eq.~(2) of maintext] 
expressed as a functional integral over the fields $\psi$ and $\w$:
\begin{equation}
\Z' = \int \DD \psi \DD\w\exp
        \left[ -\int d\rr 
\left( \frac{\left[\nabla \psi(\rr) \right]^2}{8\pi\lb} + 
\frac{\w(\rr)^2}{2v_2}\right)
        + \np\ln\Qp + \ns\ln\Qs + \nc\ln\Qc
        \right],
        \label{eq:Zprime}
\end{equation}
where $\Qp$, $\Qs$, and $\Qc$ are single-molecule partition functions
of polymer, salt ion, and counterion, respectively, which are all 
functionals of $\psi$ and $\w$:
\begin{subequations}
\label{eq:Hp0}
\begin{align}
\Qp[\psi, \w] = & \int \prod_{\tau=1}^N d\R_\tau 
\exp\Big( -\Hp\left[ \R; \psi, \w \right] \Big)  \nonumber \\
        = & \int \prod_{\tau=1}^N d\R_\tau
                 \exp\left[ -\frac{3}{2l^2}
                  \sum_{i=1}^{N-1}\left(\R_{i+1}-\R_{i}\right)^2
                - \i\sum_{i=1}^N
\left(  \sigma_i\psi(\R_i) + \w(\R_i) \right)
        \right] , \\
{\cal Q}_{\rm s,c}[\psi, \w] = & \int d\rr 
\exp\Big[ -\i\big( z_{s,c} \psi(\rr) + \w(\rr)   \big)    \Big] \; ,
\end{align}
\end{subequations}
wherein $\i$ is the imaginery unit, i.e., $\i^2=-1$.

The $\Z'$ in Eq.~(\ref{eq:Zprime}) can be analyzed via various field
theoretic approaches. Two approaches are utilized in the present work:
(i) one-loop perturbation expansion is employed to derive analytical
theories based upon random-phase-approximation (RPA), and (ii) 
field-theoretic simulation (FTS) is conducted to compute observables
numerically.
\\


\noindent
{\large\bf Renormalized-Gaussian random-phase-approximation (rG-RPA)}

As mentioned in the maintext,
sequence-specific random phase approximation (RPA) has been applied 
successfully to model electrostatic effects on the LLPSs of various 
polyampholytic IDRs\cite{linPRL,lin2017,biochemrev,SumanPNAS,Wessen2021}
to obtain behaviorial trends consistent with experiments and 
explicit-chain simulations; but
RPA is less appropriate for polyelectrolytes with large
net charge per residue (NCPR)\cite{Mahdi2000,Ermoshkin2003,Orkoulas2003} 
because of RPA's treatment of polymers as ideal Gaussian 
chains.\cite{MuthuMacro2017}
This approximation is reasonable for polyampholytes but not for
polyelectrolytes. While overall intrachain electrostatic effects in
polyampholytes can be mild because of the polymers' nearly zero net charge and
thus entail only a minor perturbation on conformational statistics, repulsive
electrostatics in polyelectrolytes with significant net charge is strong,
leading to more rod-like conformations with statistics deviating significantly
from that of Gaussian chains. Consequently, treating polyelectrolytes as
Gaussian chains can lead to large errors in theoretical intrachain and
interchain residue-residue (monomer-monomer) correlations, resulting in
drastically overestimated LLPS propensities.\cite{MuthuMacro2017}

The rG-RPA theory was put forth by some of the present authors.\cite{kings2020}
For a broad overview, we briefly summarize here the major methodological steps 
and key results of the theory. Interested readers are referred
to ref.~\citen{kings2020} for further details.
As rG-RPA has been designed and verified to tackle polyeletrolyte
conformations appropriately,\cite{kings2020} we apply it here to the
polyelectrolytic Caprin1 IDR. Because rG-RPA allows for a smooth crossover
between polyelectrolytic and polyampholytic systems, Caprin1 and pY-Caprin1 can
now be analyzed in a universal theoretical formulation without invocation of ad
hoc treatments for their different conformational statistics.

In our formulation of rG-RPA theory, simplifying assumptions are made
to the effect that excluded volume is taken into account only between 
pairs of different polymer chains (no consideration of intrachain 
excluded volume) and small ions are treated as point charges. 
Denoting the input ``bare'' Kuhn length as $l$, and the total free energy
and volume of the system as $F$ and $\Omega$ respectively,
the system free energy in units of $\kB T$ per volume $l^3$
is given by
\begin{equation}
f = \frac {Fl^3}{\kB T\Omega}= -s + \fion + \fzero + f_{\rm p} \; .
        \label{eq:f_sep}
\end{equation}
Here $s$ is translational entropy
\begin{equation}
-s = \frac{\phi_{\rm p}}{N}\ln\phi_{\rm p} + \phi_{\rm s}\ln\phi_{\rm s} 
+ \phi_{\rm c}\ln\phi_{\rm c} + \phi_{\rm w}\ln\phi_{\rm w} \; ,
\end{equation}
where $\phi_{\rm p}$, $\phi_{\rm s}$, $\phi_{\rm c}$, and 
$\phi_{\rm w} = 1-\phi_{\rm p}-\phi_{\rm s}-\phi_{\rm c}$ are,
respectively, volume fractions of polymers, salt ions, counterions, 
and solvent, with the last equality following from the incompressibility 
condition that we have stipulated.
The $\fion$ term in Eq.~\eqref{eq:f_sep} accounts for the free energy 
of the small ions via the form
\begin{equation}
\fion = -\frac{1}{4\pi}\left[ \ln\left(1+\kappa_{\rm D} l \right) 
- \kappa l + \frac{1}{2}\left( \kappa_{\rm D} l\right)^2 \right] \; ,
\end{equation}
where $1/\kappa_{\rm D} = 1/\sqrt{4\pi\lb\left( \zs^2 \rho_{\rm s} 
+ \zc^2\rho_{\rm c} \right)}$ is the Debye screening length.
The term $\fzero$ in Eq.~\eqref{eq:f_sep} is the zeroth-order 
excluded volume effect given by
\begin{equation}
\fzero = \frac{l^3}{2}v_2\rho_{\rm m}^2 \; ,
\end{equation}
where $\rho_{\rm m}=n_{\rm p}N/\Omega$ is the average monmer (residue or bead)
density of the polymers in the system
and the expression $f_{\rm p} = -(l^3/\Omega)\ln \Zp$ for the last term
in Eq.~\eqref{eq:f_sep} is derived from the polymer partition function
\begin{equation}
\Zp = \int \DD \psi \DD\w \exp
        \left[
-\int d\rr \left( \frac{\psi(\rr)
\left(-\nabla^2 + \kappa^2\right)^2 \psi(\rr)}{8\pi\lb}
                + \frac{\w(\rr)^2}{2v_2}\right)
                + \np\ln\Qp
        \right] \; .
\end{equation}
An analytical perturbative field theory may now be derived from $\Zp$ by
considering the Taylor expansion of $\ln\Qp$ up to the second order of 
$\psi$ and $\w$ while omitting terms that do not affect the relative
energies of the configurations, viz., 
\begin{equation}
\ln\Qp \approx -\frac{1}{2}\left[
\avg{  \left(\hat{c}_{\rm p} - \overline{c}_{\rm p} \right)^2}\psi^2 +
\avg{  \left(\hat{\rho}_{\rm p} - \overline{\rho}_{\rm p} \right)^2}\w^2 +
2\avg{  \left(\hat{c}_{\rm p} - \overline{c}_{\rm p} \right) 
\left(\hat{\rho}_{\rm p} - \overline{\rho}_{\rm p} \right)}\psi\w
\right] \; ,
\end{equation}
where $\overline{c}_{\rm p}$ and $\overline{\rho}_{\rm p}$
are the overall average charge and number densities, respectively, 
of the polymer [cf. Eqs.~\eqref{eq:micro_density} and 
\eqref{eq:micro_charge_density} above and Eq.~(A37) of ref.~\citen{kings2020}].
It follows that $\Zp$ can then be approximated as a Gaussian integral 
in the Fourier-transformed $\kk$-space, 
\begin{equation}
\Zp \approx \int \prod_{\kk\neq\kzero}\sqrt{\frac{\nu_k}{v_2}} \frac{d\psi_\kk d \w_\kk}{2\pi\Omega}
        \exp\left[ - \frac{1}{2\Omega}\sum_{\kk\neq\kzero}\bra{\psi_{-\kk}\; \w_{-\kk}}
        \left(
        \begin{matrix}
        \nu_k + \rho_m \gcc{\kk} & \rho_m\gmc{\kk} \\
        \rho_m\gmc{\kk} & v_2^{-1} + \rho_m \gmm{\kk}
        \end{matrix}
        \right)
        \left|
        \begin{matrix}
        \psi_\kk \\
        \w_\kk
        \end{matrix}
        \right\rangle
         \right],
         \label{eq:Zp_quadratic}
\end{equation}
where $\gcc{\kk}$, $\gmm{\kk}$, and $\gmc{\kk}$ are charge-charge, 
mass-mass (i.e., matter-matter), and mass-charge (matter-charge) 
correlation functions in $\kk$-space, and $\nu_k = k^2/(4\pi\lb)+
(\zs^2\rho_{\rm s} + \zc^2\rho_{\rm c})$. 
The free energy $f_{\rm p}$ is then given by
\begin{equation}
f_{\rm p} = \frac{l^3}{2}\int \frac{d^3 k }{(2\pi)^3}\ln\left[
        1 + \rho_m\left( \frac{\gcc{\kk}}{\nu_k} + v_2 \gmm{\kk}\right)
        + \frac{v_2}{\nu_\kk} \rho_m^2 \left( \gcc{\kk}\gmm{\kk} - \left(\gmc{\kk}\right)^2 \right)
        \right].
        \label{eq:fp}
\end{equation}
The correlation functions in Eq.~(\ref{eq:fp}) may be estimated by
various field-theory approximations. In rG-RPA, they are evaluated using
a variational approach to the single-polymer partition functon $\Qp$ by
first expressing the Hamiltonian $\Hp$ in Eq.~(3) of the maintext 
as the sum of an approximate Gaussian-chain Hamiltonian with
an effective (renormalized) Kuhn length $xl $ 
(recall $l$ is the original ``bare'' Kuhn
length) plus the remaining term:
\begin{equation}
\Hp = \Hp^0 + \Hp^1
\; ,
\end{equation}
where
\begin{subequations}
\label{eq:x}
\begin{align}
\Hp^0 = &  \frac{3}{2l^2 x}\sum_{i=1}^{N-1}
\left( \R_{i+1} - \R_i\right)^2 \; , \\
\Hp^1 = & \frac{3}{2l^2}\left(1-\frac{1}{x}\right)\sum_{i=1}^{N-1}
\left( \R_{i+1} - \R_i\right)^2
- \i\sum_{i=1}^N\left[  \sigma_i\psi(\R_i) + \w(\R_i) \right] \; 
\end{align}
\end{subequations}
are the same equations as Eqs.~(4a) and (4b) in the maintext
and adopt essentially the same form as the unrenormalized Eq.~(\ref{eq:Hp0}a).
To make progress, we take the polymer square end-to-end distance 
$\R_{\rm EE}^2$ as a key physical observable and require its 
thermodynamic average to be produced by $\Hp^0$ in good approximation. 
Based on this premise, a variational theory as described 
in ref.~\citen{kings2015} is applied to calculate the $x$ parameter in
the above Eq.~\eqref{eq:x}.
Details of the derivation are given in the Appendices of ref.~\citen{kings2020}.
Here we show only the variational equation for solving $x$:
\begin{equation}
\label{eq:Xi}
1 - \frac{1}{x} - \frac{Nl^3}{18(N-1)}
\int \frac{d^3 k }{(2\pi)^3}\frac{k^2 \Xi_k^x}{\det \Delta_k^x} = 0 \; ,
\end{equation}
where $\Delta_k^x$ is the $2\times2$ matrix in Eq.~\eqref{eq:Zp_quadratic}
with $l\rightarrow xl$ (renormalized Kuhn length), 
$|i-j|\rightarrow |i-j|/x$ (renormalized contour length) and
thus $l^2|i-j|\rightarrow l^2 x|i-j|$ 
such that the correlation functions in $\Delta_k^x$ become\cite{kings2020}
\begin{subequations}
\begin{align}
\gcc{\kk} \to  &  \gcc{k, x} =
        \frac{1}{N}\sum_{i,j=1}^N \sigma_i \sigma_j 
e^{-\frac{1}{6}(kl)^2x|i-j|} \; , \\
\gmm{\kk} \to  &  \gmm{k, x} =  \frac{1}{N}
\sum_{i,j=1}^N e^{-\frac{1}{6}(kl)^2x|i-j|} \; , \\
\gmc{\kk} \to &  \gmc{k, x} = 
\frac{1}{N}\sum_{i,j=1}^N \sigma_i e^{-\frac{1}{6}(kl)^2x|i-j|} \; ,
\end{align}
\end{subequations}
and the $\Xi_k^x$ in the integrand on the right hand side of 
Eq.~\eqref{eq:Xi} is now given by
\begin{equation}
\Xi_k^x =  \frac{\LGss_k^x}{v_2} + \nu_k \LGdd_k^x
        + \rho_m\left( \LGss_k^x \gmm{k, x} +  \LGdd_k^x \gcc{k, x} 
         - 2\LGds_k^x \gmc{k,x}\right)
\end{equation}
where
\begin{subequations}
\begin{align}
\LGss_k^x = &
        \frac{1}{N}\sum_{i,j=1}^N |i-j|^2 \sigma_i \sigma_j 
e^{-\frac{1}{6}(kl)^2x|i-j|} \; , \\
\LGdd_k^x = & \frac{1}{N}\sum_{i,j=1}^N  |i-j|^2 
e^{-\frac{1}{6}(kl)^2x|i-j|} \; ,\\
\LGds_k^x = & \frac{1}{N}\sum_{i,j=1}^N  |i-j|^2 \sigma_i 
e^{-\frac{1}{6}(kl)^2x|i-j|} \; .
\end{align}
\end{subequations}

{\bf The rG-RPA+FH formulation.}
Because the above field theory is formulated to focus only on sequence-specific 
electrostatics and excluded volume, in the form presented above it does not 
account for short-range attractions such as those arising from $\pi$-related 
and hydrophobic effects; but we need to take these effects into consideration
to arrive at a more direct comparison between rG-RPA predictions and
experiments, e.g., as those provided in figure~1c,d of the maintext. To account
for these interactions approximately in Caprin1, particularly the 
interactions involving $\pi$-electrons,\cite{LewisPNAS2021} we
introduce, as before,\cite{linPRL} a temperature-dependent Flory-Huggins (FH) 
interaction to augmented the free energy $f$ in 
Eq.~(\ref{eq:f_sep}),\cite{kings2020,SumanPNAS} resulting in an
overall total free energy 
\begin{equation}
f = -s + \fion + \fzero + f_{\rm p} 
- \left(\epsilon_h /T^* + \epsilon_s \right) \rho_{\rm m}^2 \; ,
\end{equation}
where $\epsilon_h$ and $\epsilon_s$ are the enthalpic and entropic
components, respectively, of the mean-field Flory-Huggins interaction for 
favorable non-electrostatic attraction, and $T^*=l/\lb$ is the 
reduced temperature.  With this augmented rG-RPA+FH
system free energy $f$ in hand, we solve the solute and 
solvent concentrations in dilute and condensed phases by balancing 
the chemical potentials of each solute components and the osmotic 
pressures in the two phases.
When the salt concentration is assumed for simplicity to be uniform throughout 
the system (figure~1c,d of the maintext), the system only has $\rho_{\rm m}$ 
as a variable and the binodal phase separation concentrations are readily 
obtained by solving the standard common tangent 
conditions.\cite{MiMB2023,linJML} The Flory-Huggins parameters $\epsilon_h$ 
and $\epsilon_s$ fitted to the experimental data are 
$\epsilon_h=1.0$, $\epsilon_s=0.0$ for the Caprin1 phase diagrams in figure~1c
and
$\epsilon_h=1.0$, $\epsilon_s=-1.5$ for the pY-Caprin1 phase diagrams in 
figure~1d.
When the uniform salt concentration restriction is removed to allow for fully
varying salt and polymer concentrations (figure~2 of the maintext), the final 
concentrations in the two phases depend on their initial bulk (overall) 
concentrations. The corresponding two-dimensional, polymer-salt phase 
diagram (at a given temperature $T$) is obtained by similar balancing 
conditions.\cite{njp2017} As stated in the maintext, the two-dimensional 
rG-RPA+FH phase diagrams in figure~2 are for $T=300$ K. 
\\

{\bf rG-RPA-predicted effects of counterion valency on Caprin1 LLPS.}
As discussed in the maintext, with monovalent salt (Na$^+$), rG-RPA
predicts that Caprin1 does not undergo LLPS at [Na$^+$] $=0$
when the counterion (Cl$^-$) is monovalent (maintext figure~2a,b), but 
Caprin1 LLPS is possible at [Na$^+$] $=0$ when the counterion 
is divalent (maintext figure~2e,f). As mentioned in the maintext, a likely
physical reason for this effect is the difference in
configurational entropy loss of monovalent vs divalent counterions
in the Caprin1-condensed phase.
Apparently, when [Cl$^-$] is just sufficient to balance the net positive
charge of Caprin1 (i.e., when [Na$^+$] = 0), the entropic cost 
of concentrating Cl$^-$ in a Caprin1-condensed phase
cannot be
overcome for Caprin1 LLPS to occur.  The entropic cost will be lessened (and
thus more favorable to Caprin1 LLPS) when there are more Cl$^-$ ions beyond
what is necessary to balance the net positive charge of Caprin1, corresponding
to a situation with nonzero [Na$^+$] from the added NaCl to supply the
additional Cl$^-$ ions. In comparison, when the counterion is divalent
[(ATP-Mg)$^{2-}$ in our case], the number of counterions needed for balancing
the positive net charge of Caprin1 is half of that when the counterion is
monovalent. It follows that the entropic cost of concentrating the divalent
counterion in the Caprin1-condensed phase is less and consequently, at least in
the present situation, no added salt is needed for Caprin1 LLPS.
\\


\noindent
{\large\bf Explicit-ion coarse-grained explicit-chain molecular 
dynamics (MD) simulation}\\

Coarse-grained MD simulations are performed with 
the GPU version of HOOMD-Blue software \cite{hoomd_blue_1,anderson_etal2008} 
using the slab method that has been developed recently to allow for
simulations of relatively large number of polymers\cite{panag2017} and 
applied to liquid-liquid phase separation (LLPS) of intrinsically 
disordered proteins (IDPs).\cite{dignon18} This general MD protocol has 
been utilized extensively,\cite{mittal19,Mpipi} including by 
our group.\cite{SumanPNAS,suman2} 

Within the methodological framework of this coarse-grained simulation protocol,
we introduce explicit small ions into our present simulations because they
are necessary to account for subtle experimental observations that are not 
readily reproduced by using implicit-ion electrostatic screening.
Simulations in the present study are performed with 100 chains of the
Caprin1 IDR (wildtype or variants) 
at four salt ([NaCl]) concentrations: (i) at [NaCl] = 0 
where the system is neutralized by adding appropriate number (1,300) 
of chloride ($\mathrm{Cl^{-}}$) ions, 
(ii) neutralized and at [NaCl] = 200 mM by adding 15,000 pairs of 
explicit Na$^+$ and Cl$^-$ ions, 
(iii) neutralized and at [NaCl] = 480 mM by adding the same number
of 15,000 pairs of explicit Na$^+$ and Cl$^-$ ions, 
and (iv) neutralized and at [NaCl] = 960 mM again with 
15,000 pairs of explicit Na$^+$ and Cl$^-$ ions. As will be described
below, specific small-ion concentrations in (ii), (iii), and (iv) 
are implemented by varying the size of the final simulation box.
A similar procedure is used for simulation of pY-Caprin1 IDR phase behaviors
under these four [NaCl] values. Because each pY-Caprin1 IDR has a net
$-1$ charge, the only difference with the Caprin1 IDR
simulation is that neutralization of the pY-Caprin1 chain requires 
100 Na$^+$ ions instead of 1,300 Cl$^-$ ions.
The amino acid sequences simulated using coarse-grained MD
in the present study are provided in SI Appendix-figure~1. 
Note that the experimental pY-Caprin1 sample is highly phosphorylated, 
consisting mainly of a mixture of Caprin1 IDRs with six or seven 
phosphorlations, with only a very small fraction of IDRs with five 
phosphorylations, and essentially 
no population with fewer than five phosphorylations (SI Appendix-figure~2).
As stated in the maintext, for the sake of simplicity
in our theoretical/computational models,
we use only the Caprin1 IDR with all seven 
tyrosines phosphorylated (referred to simply as pY-Caprin1 in 
SI Appendix-figure~1) to model 
the behaviors of this experimental sample, partly to avoid the combinatoric
complexity of sequences with five or six phosphorlations, which would
entail 21 and 7 possible different sequences, respectively, with
currently unknown population fractions.
\\

{\bf Coarse-grained MD interaction potentials.}
Following prior works,\cite{SumanPNAS,dignon18}
each amino acid residue is modeled by a single bead. Beads representing
different amino acid residues have different masses, sizes, and engage 
in pairwise interactions with different strengths.\cite{SumanPNAS,dignon18} 
Following the notations of our earlier simulation 
works,\cite{SumanPNAS,suman2,suman1} we consider
$n_{\rm p}$ number of polymers labeled as 
$\mu$, $\nu$ = 1, 2, ...$n_{\rm p}$, with
each polymer consisting of $N$ beads labeled
by $i,j = 1, 2, \dots, N$, and $n_{\rm c}$ counterions 
to neutralize the charged polymers.
Coarse-grained MD is readily applicable to studying variations in LLPS
properties among RtoK variants\cite{SumanPNAS} considered here
for Caprin1 (SI Appendix-figure~1). In contrast, since RtoK
substitutions do not change the sequence charge pattern of any given sequence,
rG-RPA theory as formulated above does not account for their effects on LLPS,
though polymer field theory can be extended to incorporate such effects in more
sophisticated formulations.\cite{WessenJPCB2022}

For salt-dependent LLPS ($n_{\rm s}\neq 0$), 
we consider $n_+$ small cations 
and $n_-$ small anions. 
These small cations and anions are classified as salt ions or
counterions depending on the net charge of the polymer (see maintext
as well as discussion below in this {\it SI Appendix}).
The small ions are labeled, respectively, 
by $\gamma=1,2,\dots,n_+$ and $\beta=1,2,\dots,n_-$, and they 
correspond to Na$^+$ and Cl$^-$ in the present coarse-grained MD simulations.
As stated in the maintext,
our total MD potential energy is the
sum of electrostatic (el), short-spatial-range (sr) Lennard-Jones (LJ)-type, 
and bonding (bond) interactions:
\begin{equation}
\label{eq:UT}
U_{\rm T} = U_{\rm el} + U_{\rm sr} + U_{\rm bond} \; .
\end{equation}
For our systems of interest, 
the electrostatic part is a sum of 
polymer-polymer (pp), polymer--small-ion (pi), and small-ion--small-ion (ii)
contributions:
\begin{equation}
U_{\rm el} = U_{\rm el,pp} + U_{\rm el,pi} + U_{\rm el,ii}
\; .
\end{equation}

As before,\cite{WessenJPCB2022} the polymer-polymer potential energy is given by
\begin{equation}
U_{\rm el,pp} =
\frac {e^2}{8\pi\epsilon_{0}\epsilon_{\rm r}}
{\sum_{\mu,\nu=1}^{n_{p}}\sum_{i,j =1}^{N}}
\Bigl(1 - \delta_{\mu\nu}\delta_{ij}\Bigr)
\frac {\sigma_i\sigma_j}{r_{{\mu}i,{\nu}j}}
\; ,
\end{equation}
where, as in the above field-theoretic formulation, 
$e$ is protonic charge, $\epsilon_0$ is vacuum permittivity,
and $\epsilon_{\rm r}$ is relative permittivity (dielectric constant).
Here, $r_{{\mu}i,{\nu}j}$ is the spatial distance between the $i$th residue
of the $\mu$th polymer and the $j$th residue of the $\nu$th
polymer. The Kronecker symbol $\delta$ signals exclusion
of the self-interacting $\mu=\nu$, $i=j$ terms in the summations (irrespective
of the values of these terms) because
$1-\delta_{xy}=0$ if $x=y$ and $1-\delta_{xy}=1$ otherwise.
In units of the protonic charge $e$,
$\sigma_{i}$ of unphosphorylated amino acid residue beads are taken from
ref.~\citen{dignon18}.
Except lysine and arginine ($\sigma_i=+1$), glutamic and aspartic acid
($\sigma_i=-1$), histidine ($\sigma_i=+0.5$)---but note that there is no 
histidine in the Caprin1/pY-Caprin1 sequences simulated here, and 
phosphorylated tyrosine ($\sigma_i=-2$), all other amino acid residues
are assigned zero charge. Similarly, the
interaction between polymers and small ions is given by 
\begin{equation}
U_{\rm el,pi} =
\frac {e^2}{4\pi\epsilon_{0}\epsilon_{\rm r}}
\sum_{\mu=1}^{n_{\rm p}}\sum_{i=1}^{N}
\left\{
\sum_{k=+,-}\left[
\sum_{\gamma(k)=1}^{n_\gamma(k)}
\frac {\sigma_i\sigma_k}{r_{{\mu}i,\gamma(k)}}\right]\right\}
=
\frac {e^2}{4\pi\epsilon_{0}\epsilon_{\rm r}}
\sum_{\mu=1}^{n_{\rm p}}\sum_{i=1}^{N}
\left[
\sum_{\gamma=1}^{n_+}
\frac {\sigma_i\sigma_+}{r_{{\mu}i,\gamma}}
+
\sum_{\beta=1}^{n_-}
\frac {\sigma_i\sigma_-}{r_{{\mu}i,\beta}}
\right]
\; ,
\end{equation}
where, in the term after the first equality, the summation $\sum_k$ 
(enclosed in curly brackets)
is over small ion types, summation indices $\gamma(+)$ and $\gamma(-)$
label, respectively, the positively and negatively charged small ions,
$n_\gamma(+)$ and $n_\gamma(-)$ are the total numbers of these ions, and
$r_{{\mu}i,\gamma(k)}$ is the spatial distance between the $i$th residue of the
$\mu$th polymer chain and the small ion labeled by $\gamma(k)$.  
After the second equality, the two terms in $\sum_k$ are written explicity, 
now with
$r_{{\mu}i,\gamma/\beta}$ being the spatial distance
between the $i$th residue of the $\mu$th polymer chain and the 
$\gamma/\beta$ ($\gamma$ or $\beta$)-labeled positive/negative small ion
as well as
$\sigma_+$ and $\sigma_-$ being the charges of the small positive 
and negative small ions, respectively. Following ref.~\citen{cheatham2008},
we take $\sigma_+=+1$ for Na$^+$ and $\sigma_-=-1$ for 
Cl$^-$. Depending on the net charge of the polymer, counterions
can be included in either the $n_+$ or $n_-$ count. For instance,
for the positive charged Caprin1 IDR, the total number $n_-$ of
negatively charged small ions (Cl$^-$) includes the numbers $n_{\rm c}$
counted as counterions and $n_{\rm s}$ counted as salt ions.

The interaction between the small ions is given by
\begin{equation}
\label{eq:Uelii}
U_{\rm el,ii} =
\frac {e^2}{8\pi\epsilon_{0}\epsilon_{\rm r}}
\sum_{k,k^\prime=+,-}\left[
\sum_{\gamma(k)=1}^{n_\gamma(k)}
\sum_{\gamma^\prime(k^\prime)=1}^{n_\gamma^\prime(k^\prime)}
\Bigl(1 - \delta_{kk^\prime}\delta_{\gamma(k)\gamma^\prime(k^\prime)}\Bigr)
\frac {\sigma_k\sigma_{k^\prime}}{r_{\gamma(k),\gamma^\prime(k^\prime)}}\right]
\; ,
\end{equation}
where
$r_{\gamma(k),\gamma^\prime(k^\prime)}$ is the spatial distance between two
small ions.  For the MD simulations in this work, the positively and negatively
charged small ions correspond to Na$^+$ and Cl$^-$ respectively.  
Eq.~\eqref{eq:Uelii} is equivalent to
\begin{equation}
U_{\rm el,ii} =
\frac {e^2}{4\pi\epsilon_{0}\epsilon_{\rm r}}
\left[
\sum_{\gamma=1}^{n_+-1}
\sum_{\gamma^\prime=\gamma+1}^{n_+}
\frac {\sigma_+^2}{r_{\gamma\gamma^\prime}}
+
\sum_{\beta=1}^{n_--1}
\sum_{\beta^\prime=\beta+1}^{n_-}
\frac {\sigma_-^2}{r_{\beta\beta^\prime}}
+
\sum_{\gamma=1}^{n_+}
\sum_{\beta=1}^{n_-}
\frac {\sigma_+\sigma_-}{r_{\gamma\beta}}
\right] \; ,
\end{equation}
where $r_{xy}$ is the spatial distance between a pair of small ions
labeled by $x$ and $y$.
As rationalized previously\cite{SumanPNAS} in the context of 
experimental measurements of dielectric properties of biological
systems,\cite{tros_etal2017} we use $\epsilon_{\rm r}=40$, a
value lower than the $\sim 80$ dielectric constant of bulk water,
for all simulations reported in the present work. 

Short-spatial-range non-bonded LJ interactions are similarly constituted
by three components, viz., those for polymer-polymer ($U_{\rm sr,pp}$), 
polymer--small-ion ($U_{\rm sr,pi}$), and small-ion--small-ion 
($U_{\rm sr,ii}$) interactions:
\begin{equation}
U_{\rm sr} = U_{\rm sr,pp} + U_{\rm sr,pi} + U_{\rm sr,ii} \; .
\end{equation}
Here we adopt the Kim-Hummer (KH)\cite{KH} interaction scheme for
$U_{\rm sr,pp}$. KH is based on the Miyazawa-Jernigan (MJ) statistical
potential\cite{MJ96} derived from folded globular protein structures
in the Protein Data Bank (PDB) and is therefore expected to reflect the 
energetics of polypeptides, especially the driving forces pertinent 
to protein folding, its limitations\cite{SumanPNAS} notwithstanding.
Our previous work shows that the KH potential is adequate for rationalizing
the rank ordering of LLPS propensities of the N-terminal IDR of
DEAD-box RNA helicase Ddx4 and its charge scrambled
and arginine-to-lysine (RtoK) variants. KH also rationalizes the rank 
ordering of LLPS propensities of WT and an RtoK variant of 
LAF-1.\cite{SumanPNAS} 
We therefore stipulate that the KH interaction scheme
is appropriate, at least as a first approximation, to address the
LLPS propensities of Caprin1 and its RtoK variants. 
The degree to which the differences in simulated LLPS propensity among these
Caprin1 variants are affected by how interactions involving K and R are treated
differently by the model potential function\cite{SumanPNAS,Mpipi,Kresten} 
should be further explored in the future.
As before,
$U_{\rm sr,pp}$ takes the following form:\cite{SumanPNAS,dignon18}
\begin{equation}
\begin{split}
U_{\rm sr,pp} &
= \frac {1}{2}{\sum_{\mu,\nu=1}^{n_{p}}\sum_{i,j =1}^{N}}
\Bigl(1 - \delta_{\mu\nu}\delta_{ij}\Bigr)
(U_{\rm KH})_{{\mu}i,{\nu}j} \; ,
\\
(U_{\rm KH})_{{\mu}i,{\nu}j} & 
= U_{\rm LJ} + (1-\lambda_{ij}^{\rm KH})\,\epsilon_{ij} \, 
\quad \mathrm{if} \, r \le 2^{1/6}\,a_{ij} \\
 & = \lambda_{ij}^{\rm KH} U_{\rm LJ} \, 
\quad \quad \quad \quad \quad \quad \mathrm{otherwise}
\end{split}
\end{equation}
in which
\begin{equation}
\label{eq:LJ}
(U_{\rm LJ})_{{\mu}i,{\nu}j} = 
U_{\rm LJ}(\epsilon_{{\mu}i,{\nu}j},a_{{\mu}i,{\nu}j},r_{{\mu}i,{\nu}j})=
4\epsilon_{{\mu}i,{\nu}j}
\left[\left(\frac{a_{{\mu}i,{\nu}j}}{r_{{\mu}i,{\nu}j}}\right)^{12}
-\left(\frac{a_{{\mu}i,{\nu}j}}{r_{{\mu}i,{\nu}j}}\right)^{6}\right]
\end{equation}
and $a_{{\mu}i,{\nu}j} = a_{ij}= (a_i + a_j)/2$, where
the van der Waals diameters $a_i$ and $a_j$,
depend only, respectively, on the amino acid residue type (one of twenty) for 
residue $i$ and residue $j$ (Table~S1 of ref.~\citen{dignon18}). 
In contrast, the parameters $\lambda_{ij}^{\rm KH}$ and 
$\epsilon_{{\mu}i,{\nu}j}=\epsilon_{ij}$ 
depend on both the residue types of residues $i$ and $j$.
Values for $\epsilon_{ij}$ are provided in Table~S3 of ref.~\citen{dignon18}.
The formula for $\lambda_{ij}^{\rm KH}=\pm 1$ is 
given by Eq.~(5) of ref.~\citen{dignon18}
as well as Eqs.~(S10) and (S11) of ref.~\citen{SumanPNAS}.

For the LJ interactions between polymers and small ions, we recognize
that while the coarse-grained KH parameters are based on statistical analysis
of known folded protein structures, LJ interaction parameters for 
small ions are typically scaled to match certain physical and chemical 
properties.\cite{cheatham2008} Thus it is not straightforward to postulate
an interaction scheme based upon first principles. To make progress
and to maintain simplicity of our model, we adopt the LJ form
[$U_{\rm LJ}$ in Eq.~\eqref{eq:LJ}] for $U_{\rm sr,pi}$ with a uniform 
$\epsilon_{{\mu}i,{\nu}j}=\epsilon_{ij}=0.142$ 
(denoted $\epsilon_{{\rm p}\pm}\equiv 0.142$) for all residue-small ion pairs.
This $\epsilon_{ij}=\epsilon_{{\rm p}\pm}$ value is equal to that for a pair of 
alanine residues in the KH potential and is neither too strong nor too 
weak among $\epsilon_{ij}$ values for pairwise interactions between amino 
acid residues. Accordingly,
\begin{equation}
U_{\rm sr,pi} =
\sum_{\mu=1}^{n_{\rm p}}\sum_{i=1}^{N}
\left[
\sum_{\gamma=1}^{n_+}
U_{\rm LJ}(\epsilon_{{\rm p}\pm},a_{i+},r_{{\mu}i,\gamma})
+
\sum_{\beta=1}^{n_-}
U_{\rm LJ}(\epsilon_{{\rm p}\pm},a_{i-},r_{{\mu}i,\beta})
\right] \; ,
\end{equation}
where $a_{i+}=(a_i+a_+)/2$, $a_{i-}=(a_i+a_-)/2$, 
with $a_+$ and $a_-$ being, respectively,
the van der Waals diameters of the positively and negatively charged
small ions.
For the present MD simulations, 
$a_+=a_{{\rm Na}^+}$, $a_-=a_{{\rm Cl}^-}$, and their
values are adopted from ref.~\citen{cheatham2008}.
Similarly, for small-ion--small-ion LJ interactions, 
\begin{equation}
\begin{aligned}
U_{\rm el,ii} = 
\Biggl[
\sum_{\gamma=1}^{n_+-1}
\sum_{\gamma^\prime=\gamma+1}^{n_+}
U_{\rm LJ}(\epsilon_{+},a_{+},r_{\gamma,\gamma^\prime})
+
\sum_{\beta=1}^{n_--1}
\sum_{\beta^\prime=\beta+1}^{n_-}
&
U_{\rm LJ}(\epsilon_{-},a_{-},r_{\beta,\beta^\prime}) 
\\
+
&
\sum_{\gamma=1}^{n_+}
\sum_{\beta=1}^{n_-}
U_{\rm LJ}(\epsilon_{+-},a_{+-},r_{\gamma,\beta})
\Biggr] \; 
,
\end{aligned}
\end{equation}
where 
$\epsilon_{+}$ and
$\epsilon_{-}$ are, respectively, the LJ interaction energy parameter 
for the positively and negatively charged small ions,
$\epsilon_{+-}=(\epsilon_{+}\epsilon_{-})^{1/2}$, and $a_{+-}=(a_{+} + a_-)/2$.
For the present MD simulations, the $\epsilon_{+}=\epsilon_{{\rm Na}^+}$ and
$\epsilon_{-}=\epsilon_{{\rm Cl}^-}$ values are also adopted 
from ref.~\citen{cheatham2008}.

As before, the bond-length energy term $U_{\rm bond}$ in Eq.~\eqref{eq:UT}
for chain connectivity is modeled by a harmonic potential,
\begin{equation}
U_{\rm bond} = \frac{K_{\rm bond}}{2}
\sum_{\mu=1}^{\np}\sum_{i=1}^{N-1} (r_{{\mu}i,{\nu}i+1}-l)^{2} \; .
\end{equation}
Following previous studies,\cite{SumanPNAS,dignon18} Kuhn length $l
= \mathrm{3.8\,\AA}$ is taken as the C$_\alpha$--C$_\alpha$ virtual
bond length for {\it trans} polypeptides and 
$K_{\rm bond}=10$ kJmol$^{-1}${\AA}$^{-2}$.
\\

{\bf Simulation protocol.}
In each of our coarse-grained MD simulations,
the IDR chains and ions are initially placed randomly in a sufficiently 
large cubic box of
dimensions $300\times 300\times 300$ \AA$^3$. Energy minimization is
then performed using the FIRE algorithm (available in the HOOMD-Blue
package) which includes removal of steric clashes 
among the initially placed amino acid beads. Next, the system is 
compressed at a low temperature of 100 K at 1 atm pressure for a 
period of 50 ns using the Martyna-Tobias-Klein (MTK) thermostat and 
barostat\cite{klein1994,tuckerman_etal2006}
with a coupling constant of 1 ps. The equations of motion are
integrated using velocity-Verlet algorithm with a timestep of 20 fs. 
Periodic boundary conditions are applied in all three directions. The
electrostatic interaction is computed using the PPPM 
algorithm\cite{lebard2012} available in the package. 
We use a cut-off distance of 15 $\mathrm{\AA}$ for the short-spatial-range 
non-bonded interactions.
After this initial $NPT$ step, we compress the simulation box again 
along the three dimensions for a period of 50 ns until it reaches 
a sufficiently high density, using Langevin dynamics for
an $NVT$ ensemble with a friction coefficient of $\mathrm{1 \, ps^{-1}}$. 
At the end of this compression step, the dimensions of the simulation box
for Caprin1 and its four RtoK variants
are $115\times 115\times 115$ \AA$^3$ for [NaCl] = 0 (no small ions beside
counterions) and $155\times 155\times 155$ \AA$^3$ for
[NaCl] = 200 mM, 480 mM, and 960 mM. For pY-Caprin1, the corresponding
dimensions 
are $115\times 115\times 115$ \AA$^3$ for [NaCl] = 0 (no small ions beside
counterions) and $150\times 150\times 150$ \AA$^3$ for
[NaCl] = 200 mM, 480 mM, and 960 mM.
Next, the system is expanded along one of the spatial dimensions
(taken as the $z$-axis) using isotropic linear scaling for 10 ns while 
keeping the temperature constant at 100 K. 
For Caprin1 and its four RtoK variants,
the simulation box length in the $z$-direction is expanded 14 times
for [NaCl] = 0, $33.6$ times for [NaCl] = 200 mM, 14 times for 
[NaCl] = 480 mM, and 7 times for [NaCl] = 960 mM. 
For pY-Caprin1, the expansion factors along the $z$-direction are 
10 times for 
[NaCl] = 0, $37.07$ times for [NaCl] = 200 mM, $15.47$ times for
[NaCl] = 480 mM, and $7.73$ times for [NaCl] = 960 mM.
Note that the simulation box volumes for Caprin1, its RtoK variants,
and pY-Caprin1 after this last expansion are identical for the
same [NaCl] because the same numbers of polymer chains and small ions
are used.
The practical reason for keeping the number of Na$^+$ and Cl$^-$ ions
constant for the higher salt concentration is to minimize computational 
cost. In other words, the three salt concentrations (200 mM, 480 mM 
and 960 mM) are achieved here by using different box dimensions. 
After the last box expansion, $NVT$ equilibration using the Langevin 
thermostat with a friction coefficient of 1 ps$^{-1}$ is performed 
for 2 $\mathrm{\mu}$s at various temperatures. Final
production run is then carried out for another 4 $\mathrm{\mu}$s with the same
Langevin thermostat using a much lower friction coefficient of 0.01 ps$^{-1}$
for sampling efficiency.
The snapshots are saved every 1 ns for further analysis. Detailed
descriptions of how to construct a phase diagram from simulation trajectories 
are provided in ref.~\citen{dignon18} and our previous 
works.\cite{MiMB2023,SumanPNAS}
This simulation protocol and the above-described coarse-grained MD model
are used to produce the phase diagrams, distributions, and snapshots 
in figures~4--7 of the maintext and SI Appendix-figure~3.
\\

{\bf Comparison with atomic simulations with a preformed condensate.}
As mentioned in the maintext, 
explicit-water, explicit-ion atomic simulations in the
presence of a preformed condensate of the N-terminal RGG domain of LAF-1 with a
net charge of $+4$ produce enhanced Cl$^-$ and depleted Na$^+$ in the 
IDR-condensed phase.\cite{Zhengetal-ions2020} This trend is consistent with our
implicit-water, explicit-ion MD result for Caprin1 with net charge
$+13$ (maintext figure~4c). By comparison, corresponding atomic simulations 
in the presence of a preformed condensate of the low complexity domain 
of FUS with a net charge of $-2$ produce a significant depletion of Cl$^-$ 
and a minor depletion of Na$^+$ in the IDR-condensed 
phase,\cite{Zhengetal-ions2020} which is opposite to the
trend seen here for pY-Caprin1 with net charge $-1$ in maintext figure~4d and 
SI Appendix-figure~3. Whether this difference is caused by the multiple 
phosphorylated sites with a $-2$ charge in pY-Caprin1 remains to be elucidated.
\\

\noindent
{\large\bf Field-theoretic simulation (FTS) for multiple-component LLPS}\\

Biomolecular condensates {\it in vivo} can contain hundreds of protein
and nucleic acid species. Therefore, to address their biophysical
properties and biochemical functions, theories---starting with rudimentary
constructions---are needed
for multiple-component LLPS. As a first appproximation and 
a tool for conceptualization, we find it valuable to utilize 
FTS---especially recently developed FTS approaches for biomolecular 
LLPS\cite{Fredrickson2002,Fredrickson2006,joanJPCL2019,Pal2021} 
and their extensions---to gain insights into the energetic basis of
sequence-specific spatial distributions of various biomolecular components 
in and out of phase-separated condensates. 

FTS enjoys the fundamental advantage that it is not limited 
by some of the approximations
in analytical theories such as RPA and rG-RPA because FTS accounts 
fully for all field fluctuations in principle. FTS is thus a valuable 
alternative to analytical theories, though it is computationally
more costly in practice and can be impeded by lattice-related artifacts
and limitations arising from the small FTS simulation box sizes 
necessitated by numerical tractability.
We view analytical theories and FTS as complementary.

The starting point of FTS is a statistical field theory 
[e.g., Eq.~\eqref{eq:Zprime}, which is equivalent to maintext Eq.~(5)].
To avoid numerical instabilities, we
treat polymer beads and ions as (smeared) Gaussian distributions\cite{Wang2010}
instead of the point particles stipulated by 
the Dirac $\delta$-functions in Eqs.~\eqref{eq:micro_density} and 
\eqref{eq:micro_charge_density}. 
For simplicity, this regularization is implemented 
using a common component-independent width $\bar{a}$ irrespective of
chemical species by making the general replacements 
$\delta(\rr - \rr_a) \rightarrow \Gamma(\rr - \rr_a)$ and
$\delta(\rr - \R_{\alpha,i}) \rightarrow \Gamma(\rr - \R_{\alpha,i})$,
in Eqs.~\eqref{eq:micro_density} and \eqref{eq:micro_charge_density}
where $\Gamma( \rr ) = e^{-r^2/2 \bar{a}^2} / (2 \pi \bar{a}^2)^{3/2}$,
$r^2=|\rr|^2$.

A general field-theoretic Hamiltonian applicable
to a system comprising of one or more charged
polymer species including Caprin1, pY-Caprin1, (ATP-Mg)$^{2-}$
(maintext figure~8a), ATP$^{4-}$ and small ions such as
Na$^+$, Cl$^-$, and Mg$^{2+}$ is given by 
\begin{equation} \label{Seq:H_FTS}
H[\w, \psi] = \int d\rr \left( \frac{\left(\nabla \psi(\rr) \right)^2}{8\pi\lb} + \frac{\w(\rr)^2}{2v_2}\right)
        - \sum_m n_m \ln {\cal Q}_m[\breve{\w}, \breve{\psi}] \; ,
\end{equation}
where the ${\cal Q}_m$ functionals ($m$ labels system components) 
are in general complex when evaluated beyond quadratic order in the
fields.  Equation~\eqref{Seq:H_FTS} above is identical to Eq.~(5) of 
maintext and formally equivalent to the Hamiltonian in Eq.~\eqref{eq:Zprime}.
Because of the above-described Gaussian smearing, the fields 
in the arguments of ${\cal Q}_m$ are now convoluted with
$\Gamma$, i.e.~$\phi(\rr)\rightarrow\breve{\phi}(\rr)=
\Gamma \star \phi(\rr) \equiv \int d \rr'
\Gamma(\rr - \rr') \phi(\rr')$, where the generic $\phi = \w, \psi$.
\\

{\bf Complex Langevin evolution in fictitious time.}
A simulation approach developed in the 1980s to handle the complex
nature of $H[\w, \psi]$ and obtain statistical (Boltzmann) averages 
is the Complex Langevin (CL) 
method,\cite{Parisi1983,Klauder1983} which analytically continues the 
fields $\w$ and $\psi$ into their respective complex planes and introduces 
an fictitious (artificial, unphysical) time-coordinate $t$ on which 
$\w(\rr,t)$ and $\psi(\rr,t)$ now depend. The CL time evolution is 
governed by stochastic Langevin differential equations
[maintext Eq.~(6)], as follows:
\begin{equation} 
\label{Seq:CL_evolution}
\frac{\partial \phi(\rr,t)}{\partial t} = 
- \frac{\delta H}{\delta \phi(\rr,t) } + \eta_{\phi}(\rr,t) \; , 
\quad \phi=\w,\psi \; ,
\end{equation}
where $\eta_{\phi}(\rr,t)$ is real-valued Gaussian noise with zero mean:
\begin{equation}
\langle \eta_{\phi}(\rr,t) \eta_{\phi'}(\rr',t') \rangle = 
2 \delta_{\phi,\phi'} \delta(\rr - \rr') \delta(t-t') \; .
\end{equation}
Thermal averages of any thermodynamic observable 
$\hat{\cal O}[\R , \rr]$ can then be computed in the field picture 
(indicated by ``$\langle\dots\rangle_{\rm F}$'' with subscript ``F'') 
using a corresponding field operator
$\tilde{ \cal O}[\w,\psi]$ through averages over all possible equilibrium 
field configurations, which in turn translate into asymptotic CL time averages
with no final dependence on the fictitious time variable $t$, i.e.,
\begin{equation}
\langle \tilde{ \cal O}[\w,\psi]  \rangle_{\rm F} 
\equiv \frac{\int \DD \w \int \DD \psi \, 
\tilde{\cal O}[\w,\psi]  \, e^{-H[\w,\psi] }} 
{ \int \DD \w \int \DD \psi \, e^{-H[\w,\psi] }} 
= \lim_{t_{\rm max} \rightarrow \infty} \frac{1}{t_{\rm max}} 
\int_0^{t_{\rm max}} d t \, \tilde{\cal O}[\w(\rr,t),\psi(\rr,t)] \; .
\end{equation}
The Langevin Eq.~\eqref{Seq:CL_evolution} involves functional derivatives 
of the Hamiltonian with respect to the complex fields, which are
formally evaluated as
\begin{subequations}
\begin{align}
\frac{\delta H}{\delta \w(\rr ) } = \, 
& \i \sum_m \tilde{\rho}_m(\rr) + \frac{1}{v_2} \w(\rr) \; , \\
\frac{\delta H}{\delta \psi(\rr ) } = \,
& \i \sum_m \tilde{c}_{m}(\rr) - \frac{1}{4 \pi \lb } \nabla^2 \psi(\rr) \; ,
\end{align}
\end{subequations}
where
\begin{subequations}
\begin{align}
\tilde{\rho}_m(\rr) = \, 
& \i n_m \frac{\delta \ln {\cal Q}_m[\breve{\w},\breve{\psi}] }{\delta\w(\rr)}
\;  , \\
\tilde{c}_{m}(\rr) = \,
& \i n_m 
\frac{\delta \ln {\cal Q}_m[\breve{\w}, \breve{\psi}] }{\delta \psi(\rr) }
\end{align}
\end{subequations}
are field operators corresponding, respectively, to 
number- and charge density of chemical component $m$. 
\\

{\bf Number density correlation functions.}
Information about the polymer-polymer, polymer-ion, ion-ion association
and ion partitioning into the condensate can be gleaned from number 
density-number density correlation functions [maintext Eq.~(7)]
\begin{equation}
\label{eq:Gmn}
G_{m,n}(| \rr - \rr' |)=\langle \hat{\rho}_m(\rr )\hat{\rho}_n(\rr ') \rangle 
\, ,
\end{equation}
which can be computed in field theory \cite{Pal2021} as
\begin{subequations}
\begin{align}
G_{m,n\neq m}(| \rr - \rr' |) = \,
&  
\langle \tilde{\rho}_m(\rr ) \tilde{\rho}_n(\rr ') \rangle_{\rm F} \; , \\
G_{m,m}(| \rr - \rr' |) = \,
& \frac{\i}{v_2} 
\langle \tilde{\rho}_m(\rr ) \w(\rr ') \rangle_{\rm F} 
-\sum_{n \neq m}\langle\tilde{\rho}_m(\rr )\tilde{\rho}_n(\rr')\rangle_{\rm F} 
\; .
\end{align}
\end{subequations}
The $G_{m,n}$ functions are useful for assessing Caprin1 and pY-Caprin1 phase
separation and the colocalization of ATP-Mg with the polymer condensed droplet
(maintext figure~8b--e). 
Information with higher spatial resolution can also be provided
by $G_{m,n}$ if we identify component $m$ with individual polymer bead (labeled
by $i$) along a chain sequence.

In some situations, the physical implications of density-density 
correlation functions $G_{m,n}(r)$ are more apparent when normalized by 
the component bulk (overall) densities $\rho_m^0$ and
$\rho^0_n$, as discussed in the maintext
in connection with the correlation functions shown in figure~8. 

For small ions that are each represented by a single Gaussian distribution,
the density operator is given by
\begin{equation}
\tilde{\rho}_m(\rr) = \frac{n_m}{\Omega{\cal Q}_k} 
\Gamma \star e^{-\i [\breve{\w} (\rr) + z_{m} \breve{\psi}(\rr)]},
\end{equation}
and the charge density operator is 
$\tilde{c}_m(\rr )= z_{m} \tilde{\rho}_m(\rr )$, where $z_m$ is the charge 
of ion species $m$ and, as defined above, $\Omega$ is the system
volume. For polymers (denoted ``p''), the density and 
charge-density operators are calculated using forward 
(subscript ``$F$'') and backward (subscript ``$B$'') chain propagators 
$q_{F}(\rr,i)$ and $q_{B}(\rr,i)$ 
as follows, with $i$ being the label for the beads/monomers along the
polymer chain:
\begin{subequations}
\begin{align}
\Qp = \,
& \frac {1}{\Omega}\int d \rr \; q_{F}(\rr,N)  \; , \\
\label{eq:ops}
\tilde{\rho}_{\rm p}(\rr) = \,
& \frac{\np}{\Omega\Qp} \Gamma \star \sum_{i = 1}^N q_{F}(\rr,i) 
q_{B}(\rr,i) e^{\i[\breve{\w}(\rr) + \sigma_i \breve{\psi}(\rr) ] } \; , \\
\tilde{c}(\rr) = \,
& \frac{\np}{\Omega\Qp} \Gamma \star \sum_{i=1}^N q_{F}(\rr,i) q_{B}(\rr,i) 
e^{\i[\breve{\w}(\rr) + \sigma_i \breve{\psi}(\rr) ] } \sigma_i \; ,
\end{align}
\end{subequations}
and the chain propagators are constructed iteratively as
\begin{subequations}
\begin{align}
q_{F}(\rr,i+1)= \,
& e^{-\i[\breve{\w}(\rr)+\sigma_{i+1}\breve{\psi}(\rr)] } 
\left( \frac{3}{2 \pi b^2} \right)^{3/2} 
\int d \rr' e^{-3(\rr-\rr')^2/2b^2}  q_{F}(\rr',i) \; ,\\
q_{B}(\rr,i-1)= \,
& e^{-\i[\breve{\w}(\rr) + \sigma_{i-1}\breve{\psi}(\rr) ]} 
\left( \frac{3}{2 \pi b^2} \right)^{3/2} 
\int d \rr' e^{-3(\rr-\rr')^2/2b^2}  q_{B}(\rr',i) 
\end{align}
\end{subequations}
with the starting $q_{F}(\rr,1)= e^{-\i[\breve{\w}(\rr) + \sigma_{1}
\breve{\psi}(\rr) ] }$, $q_{B}(\rr,N) = e^{-\i  [\breve{\w}(\rr) +
\sigma_{N} \breve{\psi}(\rr) ] } $, and
we use $b$ for Kuhn length ($b=l$) in the present
FTS formulation to conform to the notation in our published 
FTS studies.\cite{MiMB2023,WessenJPCB2022,Pal2021}
In the present work, the correlation functions in maintext figure~8 and 
SI Appendix-figures~4--6 are computed by integrating pertinent CL 
fictitious-time evolution equations defined in Eq.~\eqref{Seq:CL_evolution} 
using the first order semi-implicit method of ref.~\citen{Lennon2008}.
\\

{\bf Residue-specific Caprin1--(ATP-Mg) association.}
As outlined in {\it Materials and Methods} of the maintext, residue-specific
properties of the polymers in our FTS systems can be gleaned from
the $G_{m,n}$ function in Eq.~\eqref{eq:Gmn} by identifying $m$ 
as individual polymer beads (indexed by $i$) along the polymer chain sequence,
viz., define $G^{(i)}_{\mathrm{pq}}(| \rr - \rr' |) \equiv
\left\langle \hat{\rho}_{\mathrm{p},i}(\rr)
\hat{\rho}_{\rm q}(\rr') \right\rangle$ where 
$\hat{\rho}_{\mathrm{p},i}(\rr) \equiv
\sum_{\alpha=1}^N \Gamma (\rr - \R_{\alpha,i})$, the corresponding
operator $\tilde{\rho}_{{\rm p},i}(\rr)$ being equal to the $i$th term 
in the summation in Eq.~(S43b),
and q is another component in the FTS system. 
Accordingly [maintext Eq.~(9)],
\begin{equation}
\label{Seq:calG}
\mathcal{G}^{(i)}_{\mathrm{pq}} \equiv
4\pi\int_{0}^{r_{\rm contact}} dr\, r^2  G^{(i)}_{\mathrm{pq}}(r)
\end{equation}
with a reasonably small residue-q distance $r_{\rm contact}$
(spatial separation between residue $i$ and the positions of
particles belonging to component q) can be used to represent
residue-specific relative residue-q contact frequencies. A normalized
version of this quantity is defined by
\begin{equation}
\label{eq:gipq}
\frac {\mathcal{G}^{(i)}_{\mathrm{pq}}}{\rho^0_{{\mathrm{p}},i}\rho^0_{\mathrm{q}}}
=
4\pi\int_{0}^{r_{\rm contact}} dr\, r^2  g^{(i)}_{\mathrm{pq}}(r)
\;, \quad
g^{(i)}_{\mathrm{pq}}(r)\equiv
\frac {G^{(i)}_{\mathrm{pq}}(r)}{\rho^0_{{\mathrm{p}},i}\rho^0_{\mathrm{q}}}
\; ,
\end{equation}
where $\rho^0_{\mathrm{q}}$ and $\rho^0_{\mathrm{p},i}$ 
are the bulk (overall) densities, respectively, of the q-component and
the $i$th residue along the given polymer species.
Values of $\mathcal{G}^{(i)}_{\mathrm{pq}}/\rho^0_{{\mathrm{p}},i}\rho^0_{\mathrm{q}}\,$
in the above Eq.~\eqref{eq:gipq}
for ${\rm p}=$ Caprin1 and ${\rm q}=$ (ATP-Mg)$^{2-}$, Na$^+$, or Cl$^-$ 
under the simulation conditions we considered are provided in
maintext figure~8f. The variation in 
$\mathcal{G}^{(i)}_{\mathrm{pq}}/\rho^0_{{\mathrm{p}},i}\rho^0_{\mathrm{q}}\,$
for (ATP-Mg)$^{2-}$ with residue position $i$ is largely consistent with 
the experimental trend of NMR-measured volume ratios on Caprin1-ATP
association in ref.~\citen{LewisPNAS2021}.

In all the FTS simulations in this study except for a part of the model
with no (ATP-Mg)$^{2-}$ described immediately below, we use a cubic simulation 
box of length $L=N_{\rm L}\Delta x$ where $N_{\rm L}=32$ 
(i.e., a $32\times 32\times 32$ lattice) 
and $\Delta x=b/\sqrt{6}$ is the lattice resolution. 
In view of the periodic boundary conditions implemented for all three 
spatial dimensions, 
the maximum possible physical distance between two volume elements 
is $\sqrt{3}L/2$. All possible physical distances $r=r_{i,j,k}$
on this cubic simulation box satisfy the relation
$r_{i,j,k}^2 = [\min\{i,N_{\rm L}-i\}^2 + 
\min\{j,N_{\rm L}-j\}^2 + \min\{k,N_{\rm L}-k\}^2]\Delta x^2$,
for some $i,j,k=0,1,2,\dots,31$. 
There is thus a finite number of discretized distances between $0$ 
and $\sqrt{3}L/2$. 
One of these discretized distances, $r=1.47b\approx 1.50b$ is used for the
$\int_0^{r_{\rm contact}}dr$ integrations in
maintext Eq.~(9) and Eqs.~\eqref{Seq:calG} and \eqref{eq:gipq} above.



Further details of the main FTS model utilized for the results 
in maintext figures~8 and 9 as well as alternate 
FTS models discussed under the heading ``Robustness of general trends 
predicted by field-theoretic simulation'' in the maintext are provided 
below in ascending order of number of components treated by the model:
\\

{\bf FTS models of Caprin1/pY-Caprin1 with Na$^{\bf +}$ and Cl$^{\bf -}$
but no ATP-Mg.}  
FTS is conducted for Caprin1 and pY-Caprin1 at various concentrations 
of explicit Na${}^+$ and Cl${}^-$ ions. For all systems, polymer bead
concentration is fixed at $\np N/\Omega = 0.4 b^{-3}$. 
The salt concentrations, here
referring to the concentration of the small ion species with same sign charge
as the net charge of the polymer, are set to ${\rm [NaCl]} \, b^3 = 0, 10^{-6},
10^{-5}, \dots 10^{-1}, 10^{0}$. Additional small ions of charge opposite to
the polymer net charge are added to achieve overall electric 
neutrality of the system. For the results in SI Appendix-figure~4e--h,
simulations are performed in an elongated simulation box on 
a $24\times24\times80$ lattice with lattice spacing set to
the Gaussian smearing length, $\Delta x = \bar{a} = b/\sqrt{6}$.
The Complex-Langevin (CL) evolution equations [Eq.~\eqref{Seq:CL_evolution}] 
are integrated using a time-step $\Delta t = 10^{-3}b^3$
for $6 \times 10^{4}$ steps and the system is sampled every 50th step. An
equilibration period of $3\times 10^4$ CL steps, determined by monitoring the
equilibration of the chemical potentials for each molecular species, is
discarded from each trajectory. Eight independent simulations are run for each
combination of salt concentration and Caprin1 or pY-Caprin1. All simulations 
are performed at $l_{\rm B} = 7 b$ and $v_2= 0.0068 b^3$. The density profiles 
shown in SI Appendix-figure~4e,f 
are obtained by averaging the real part of the field 
theoretic polymer bead density operator over the $x$- and $y$ dimensions 
(i.e.~the dimensions corresponding to the short sides of the simulation box). 
The resulting one-dimensional density snapshots were then individually centred
around their center-of-mass $z$-coordinate $z_{\rm c.o.m.}$ before taking the
trajectory average to give the profiles in SI Appendix-figure~4e,f. 
The shaded uncertainty
bands in these plots indicate the root-mean-squared difference among 
the eight independent runs. The density profiles are then used to estimate 
the coexisting condensed and dilute phases shown in SI Appendix-figure~4g,h. 
Here, the condensed-phase concentration is obtained as the average density 
at $z_{\rm c.o.m.}$, whereas the dilute-phase
concentrations are estimated as the average density among the 10 and 60
$z$-coordinates (for Caprin1 and pY-Caprin1, respectively) furthest from
$z_{\rm c.o.m.}$.
Consistent with rG-RPA and explicit-MD, these FTS models show reentrant 
behavior---albeit subtle---for Caprin1 at low [protein]s (an LLPS
region is seen in SI Appendix-figure~4g 
at intermediate [NaCl]s but not at higher or lower [NaCl]s) but not 
for pY-Caprin1 (no such feature in SI Appendix-figure~4h).
The elongated box field-theoretic simulations were performed using the 
BioFTS Python package, publicly available at 
{\tt https://github.com/jwessen/BioFTS} under an MIT open-source license.
\\

{\bf FTS models for Caprin1/pY-Caprin1 with (ATP-Mg)$^{\bf 2-}$ and either
Na$^{\bf +}$ or Cl$^{\bf -}$.}
Simulations are performed at $l_{\rm B} = 5b$ on a periodic
$32\times32\times32$ grid (see above) with CL time step $\Delta t=0.002$.
All ATP$^{4-}$s and Mg$^{2+}$s are assumed to be in the complex 
(ATP-Mg)$^{2-}$ form with charge sequence (-1 -1 -1 -1 +1 +1) 
as depicted in maintext figure~8a.
Bulk Caprin1 and pY-Caprin1 bead densities in the their respective 
simulation systems are both set at $0.1b^{-3}$. 
For Caprin1, depending on the concentration of (ATP-Mg)$^{2-}$, 
counterions Na$^+$ or Cl$^-$ (but not both) are added
to maintain overall electric neutrality of the FTS system.
For pY-Caprin1, Na$^+$ is added as counterions to maintain overall
electric neutrality. Results from this set of models are provided 
in SI Appendix-figure~5.
The bands representing sampling uncertainties in the correlation
function plots in maintext figure~8b--e and 
SI Appendix-figure~5 (top) and SI Appendix-figure~6a,b 
are standard deviations across eight 
independent simulations.
If we take the model Kuhn length $b$ in FTS as the C$_\alpha$-C$_\alpha$ 
virtual bond length $\approx 3.8$ \AA~of polypeptides, 
a unit bead concentration of $b^{-3}$ is equivalent to $\approx$ 30 M. 
Because (ATP-Mg)$^{2-}$ is modeled by six beads (figure~8a of maintext), 
a model bead concentration of (ATP-Mg)$^{2-}$
(denoted as [(ATP-Mg)$^{2-}$] in our FTS results) which is 
equal to $b^{-3}$ reported
for the present FTS results is equivalent to a molar concentration of 
$\approx 5$ M of (ATP-Mg)$^{2-}$. As mentioned above, since excluded volume 
is often significantly underestimated in FTS,\cite{Pal2021}
we do not directly compare FTS model (ATP-Mg)$^{2-}$ concentrations
with experimental (ATP-Mg)$^{2-}$ concentrations, which tend to be 
substantially lower. Instead, physical insights are gleaned from
the trend of variation of model concentrations.
\\

{\bf FTS models for Caprin1/pY-Caprin1 with 
(ATP-Mg)$^{\bf 2-}$, Na$^{\bf +}$ and Cl$^{\bf -}$.}
We use this set of models for the results in maintext figures~8 and 9. 
Simulations are performed at $l_{\rm B} = 7b$ on a periodic
$32\times32\times32$ grid (see above) with CL time step $\Delta t=0.005$.
Again, all ATP$^{4-}$s and Mg$^{2+}$s are assumed to be in the complex
(ATP-Mg)$^{2-}$ form with charge sequence (-1 -1 -1 -1 +1 +1)
as depicted in maintext figure~8a.
Bulk Caprin1 and pY-Caprin1 bead densities in the their respective 
simulation systems are both set at $0.1b^{-3}$.
Three concentrations of (ATP-Mg)$^{2-}$ are studied:
[(ATP-Mg)$^{2-}$]= $0.0001b^{-3}$, $0.03b^{-3}$, and $0.5b^{-3}$. 
With overall electric neutrality of the simulation system in mind,
for Caprin1 (WT), the bulk (overall) densities for Na$^+$ and Cl$^-$ are
[Na$^+$] = 4[(ATP-Mg)$^{2-}$]/6 and 
[Cl$^-$] = 2[(ATP-Mg)$^{2-}$]/6 + 13[Caprin1]/103.
For pY-Caprin1,
[Na$^+$] = 4[(ATP-Mg)$^{2-}$]/6 + [pY-Caprin1]/103 and
[Cl$^-$] = 2[(ATP-Mg)$^{2-}$]/6.
\\

{\bf FTS models for Caprin1/pY-Caprin1 with ATP$^{\bf 4-}$, 
Mg$^{\bf 2+}$, Na$^{\bf +}$ and Cl$^{\bf -}$.}
In contrast to the above models, here
we consider ATP$^{4-}$ and Mg$^{2+}$ as independent components. That is,
they can freely dissociate if the favorable electric interaction between
them is insufficiently strong. In this set of models,
ATP$^{4-}$ is taken as a four-bead charge sequence (-1 -1 -1 -1)
whereas Mg$^{2+}$ is modeled by a single bead with charge $2+$ [instead
of the two $-1$ beads in the (ATP-Mg)$^{2-}$ model in maintext figure~8a]. 
As before, bulk (overall) 
Caprin1 and pY-Caprin1 bead densities in the their respective
simulation systems are both set at $0.1b^{-3}$, and the same three
[(ATP-Mg)$^{2-}$] = $0.0001b^{-3}$, $0.03b^{-3}$, and $0.5b^{-3}$ are
studied.
For Caprin1 (WT), the bulk (overall) 
densities for Mg$^{2+}$, ATP$^{4-}$, Na$^+$ 
and Cl$^-$ are given by [Mg$^{2+}$] = [ATP$^{4-}$]/4, 
[Na$^+$] = [ATP$^{4-}$], and [Cl$^-$] = 2[Mg$^{2+}$] + 13[Caprin1]/103.
For pY-Caprin1, [Mg$^{2+}$] = [ATP$^{4-}$]/4, 
[Na$^+$] = [ATP$^{4-}$] + [pY-Caprin1]/103, and [Cl$^-$] = 2[Mg$^{2+}$].
Results from this set of FTS models are provided in SI Appendix-figure~6.

\vfill\eject

\setcounter{figure}{0}
\renewcommand{\figurename}{{\bf SI}}
\renewcommand{\thefigure}{{\bf Appendix-figure\ }{\bf \arabic{figure}}}

\begin{figure}[ht]
\centerline{\Large\bf Supporting Figures}
$\null$\\
{\includegraphics[width=\columnwidth,angle=0]{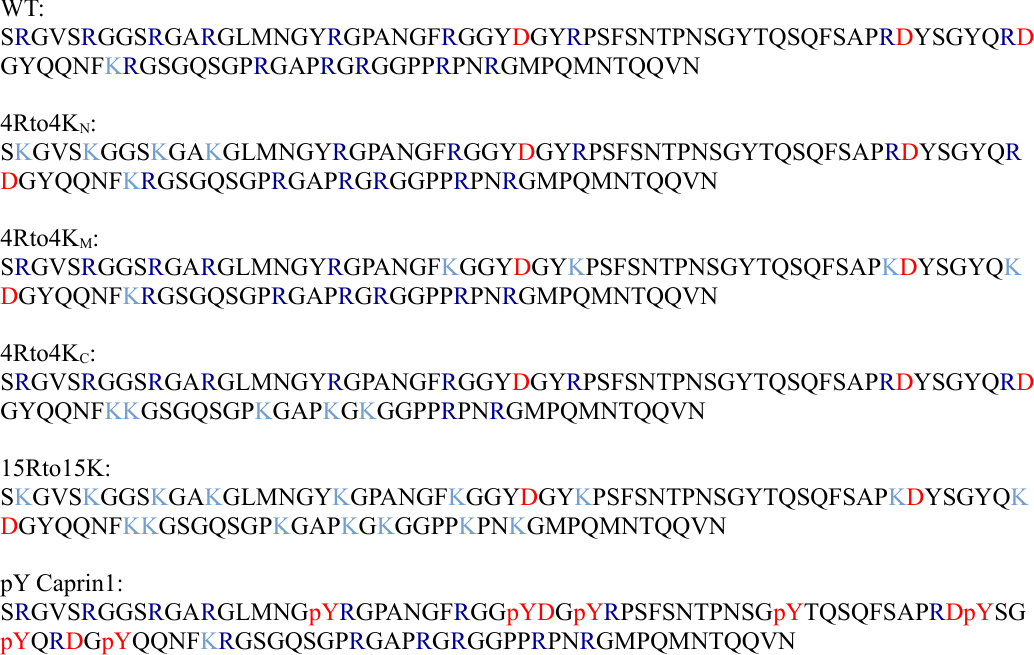}}
\caption{Sequences of wildtype (WT) and variant Caprin1 IDRs studied
in this work. Positively charged arginine (R) and lysine (K) residues
are shown, respectively, in dark and light blue, negatively charged
aspartic acid (D) residues and phosphorylated tyrosines (pY) are shown
in red. Other residues are in black.
}
\label{figS1}
\end{figure}

\vfill\eject

\begin{figure}[ht]
{\includegraphics[width=0.5\columnwidth,angle=0]{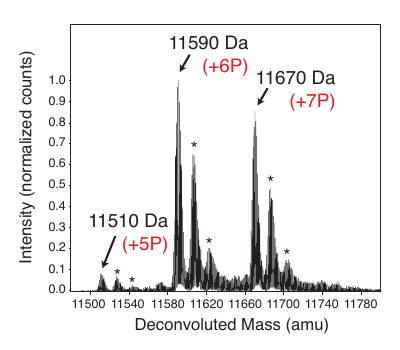}}
\caption{Mass spectrometry analysis of pY-Caprin1.  The graph plots 
deconvoluted mass (in atomic mass units, amu) on the horizontal axis 
against intensity (normalized counts) on the vertical axis. Peaks 
are observed at 11510 Da (+5 phosphate groups, +5P), 11590 Da
(+6P), and 11670 Da (+7P). Asterisks mark the peaks of pY-Caprin1
with oxidized methionine residues. 
}
\label{figS2}
\end{figure}

\vfill\eject

\begin{figure}[ht]
{\includegraphics[width=\columnwidth,angle=0]{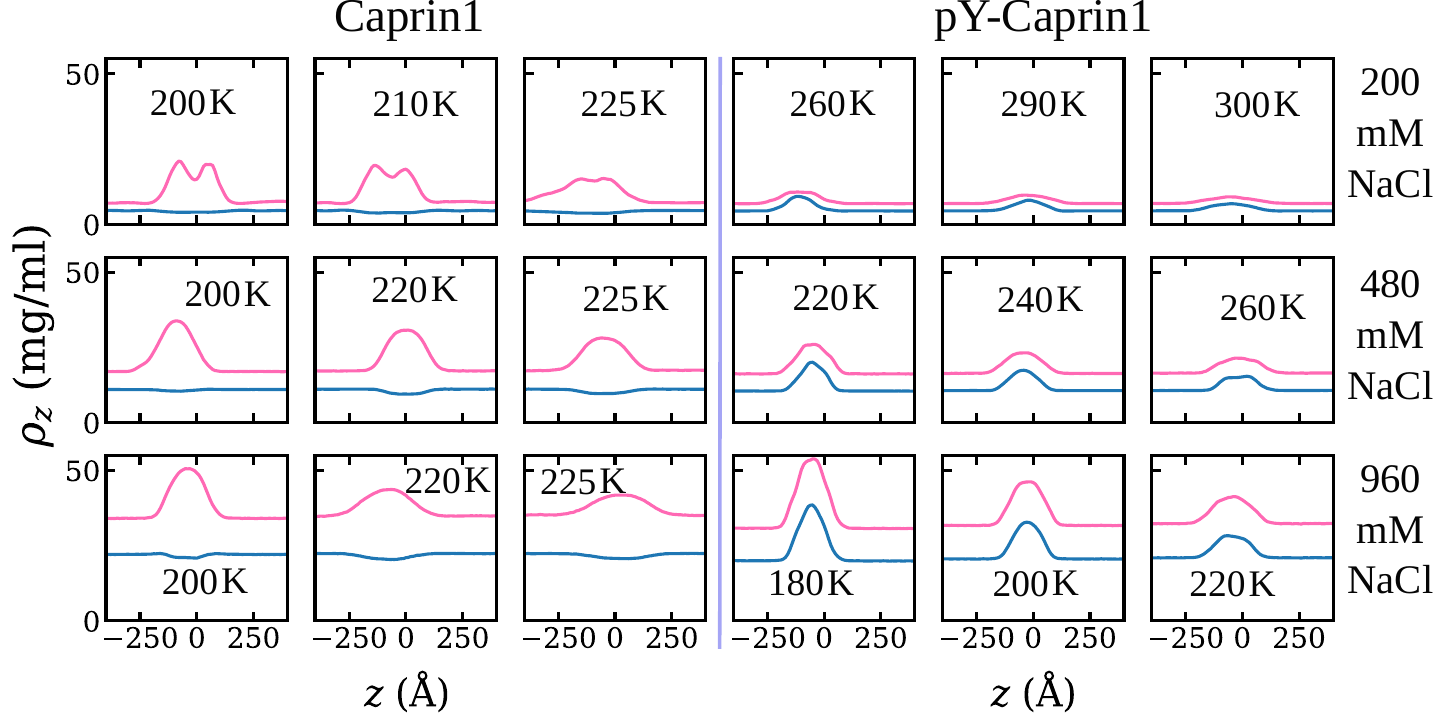}
}
\caption{Explicit-ion coarse-grained molecular dynamics simulation of
salt and counterion mass density profiles in protein-dilute and
protein-condensed phases of Caprin1 and pY-Caprin1.
Mass density profile $\rho_z$ for Na$^+$ (blue) and Cl$^-$ (red)
in units of mg/ml
for the Caprin1 (three left columns) and pY-Caprin1 (three right columns)
systems are shown as in figure~4e,f of the maintext. Regions with elevated
[Cl$^-$] here coincide with positions of the condensed protein
droplets. Overall [NaCl] values used for the simulations are provided
on the right.
}
\label{figS3}
\end{figure}

\vfill\eject

\begin{figure}[ht]
{\includegraphics[width=0.75\columnwidth,angle=0]{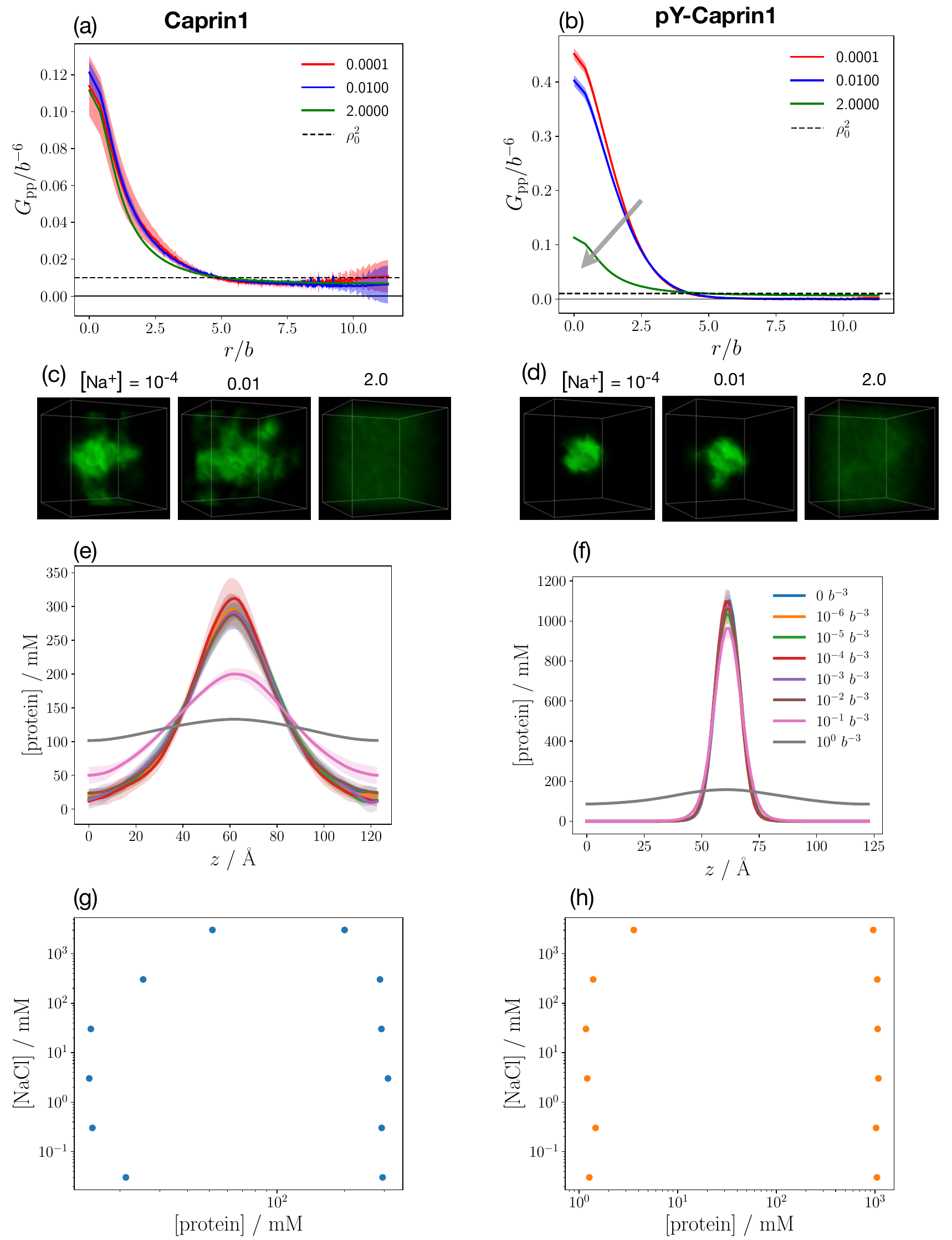}
}
\caption{FTS models for Caprin1 and pY-Caprin1 with only Na$^+$ and Cl$^-$
but no ATP-Mg.
(a,b) Protein-protein correlation functions 
[maintext Eq.~(7)] for Caprin1 (a) and
pY-Caprin1 (b) at three different [Na$^+$]s, color coded in units
of $b^{-3}$ as provided.
In each figure, the baseline value of protein-protein correlation
function $(\rho^0_{\rm p})^2$ 
(where $\rho^0_{\rm p}$ is overall protein concentration) is
marked by the horizontal dashed line. Phase separation is indicated
by large-$r$ correlation function values falling below this baseline.
The grey arrow in (b) marks the direction of increasing [Na$^+$].
(c,d) Field snapshots for the Caprin1 (c) and pY-Caprin1 (d) systems
at different [Na$^+$]s.
The above results are obtained at Bjerrum length $l_{\rm B}=7b$.
(e--h) Results from an alternate FTS model using an elongated simulation
box similar to that utilized for our explicit-ion coarse-grained MD.
(e,f) Protein concentration profiles computed at different NaCl concentrations
for Caprin1 (e) and pY-Caprin1 (f) [color code for density profiles
provided in (f)]. (g,h) salt-protein phase diagrams obtained from
the concentration profiles in (e,f) for Caprin1 (g) and pY-Caprin1 (h).
}
\label{figS4}
\end{figure}

\vfill\eject

\begin{figure}[ht]
{\includegraphics[width=\columnwidth,angle=0]{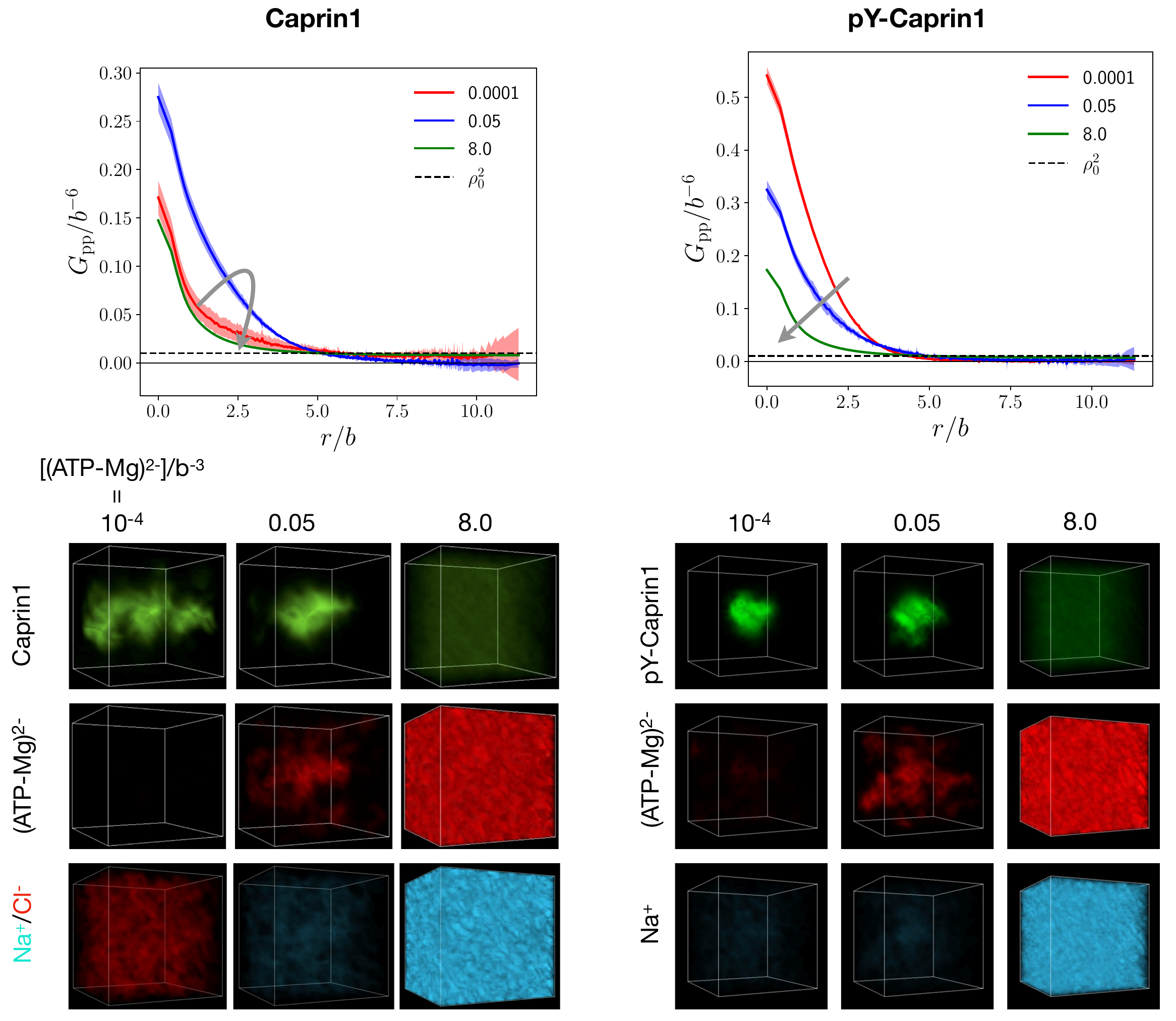}
}
\caption{Alternate FTS models for Caprin1 and pY-Caprin1
with (ATP-Mg)$^{2-}$ and either Na$^+$ or Cl$^-$
(but not both) to maintain overall electric neutrality.
For Caprin1, which has a net positive charge, (ATP-Mg)$^{2-}$ is the counterion.
Depending on [(ATP-Mg)$^{2-}$], either Cl$^-$ is included as an additional
counterion (when [(ATP-Mg)$^{2-}$] 
is insufficient to balance the positive charges
on Caprin1), or Na$^+$ is included as salt ion (when [(ATP-Mg)$^{2-}$]
overcompensates the positive charges on Caprin1). Na$^+$ and Cl$^-$ are
not included together in this simplified formulation. For pY-Caprin1,
which has a net negative charge, Na$^+$ is used as counterion, and its
concentration depends on [(ATP-Mg)$^{2-}$] in such a way that electric
neutrality of the entire system is maintained.
Top panels: protein-protein correlation functions at different
concentrations of (ATP-Mg)$^{2-}$ (color coded in units of $b^{-3}$ 
as provided).
Horizontal dashed lines are $(\rho^0_{\rm p})^2$ baselines as in 
SI Appendix-figure~4a,b. Grey arrows indicate increasing [(ATP-Mg)$^{2-}$].
Bottom panels: Field snapshots for system components at different
[(ATP-Mg)$^{2-}$] (as indicated) for the Caprin1 (left panels)
and pY-Caprin1 (right panels) systems.
Results here are obtained at $l_{\rm B}=7b$.
}
\label{figS5}
\end{figure}

\vfill\eject

\begin{figure}[ht]
{\includegraphics[width=\columnwidth,angle=0]{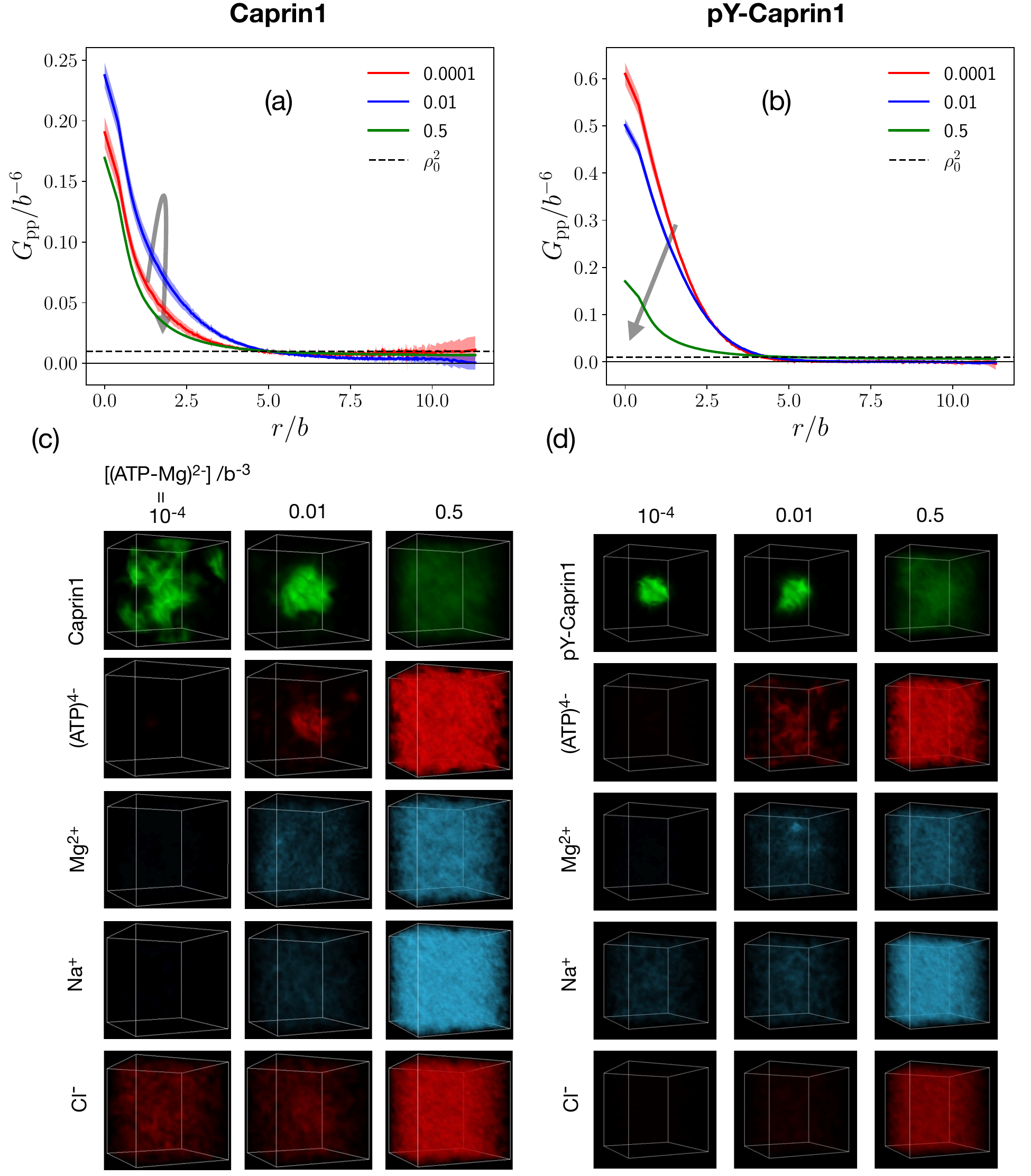}
}
\caption{Alternate FTS models for Caprin1 or pY-Caprin1 with
ATP$^{4-}$, Mg$^{2+}$, Na$^+$ and Cl$^-$, wherein (ATP-Mg)$^{2-}$ is assumed to
be fully dissociable.
Results are obtained for $l_{\rm B}=7b$ and presented in the same style
as that in figure~8b,c and figure~9 of
the maintext as well as SI Appendix-figure~5.
}
\label{figS6}
\end{figure}

\vfill\eject


\vfill\eject

\end{document}